\documentclass{article}

\usepackage{arxiv}

\usepackage[utf8]{inputenc} % allow utf-8 input
\usepackage[T1]{fontenc}    % use 8-bit T1 fonts
\usepackage{newtxtext,newtxmath}
\usepackage{hyperref}       % hyperlinks
\usepackage{url}            % simple URL typesetting
\usepackage{booktabs}       % professional-quality tables
\usepackage{amsfonts}       % blackboard math symbols
\usepackage{microtype}      % microtypography
\usepackage{amsmath}
\usepackage{cleveref}       % smart cross-referencing
\usepackage{graphicx}
\usepackage[square,numbers]{natbib}
\usepackage{doi}
\usepackage{xcolor}
\usepackage{multirow}

\usepackage{makecell}

\title{DRUG-TARGET INTERACTION/AFFINITY PREDICTION: DEEP LEARNING MODELS AND ADVANCES REVIEW}

\author{Ali Vefghi \\
        Department of Mathematics and Computer Science\\
        Amirkabir University of Technology\\
        Tehran, Iran\\
        \texttt{alivefghi@aut.ac.ir} \\
        \And
        Zahed Rahmati\thanks{Corresponding author} \\
        Department of Mathematics and Computer Science\\
        Amirkabir University of Technology\\
        Tehran, Iran\\
        \texttt{zrahmati@aut.ac.ir} \\
        \And
        Mohammad Akbari \\
        Department of Mathematics and Computer Science\\
        Amirkabir University of Technology\\
        Tehran, Iran\\
        \texttt{akbari.mo@aut.ac.ir} \\
}

\renewcommand{\headeright}{\textit{arXiv} Template}
\renewcommand{\undertitle}{}
\renewcommand{\shorttitle}{DTI/DTA Prediction: Deep Learning and Advances Review}

\hypersetup{
pdftitle={DRUG-TARGET INTERACTION/AFFINITY PREDICTION: DEEP LEARNING MODELS AND ADVANCES REVIEW},
pdfsubject={cs.LG},
pdfauthor={Ali Vefghi, Zahed Rahmati, Mohammad Akbari},
pdfkeywords={drug-target interaction prediction, drug-target affinity prediction, Drug discovery},
}

\begin{document}
\maketitle

\begin{abstract}
Drug discovery remains a slow and expensive process that involves many steps, from detecting the target structure to obtaining approval from the Food and Drug Administration (FDA), and is often riddled with safety concerns. 
Accurate prediction of how drugs interact with their targets and the development of new drugs by using better methods and technologies have immense potential to speed up this process, ultimately leading to faster delivery of life-saving medications.
Traditional methods used for drug-target interaction prediction show limitations, particularly in capturing complex relationships between drugs and their targets.
As an outcome, deep learning models have been presented to overcome the challenges of interaction prediction through their precise and efficient end results.
By outlining promising research avenues and models, each with a different solution but similar to the problem, this paper aims to give researchers a better idea of methods for even more accurate and efficient prediction of drug-target interaction, ultimately accelerating the development of more effective drugs.
A total of 180 prediction methods for drug-target interactions were analyzed throughout the period spanning 2016 to 2025 using different frameworks based on machine learning, mainly deep learning and graph neural networks. 
Additionally, this paper discusses the novelty, architecture, and input representation of these models.
\end{abstract}


%% main text
\section{Introduction}

\begin{figure}
  \centering
  \includegraphics[width=\columnwidth]{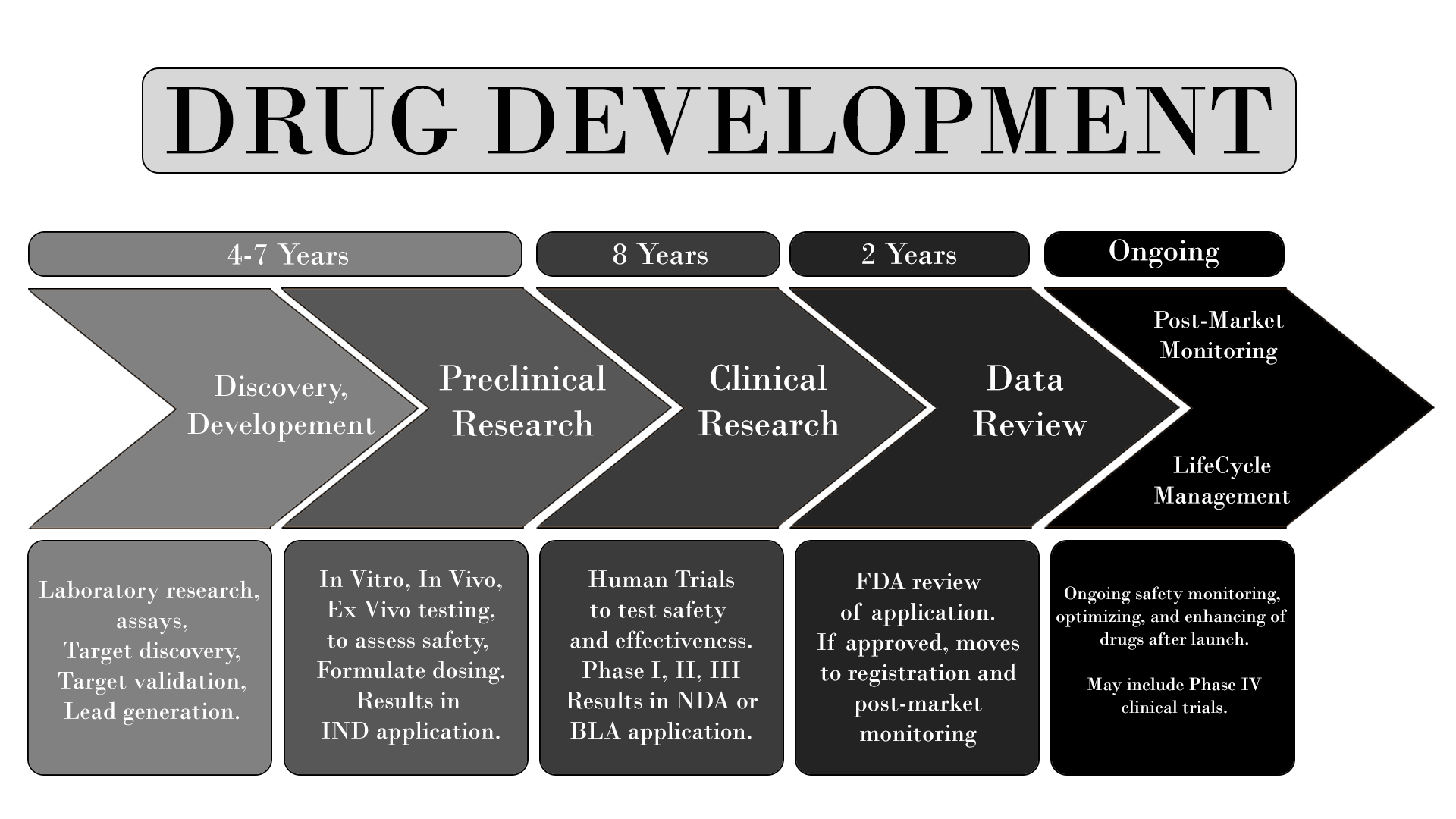}
  \caption{Overview of drug development process}
  \label{fig:fig1}
\end{figure}

Developing drugs is a slow and complex process which involves multiple stages of research, development, and testing \footnote{\href{https://www.bioagilytix.com/solutions/phases/discovery-phase-drug-development/}{bioagilytix.com}}. 
These stages of development aims to ensure the safety, efficacy, and quality of pharmaceutical products before public access and patient use \cite{10.3389/fddsv.2023.1201419}. 
The diagram presented in Figure \ref{fig:fig1} shows this process and its phases.
The initial step of drug development is usually drug discovery, where potential drug candidates are identified through different stages like identifying a specific target that binds to a chemical compound, determining the lead compound of the binding chemical and lead optimization to enhance the efficiency as well as the specificity \cite{lavecchia2013virtual, yao2010novel}. 
A discovery leading to approved medicine takes about 12 to 15 years and requires total investment of 1.8 billion US dollars, and many potential drugs never make it to market due to safety concerns, lack of efficacy, or other factors \cite{deore2019stages, paul2010improve, chen2013semi}.
On average, a million molecules undergo screening but the final product ends up being just one drug that reaches late-phase clinical trials and becomes available for patients.
Once promising candidates are identified, pre-clinical testing assesses safety, efficacy, and pharmacokinetics in animal models. 
Clinical trials become the next step once the drug demonstrates potential for approval by testing it on human volunteers. 
These trials are typically divided into three phases: Phase I evaluates safety and tolerability, Phase II assesses efficacy and determines the optimal dosage, and Phase III confirms effectiveness and safety in a larger patient population. 
The drug manufacturer sends the New Drug Application (NDA) to the FDA for review after concluding successful clinical trials.  
The FDA evaluates the NDA, including data from pre-clinical and clinical studies, to determine whether the drug meets safety and efficacy standards. 
If approved, the drug can be marketed and prescribed to patients \cite{https://doi.org/10.1111/j.1476-5381.2010.01127.x}.  

Drug-target specificity is a crucial factor in drug design which refers to the ability of a drug to be selectively bind to its intended target, while minimizing interactions with other targets.
Multiple research suggests that certain drugs interact with more than one target sites, known as poly-pharmacology \cite{lu2012multi} and also some diseases have shown that they need multiple factors to be manipulated simultaneously in order to be treated effectively, which has led to the development of drug repositioning.

Drug re-purposing or repositioning \cite{dudley2011exploiting, swamidass2011mining, moriaud2011identify, kumar2025knowledge} is a promising technique that helps significantly speed up the drug development process, reduce time-to-market, and lower financial costs by finding novel uses outside the scope of the original medical indication for existing drugs after FDA approval or abandoned drugs before FDA approval.
The main point is that the already known drugs or approved drugs mostly have identified side effects and safety of their usage is known, therefore it can accelerates the study of these drugs.
Understanding how marketed drugs interact with novel targets and predicting these interactions with high precision can guide researchers in selecting the best potential candidates and optimizing their use for specific applications. \cite{strittmatter2014overcoming}

The pharmacological principle also suggests that a medicine which has therapeutic effects on one target can have side effects on multiple other targets at the same time.
Studying these phenomenon lead to a method called Drug side effect profiling \cite{lounkine2012large, pauwels2011predicting} which is a critical research area according to \cite{giacomini2007good}.

Drug-target interaction, affinity prediction (DTI/DTA), identifying binding sites and interaction types are crucial tasks in drug discovery, facilitating the identification of new therapeutic agents, optimizing existing ones, and assessing the interaction potential of various molecules for disease treatment \cite{hopkins2009predicting}. 
Desired therapeutics, target specificity, long residence, and drug resistance delay can be obtained from the intensity of binding between a drug and its target protein, therefore, its prediction is considered a very important task in drug discovery \cite{copeland2006drug}. 

Classic experimental methods are time-consuming and expensive, making computational approaches increasingly valuable. 
Machine learning algorithms and AI techniques have emerged as promising tools for DTI prediction, leveraging large-scale datasets and complex computational models to accurately predict interactions between components and their targets \footnote{\href{https://blog.biostrand.ai/ai-in-early-phase-drug-development}{biostrand.ai}}.

Many other studies have been conducted to review or survey the literature on DTI/DTA prediction.
Some of these studies are focused on specific methods like graph neural networks (GNNs) \cite{zhang2022graph, zhang2023survey}, while others are focused on broader scopes like machine learning and computational methods \cite{shi2023review, suruliandi2024drug, chen2016drug}.
Zhang et al. \cite{zhang2024benchmark} proposed a combo of models for DTI prediction and compared various models on different datasets.
Sachdev et al. \cite{sachdev2019comprehensive} reviewed the feature-based methods for DTI prediction.
Jung et al. \cite{jung2020survey} reviewed the literature on network-based DTI prediction methods.
Some other studies are focused on component-protein interactions (CPI) \cite{lim2021review} or protein-ligand interaction (PLI) \cite{zhao2022brief, abdelkader2024advances, michels2024natural, wang2024prediction}.
Closest study to our work are the studies by Zeng et al. \cite{zeng2024comprehensive} and Abbasi et al. \cite{abbasi2021deep} (focused on the architectures) which are comprehensive reviews of the literature on deep learning models for prediction of drug target affinity.
Our paper expands on their work by providing a more comprehensive review of the literature on DTI/DTA prediction, focusing on the input representations and more deep learning models used throughout the history of DTI/DTA prediction research.

This paper offers a clear perspective and categorization of nearly over 180 methods presented from 2016 to 2025 on the DTA/DTI problem and serves as a preliminary overview guide for researchers who want to have a general idea of what the models and solutions to DTA/DTI are.
Models are categorized into different categories based on their input representations including sequence-based, structure-based, sequence-structure-based, complex-based, and utility-based models.
Supplementary data which includes Metrics, datasets, baselines, and other information of all models are available here.\footnote{\href{https://github.com/agmlcenter/Bioinformatics/blob/main/DrugTargetInteraction/Review}{github.com/agmlcenter/Bioinformatics}}

The structure of paper is as follows: 

\begin{itemize}

\item Section 1 provides an introduction to the drug discovery process and the importance of DTI/DTA prediction.
\item Section 2 provides an overview of input representations and the definition of drugs, proteins, and the problem used in the literature. 
\item Section 3 presents the models used in the literature in different categories. 
\item Section 4 discusses the datasets and evaluation metrics which are most used in the literature. 
\item Section 5 discusses the challenges and future directions in the field and concludes the paper.

\end{itemize}

\section{Input Representations}

This section provides an overview of various input representations used in the literature from two aspects of the problem which are proteins and drugs.
The proper representation is a crucial part in analyzing and extracting the right features from proteins and drugs for learning the best pattern that the selected features can offer.
Existing methods may use different representations for drugs and proteins or both can be represented in the same way. 
For example, One side can use sequence representation while the other side use graph representation, and then train a model on both representations.

\begin{figure}
  \centering
  \includegraphics[width=\columnwidth]{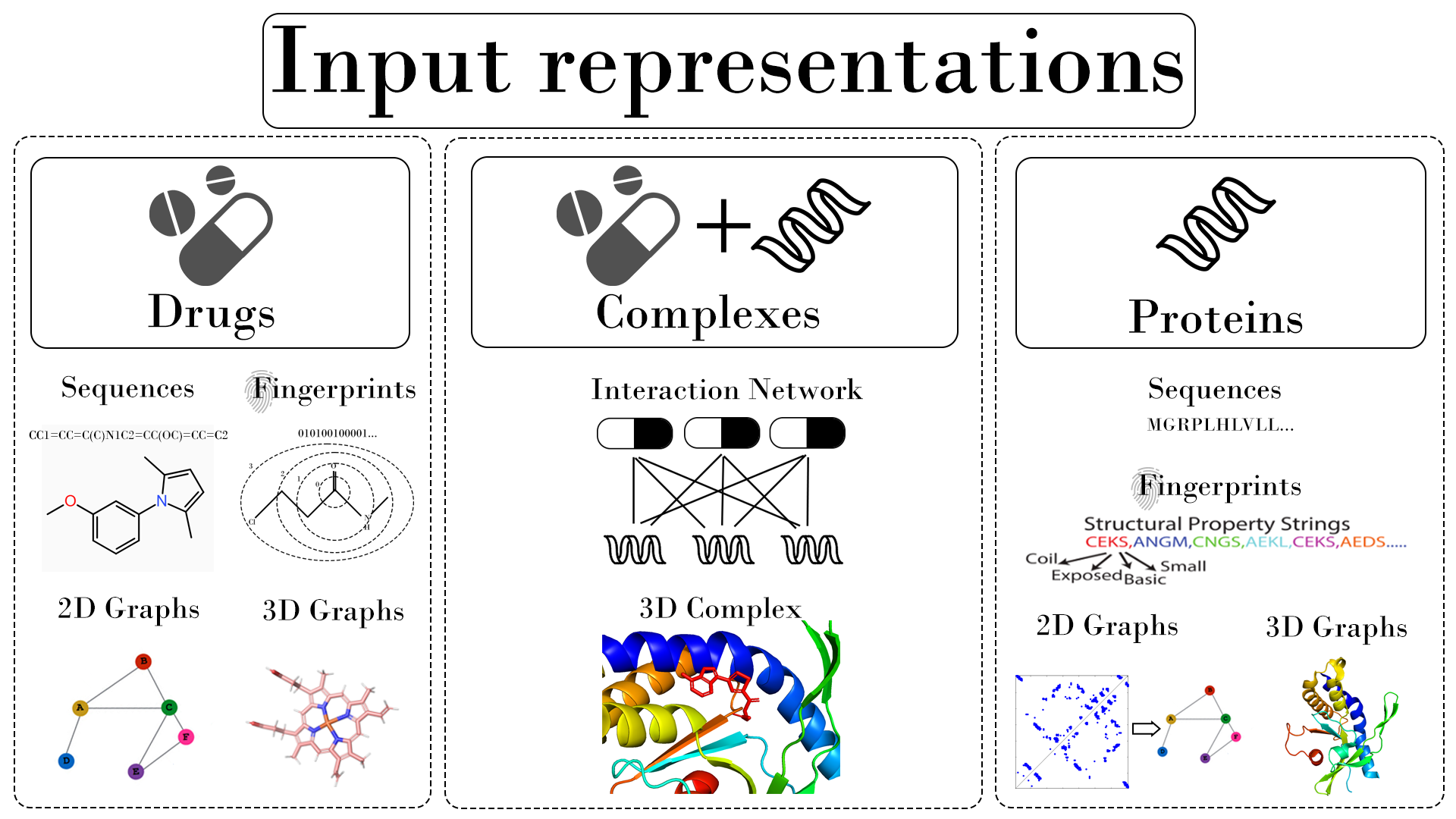}
  \caption{Overview of various input representations}
  \label{fig:fig2}
\end{figure}

As Figure. \ref{fig:fig2} shows, inputs of DTI/DTA models can be string of sequences, graphs, vectors of information (e.g. physico-chemical properties) or any other form of data.
The most common input representations for drugs are Simplified Molecular Input Line Entry System (SMILES), fingerprints, and molecular graphs, while for proteins, the most common representations are primary sequences, fingerprints, and protein graphs.
To feed these inputs into the interaction learning models, encoding modules exist in most of models to convert the input data into a form that the model can understand.

% In some approaches, encoding the structure and features of drug molecules or proteins are encoded as points in low-dimensional vectors which are obtained through training deep learning models. 
% These learning models try to learn high level features, nonlinear relations, and structural information from drug molecules and represent them in a lower dimension embedding space.
% This data-driven approach allows the models to learn and generalize to a variety of molecular structures, resulting in broadening their scope.

There are lots of programs and software that can be used to manipulate and visualize drugs or proteins, e.g. PyMOL \cite{delano2002pymol} is a commonly used software.

\subsection{Drugs as ligands / compounds}

Drugs are often micro-molecules that interact with cellular constituents by causing a change in constituent function that results in an overall physiological response intended for therapeutic reasons.
Drugs can activate a particular cell function in the body (Agonist), block a function (Antagonist), or even do the opposite of a function (Inverse agonist).
Drug action describes the influence of a drug molecule on our cells in which it binds with at least one of the target's binding sites.
Individual drug classes bind to specific targets and individual targets recognize only specific classes of drugs, while no drug acts with complete specificity.
Drugs are also referred as ligands which is just an expression for the small drug molecules attaching to proteins.
In general, Molecules that bind to a target are called ligands, therefore, all drugs are ligands, but not all ligands are drugs.
Examples of other types of ligands other than drugs are endogenous ligands (like hormones, neurotransmitters, and other signaling molecules), toxins, and environmental pollutants (like heavy metals).

\subsubsection{Drug sequences}

Simplified Molecular Input Line Entry System (SMILES) \cite{weininger1988smiles} is a commonly used representation for drugs which can be obtained and processed via the RDKit \cite{landrum2006rdkit} library.
SMILES is a common notation system for drugs that uses a canonicalization algorithm to generate unique ASCII string notations for each drug in respect to the three-dimensional structure of drugs.
In another words, SMILES of a drug consists of a series of characters representing atoms and bonds. 
With its compact nature and simple data handling system, SMILES provides standardized and unique molecular descriptions.
% There can be several equally valid SMILES strings for a molecule, e.g. CCO, OCC and C(O)C all specify the structure of ethanol. 
% SMILES is unique for each structure, although dependent on the canonicalization algorithm used to generate it. 
The canonicalization algorithms first convert the SMILES to an internal representation of the molecular structure and then produces a unique SMILES string with another algorithm from that structure.
One way of acquiring SMILES is through PubChem compound database based on the drug PubChem CIDs(or CHEMBL IDs) \cite{bolton2008pubchem}.
 
SELF-referencIng Embedded Strings (SELFIES) \cite{krenn2020self} are another notation system which are introduced so that all of the tokens or words of the SELFIES correspond to chemically valid molecules, while SMILES respective characters do not always correspond to chemically valid molecules.

SMILES Arbitrary Target Specification (SMARTS) \cite{schmidt2019comparing} is a powerful extension of the SMILES notation system, enabling the definition of molecular patterns beyond the representation of individual molecules. 
While SMILES provides a linear representation of a complete chemical structure, SMARTS allows for the specification of sub-structural motifs, functional groups, and other chemical features within a molecule. 

Embedding of SMILES strings can be obtained with pre-training Word2Vec \cite{mikolov2013distributed}, use a pre-trained model like Mol2Vec \cite{jaeger2018mol2vec}, or simply encode SMILES with One-hot encoding.
Mol2Vec is an unsupervised learning method that generates vector representations of molecular substructures, drawing inspiration from word embedding in natural language processing. 
By training on a large corpus of molecules, it learns to encode chemical information, capturing relationships between substructures within a continuous vector space.

\subsubsection{Structural keys: Fingerprints}

Molecular fingerprints are conventional methods for encoding chemical compounds. 
They represent molecules as unique identifiers based on the presence or absence of specific molecular features within a defined radius around each atom. 
These features can include Motifs (specific substructures), functional groups, and other chemical properties.
One common approach is to represent these features as a binary vector, where each bit corresponds to a specific feature. 
If a molecule possesses a particular feature, the corresponding bit is considered one; otherwise, it is zero.
Fingerprint methods can also incorporate more complex representations, such as feature counts or continuous values. 
Some commonly used fingerprints are Extended-Connectivity Fingerprints (ECFPs)/Functional-Class Fingerprints (FCFPs)/Morgan fingerprint \cite{rogers2010extended}, ErG fingerprint \cite{stiefl2006erg}, Daylight fingerprint \cite{toolkit2007daylight}, MACCS Keys \cite{durant2002reoptimization}, and The PubChem fingerprint \cite{pubchem} which can be acquired via RD-Kit.

Morgan fingerprints are a class of circular fingerprints that encode the local chemical environment surrounding each atom within a molecule.
These fingerprints are generated by iteratively expanding a circular neighborhood around each atom, considering neighboring atoms and their bond types.
At each iteration, the atom's environment is hashed into an integer, representing a unique fingerprint. 
This process captures the structural complexity of the molecule within a fixed-size bit vector, enabling efficient comparison and analysis. 

ECFPs are a specific type of Morgan fingerprint that has become a widely used standard in chemo-informatics.
ECFPs are generated by extending the circular neighborhood around each atom to a specified radius, typically two or three bonds. 
This approach captures more extensive structural information compared to simpler fingerprints, including longer-range interactions and more complex sub-structural motifs.

Another fingerprint that can be used for compounds is algebraic graph-based fingerprints (AG-FPs), which generate low-dimensional molecular representations with   element-specific weighted colored algebraic graphs while preserving essential physical/chemical information and physical insight \cite{chen2021algebraic}.

PaDEL-Descriptor \cite{yap2011padel} is a software using Java language to calculate the  chemical small molecules descriptors.
It includes 134-dimensional 3D descriptors, 1444-dimensional 1D, 2D descriptors, and 10 various fingerprints.

\subsubsection{Two-dimensional molecular topology graphs}

In this kind of representation, a 2D molecular graph is used which captures both the chemical and topological information (molecular formula, atomic type, bond type, and spatial arrangement). 
Atoms and chemical bonds are represented as nodes and edges of the graph in this form.
To generate the node feature vectors in the drug graph physical-chemical properties of atoms or SMILES can be used.
Graphs can be fed into the models in the form of adjacency matrices, node features, and edge features.

Some studies use R-radius sub-graphs as input representation of drugs \cite{costa2010fast}. 
R-radius sub-graphs are graphs where the graph's vertices are the aggregation of all neighboring vertices and edges information within a radius r.

Chemistry software tools such as DGL-lifeSci \cite{li2021dgl} (or DGLGraph \cite{wang2019deep}), Open Babel \cite{o2011open}, and RDKit can be employed to convert SMILES to molecular graphs (using structure diagram generation (SDG) algorithms \cite{helson1999structure}) and extract useful features including, chemical bonds, hydrogen presence, electron properties, whether the atom is aromatic, and so on.

\subsubsection{Three-dimensional spatial structures}

In 3D structural format, atoms serve as vertices and bonds act as edges in a 3D graph.
Like drug 2D graphs, node features can be calculated from SMILES or other properties.

Researchers aim for representing drug molecules to be as accurate as possible while capturing their structural information. 
In these methods, the mesh representation splits 3D space into regular mesh units that function as volume elements.
With the point cloud \footnote{A point cloud is a set of data points in a three-dimensional coordinate system defined by X, Y, Z coordinates.} representation method the three-dimensional molecular structure gets converted into a collection of discrete points where each point shows the position of atom elements.
Voxel representation divides three-dimensional space into a series of small cube units, which are called voxels and maps atoms in molecules to the nearest voxel. 

Also, there are new ways to represent molecules in three-dimensional space like in \cite{MILON2024108117} which is a novel representation of drug 3D structures and enhancement of the TSR-based method for probing drug and target interactions.

\subsection{Proteins as targets}

Drug-targets are mostly proteins macro-molecules like receptors (controlling signaling in the body), enzymes (catalyzing chemical reactions), carrier molecules (transporters like hemoglobin), and Ion channels. 
In this paper, we consider targets as proteins, which are the most common drug targets.
Proteins exist as linked amino acids arranged in particular orders like a chain. 
The molecular unit called amino acid functions as the basic structural component that builds proteins.
Twenty main types of amino acids are found in the proteins of living things, and the properties of a protein are determined by its particular amino acid sequence.

Proteins have four levels of structure: primary, secondary, tertiary, and quaternary.
The simplest level of protein structure, known as primary structure, is simply the sequence of amino acids in a poly-peptide chain. 
The sequence of a protein is determined by the DNA of the gene that encodes the protein.
The next level of protein structure, secondary structure, refers to local folded structures that form within a poly-peptide due to interactions between atoms of the backbone. 
The most common types of secondary structures are the alpha helix and the Beta pleated sheet.
The overall three-dimensional structure of a poly-peptide is called its tertiary structure. 
The tertiary structure is primarily due to interactions between the R groups of the amino acids that make up the protein.
Lastly, protein sub-units come together and they give the protein its quaternary structure.

Proteins are a lot more complicated to represent comparing to drugs due to various reasons.
Some reasons are the lengthy sequence of proteins, different foldings, different conformations \footnote{the spatial arrangement of its constituent atoms, which determine the overall shape of the macro-molecule.}, and vastness of combinations for proteins which are composed of different amino acids.
Another reason is the unknown protein structures which nowadays is to some extent handled via structure prediction models like Alpha-fold \cite{jumper2021highly}.  

A protein pocket is a concave surface region on a protein where small molecules, such as drugs, can bind. 
These pockets are essential for the protein's biological function. 
They often serve as active sites for enzymatic reactions or as binding sites for small molecules, including substrates, co-factors, and inhibitors.
Some pocket detection models are \cite{le2009fpocket, volkamer2010analyzing, zhu2011mspocket, abibi2014role, rooklin2015alphaspace, saberi2014simple} which are using 3D structure or other structures of proteins.
The binding sites generally possess large and flat areas with high hydrophobicity. \cite{katigbak2020alphaspace}. 
Additionally, large conformational changes in the binding region, such as the transition between open and closed states of the binding pocket, and the local secondary structure changes induced by the ligand, are associated with the ligand binding affinity \cite{yasuda2022differences}.
Some studies have used probe points named "anchors" describing protein pockets for better prediction of DTI \cite{li2023pocketanchor}.

\subsubsection{Protein sequences}

The most common approach of representing proteins involves the Protein sequences representation in which each protein is represented using the ordered text string of its different unique characters for each amino acid in the linear poly-peptide chain.
No further information is accessible through this representation other than the names of amino acids in text form and their order in the protein chain sequence, therefore there is a need for text feature extraction including natural language processing techniques in these representations \cite{michels2024natural}.
These methods take the amino acid sequence as input and try to capture sequence and structural patterns by learning the relative positions between amino acids in 1D representation to produce the best sequence embedding \cite{zhang2019review}.

Common pre-processing for Protein sequences, involves converting amino acid sequences into N-gram segments (biological words) \cite{pan2023submdta, dong2006application} or integers \cite{nguyen2021graphdta} sequences inspired by methods in NLP \cite{wang2020dipeptide}. 
Protein sequences are commonly encoded as one-hot or labels or other novel methods \cite{eddy2004did}.
One example of sequences is proposed by \cite{lipman1985rapid} which is called FASTA sequences (sequences of amino acids represented with specific characters).

For capturing informative features from the protein sequences different algorithms prior to the main model can be used.
For example, the Smith-Waterman algorithm compares these amino acid sequences to identify regions of similarity, which can indicate functional or evolutionary relationships between proteins.
Alignment-free algorithms like Lempel Ziv Markov chain algorithm (LZMA) \footnote{\url{http://www.7zip.org/7z.html}} can also be used.

There are other traditional methods to process and extract features from the Protein sequences of proteins such as position-specific score matrix (PSSM) \cite{altschul1997gapped}, and Hidden Markov Matrix (HMM) \cite{remmert2012hhblits}.
A PSSM is a matrix that represents the probability of amino acid occurrences at each position in protein sequences. \cite{mousavian2016drug}
It's derived from a multiple sequence alignment (MSA) of homologous proteins.
HMMs are statistical models that can be used to model biological sequences, including proteins.
They are particularly effective for identifying protein classes, families and domains, which are often defined by specific patterns in the Protein sequences.
Other popular protein sequence descriptors include Auto-correlation, CTD (Composition, Transition and Distribution) descriptor, Quasi-sequence order, and Protein Sequence Composition (PSC) \cite{cao2013propy}.

Some models leverage pre-trained Word2Vec \cite{mikolov2013distributed, quan2019graphcpi, wang2023ammvf, li2022bacpi, tsubaki2019compound, lin2020deepgs, cheng2022iifdti, xia2022drug} or pre-trained protein language models \cite{kim2021bayesian, bal2023pgraphdta} to make the embeddings more expressive.
Pre-trained Protein sequences protein models are trained on massive datasets of protein sequences, therefore, it is capable of predicting protein structures with high accuracy, even for challenging protein families \cite{elnaggar2021prottrans}.
ESM-2 \cite{lin2022language}, ProtBERT (BERT-based pre-trained model on protein sequences using a masked language modeling objective.) \cite{brandes2022proteinbert}, Prot2Vec \cite{asgari2015continuous}, TAPE (Tasks Assessing Protein Embeddings) \cite{rao2019evaluating}, and UniRep \cite{alley2019unified} are examples of these pre-trained Protein sequences protein models.

One of the access points for Protein sequences is UniProt protein database based on gene names/RefSeq accession numbers (or UniProt IDs) \cite{uniprot2018uniprot}.

\subsubsection{Protein fingerprints}

A protein fingerprint is a unique pattern that characterizes a specific protein or a group of related proteins. 
This fingerprint can be derived from various properties of the protein, including its amino acid sequence, its 3D structure, or its biochemical properties.
PROSITE and Pfam are two datasets that contain protein fingerprints.
PROSITE contains protein domains, families, and functional sites represented as patterns and profiles \cite{hulo2004recent}. 
Pfam contains a large collection of protein families, represented as multiple sequence alignments and hidden Markov models (HMMs) \cite{bateman2004pfam}.
Also, Structural property sequences (SPS) (Secondary Structure, Solvent Accessibility, Hydrophobicity/Hydrophilicity, Charge, Flexibility/Rigidity) can be generated via SSPro \cite{magnan2014sspro} from protein sequences.

Secondary Structure Elements (SSEs)  are the localized, regular arrangements of the poly-peptide backbone within a protein.
The most common SSEs are alpha helices and beta sheets, which are stabilized by hydrogen bonds between the backbone atoms.
SSEs and Physicochemical characteristics of Amino Acids (e.g. non-polar, polar, acidic, basic) can be used as protein fingerprints.

\subsubsection{Protein feature Graphs}

These methods represents proteins in graphical form and visual information, where nodes represent amino acids, atoms, or residues, and edges represent their real or custom made relationships.

One common graph that is used instead of the whole 3D graph of proteins is the contact map graph.
The whole protein structure produces complex graph topology together with prolonged training duration because proteins contain numerous atomic elements. 
Instead, a contact map (a form of adjacency matrix), which reflects the high-dimensional structure of a protein, can be used to reveal the interactions between protein residues (as they are less in number, only a few hundred) \cite{jiang2020drug, wang2022structure}. 
Generally, two Cb atoms of residues in Euclidean space are in contact if the distance between them is less than a certain threshold.
The structural features of Protein can directly be extracted from contact maps through CNN models or contact maps can be transformed into network topology to utilize GNNs.
These interactions between residues make the contact map, which can be acquired using Pconsc4 \cite{michel2019pconsc4} (uses a U-net architecture, which operates on the 72 features calculated from each position in the MSA) or evolutionary scale modeling (ESM) models \cite{rao2020transformer}.
ESM models can save alignment time by predicting contact maps without requiring protein sequence alignment.
After predicting the contact map a threshold is used to create the edges in graph.

Another graph that can be used is 2D pocket graphs which are constructed based on distances of amino acids in pocket regions of proteins.
These graphs are subsets of the whole protein graph and only contain the pocket area of the protein.

\subsubsection{Protein 3D structure}

Three dimensional graph of proteins can be used to represent proteins in a more detailed way.
Each node of the graph can be amino acids or atoms and edges can be the bonds between them, but this representation is not common due to the complexity of the structure and the vast number of atoms in proteins.

Protein 3D structure prediction is the process of predicting the three-dimensional structure of a protein from its amino acid sequence which is a challenging problem in drug discovery.
Traditionally, for acquiring the 3D structure of proteins, researchers use X-Ray Crystallography, Nuclear Magnetic Resonance (NMR), or 3D Electron Microscopy \cite{maveyraud2020protein, kaptein1988protein, esquivel2013computational}.
However, these methods are too slow and costly. 
Other computational approaches tried to solve this problem like homology modeling techniques (comparative modeling, protein threading and segmenting and etc.).
Crystallography makes better resolution than NMR but some proteins cannot be crystallized; therefore, NMR is needed which can present different conformations of proteins. 
The state-of-the-art computational method, Alpha-fold model, uses deep learning approaches to predict the 3D structure of proteins by utilizing attention heads and Multiple sequence alignment (MSA) on evolutionary similar Protein sequences \cite{jumper2021highly}.
Protein structure prediction with pre-trained models provides key information for CPI and DTA prediction, however, selecting a fast and accurate method to improve the efficiency is very important.

Another graph that is used recently is pocket graph representation of proteins.
In this representation, the protein is represented as a graph just in the pocket area and not the whole graph, therefore, the construction of graphs can be more manageable.

\subsection{Drug-target complexes}

In these form of representations drugs and targets are not treated individually, instead, they are considered as complexes and structures from crystallized forms.
Approaches based on this representation are limited to known protein ligand complex structures.
As these structures are often unavailable, they often need to be first predicted by docking individual structures of proteins and compounds together.

Features like each atom's element type, hybridization, bonding information, structural properties (hydrophobicity, aromaticity, hydrogen bonding), partial charge, molecule type (protein or ligand), and van der Waals radius can be extracted and used from the 3D structure of the complexes.
Open-Babel cheminformatic tool \cite{o2011open} can be used to extract these features from the 3D structure of the complexes.

\subsubsection{Interaction network graph representation}

The construction of interaction network graphs relies on three-dimensional drug-target complex structures. 
Researchers generate the graph by choosing particular atoms between the drug molecules and target proteins. 
Typically, they choose the carbon alpha atoms of the amino acids in the target protein, as well as all atoms from the drug molecule.
An edge is connected between the corresponding nodes in the graph if the distance between two atoms is less than a predefined threshold.
Any atoms or amino acids that are not involved in any interactions are removed.
This graph can then be used as input for models to predict the binding affinity or other properties of the drug-target interaction.

\subsubsection{3D structural spatial grid representation}

While interaction network graphs offer valuable insights, they may overlook certain interactions by focusing on a subset of atoms. 
To capture the complete spatial arrangement of all atoms, 3D spatial grid representations are employed. 
These representations, coupled with 3D Convolutional Neural Networks (3D-CNNs), can effectively extract key structural features important for drug-target binding.

\subsection{Formulating the drug-target Problem}

Given a set of drug molecules and a set of potential target proteins, predict whether a drug molecule will bind to a target protein and, if so, with what affinity.
For a regression task, if we have $ D = \{ d_1, d_2, ... , d_n \}$ for drug molecules and $ T = \{ t_1, t_2, ... , t_n \}$ for target proteins, for each $ i \in n $ and $ j \in m $, the aim is to predict $ A (d_i,t_j)$, which represents the binding affinity of drug $d_i$ to target protein $t_j$. 
A higher value indicates a stronger binding affinity. This regression task is called drug-target affinity prediction.
The problem can also be formulated as a binary classification task, where the objective is to predict whether a drug molecule will bind to a target protein $ A(d_i, t_j) > threshold $ or not $ A(d_i, t_j) <= threshold $. 
This task is called Drug-Target Pair prediction or Drug-Target Interaction prediction.
In Figure \ref{fig:fig3} binding affinity is a continuous number and binary interaction is a binary number which can be zero or one.

Binding affinities may be subject to biological constraints, such as steric hindrance or electrostatic interactions. 
It is more informative but also more challenging to predict the strength of the binding between a drug and its target. 
If the strength is not strong enough, such a drug-target pair may not be useful.

Protein-ligand interaction prediction is a related problem that focuses on predicting the interaction between a protein and a ligand, which is a small molecule that binds to a protein.
Component-protein interaction (CPI) is another related problem that focuses on predicting the interaction between a component and a protein, where the component can be a drug, a ligand, or a small molecule.
These tasks can be formulated as binary classification or regression tasks, similar to the drug-target interaction problem.

\subsubsection{Experimental settings}

\begin{figure}
  \centering
  \includegraphics[width=\columnwidth]{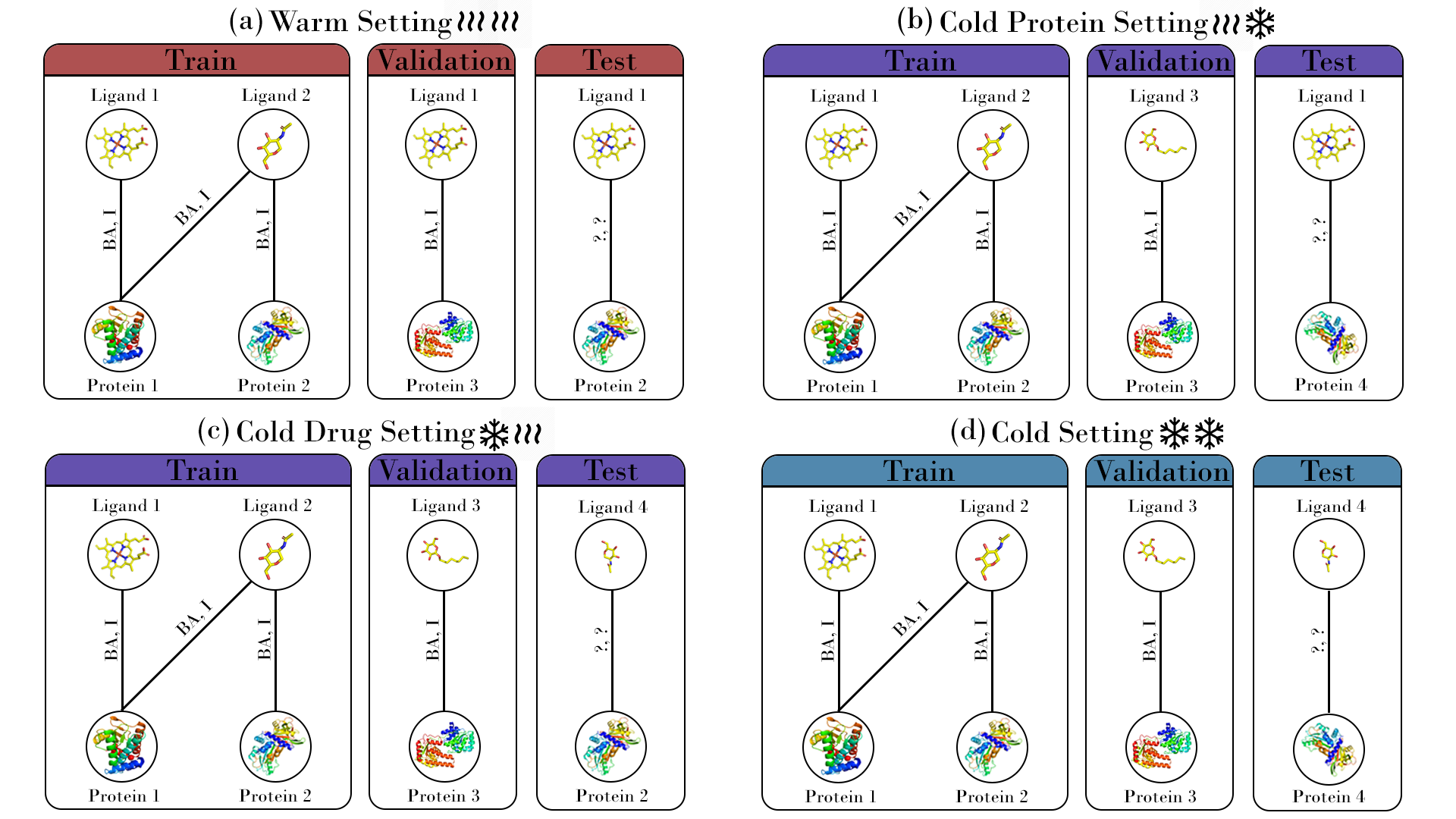}
  \caption{Overview of drug-target interaction problem in different settings. BA: binding affinity, I: binary interaction. (a) Warm setting, (b), Cold protein setting (c), Cold drug setting, (d) Cold drug-protein setting.}
  \label{fig:fig3}
\end{figure}

As shown in Figure \ref{fig:fig3}, there are four different experimental settings \cite{pahikkala2015toward} for drug-target interaction.
Warm setting is the same as the traditional setting where the training and test sets are randomly split.
Cold setting is the most challenging setting where the test set contains drug-target pairs that are not present in the training set.
Methods tried to test their frameworks in cold settings to evaluate the robustness and generalizability of their models. 

Using label reversal dataset \cite{zhao2022cpgl} is another technique to show how models generalize in unseen situations.
A ligand in the training set appears only in one class of interaction, whereas, in the test set, the same ligand appears only in the opposite class of interaction.

\begin{itemize}

\item Warm setting (Random split): Both drug and protein of pairs in test and validation sets can also be seen in the training samples.
Any protein or ligand may be repeated; however, interactions are not duplicated across the two splits.
\item Cold target setting (Orphan-target split): Test protein targets are not observed in the training set.
\item Cold drug setting (Orphan-drug split): Test drug compounds are not observed in the training set.
\item Cold setting: Both test drugs and protein targets are not observed during the training phase.

\end{itemize}

\subsubsection{Negative samples from DTA to DTI}

The designation of the negative (not-binding) samples is an important step that affects the model's performance (recognizing the best and real negative samples is challenging)\cite{lin2019learning}.
In the literature of drug discovery positive samples (pairs that exhibit interactions or have high affinity) are also referred to as active components and the negative samples (pairs that do not exhibit interactions or have low affinity) are sometimes referred to as decoys.
Several DTI studies use four major datasets (Enzymes (E), Ion-Channels (IC), G-protein-coupled receptors (GPCR), and nuclear receptors (NR) which drug-target interactions are acquired from KEGG BRITE, BRENDA, SuperTarget, and DrugBank) in which DT pairs with no known binding information are treated as negative samples \cite{yamanishi2008prediction}.
Some DTI studies that rely on databases with binding affinity information have been providing more realistic datasets created with a specific binding affinity threshold \cite{wan2016deep}. 
Formulating the drug-target prediction task as a binding affinity prediction problem enables the creation of more realistic datasets, where the binding affinity values are directly used and the challenge of selecting the best threshold is avoided.

\section{Methods}

In this section, an overview of the models used in the literature for DTI/DTA prediction is provided.
First, some traditional methods and categorizations are discussed, then the machine learning and deep learning models are presented.

\subsection{A brief history of traditional approaches}

The field of drug-target interaction has developed significantly throughout time.
Early research into drug-target interactions often relied on classical methods such as Chemical-biological experiments including wet lab experiments \footnote{A wet lab is one where drugs, chemicals, and other types of biological matter can be analyzed and tested by using various liquids.}. 
These methods contains laboratory experiments $in-vivo$ \footnote{Research done on a living organism} and $in-vitro$ \footnote{Research done in a laboratory dish or test tube}.
Most of these experiments are well known to be accurate but time consuming and costly, therefore Computational methods, in-silico and dry lab experiments are more desirable, as they can reduce the search space and lessen the required time and cost. \cite{ezzat2019computational}
As more compounds are being synthesized and new diseases are being introduced, the need for more accurate, powerful and efficient methods to handle large amounts of datasets and component properties rises.
The stage of drug development and the type of data are the main factors in determining in-silico methods that can be used. 
For example, Quantitative Structure Activity Relationships (QSAR) studies are conducted to determine the drug properties of the lead compound, by analyzing the relationship between the chemical structure and the biological activity. \cite{roy2015understanding}

\subsection{Computational methods}

Computational methods can be based on machine learning approaches and non-machine learning methods or hybrids of them.
Furthermore, The computational methods based on machine learning can be classified into traditional machine learning methods and deep learning methods.
Traditional machine learning methods use linear regression, random forest, nearest neighbor, and support vector machine to predict \cite{ballester2010machine, li2015low}.
These methods cannot learn the high-level features automatically, however, deep learning models can overcome this limitation \cite{rube2022prediction}.

Scoring models estimate the likelihood of a drug-target interaction. 
Models like AGL-Score (Algebraic Graph Learning Score) \cite{nguyen2019agl}, X-Score (based on empirical force field calculations) \cite{wang2002further}, ChemScore, ChemPLP \cite{khamis2015comparative, korb2009empirical}, AutoDock Vina, and AutoDock (calculates score by docking simulation) \cite{gaillard2018evaluation} are classic scoring models.
Newer scoring methods are RF-Score (Random Forest Score) \cite{stepniewska2018development}, kNN-Score, OnionNet, K-DEEP, Pafnucy.
Some of these models require significant computational resources and are limited.

Because of the simplicity of calculation and their close relationship with physics-based interactions, empirical scoring functions are still used. 
The empirical scoring functions have been implemented to various docking programs: DOCK \cite{moustakas2006development, lorber2005hierarchical}, AutoDock \cite{morris2009autodock4}, AutoDock Vina \cite{trott2010autodock}, Glide \cite{friesner2004glide}, GOLD \cite{jones1997development}, FlexX \cite{rarey1996fast},and Surflex-DOCK \cite{jain2007surflex}.

Knowledge-based scoring functions basic assumption is highly frequent atomic pairs (distances) contribute more to a binding affinity than the less frequent ones. 
The advantage of knowledge-based scoring functions is the computing cost since they only require distance calculation. 
Examples of knowledge-based scoring functions are DrugScore \cite{velec2005drugscorecsd}, IT-Score \cite{huang2010inclusion}, SMoG \cite{ishchenko2002small}, DFIRE \cite{zhang2005knowledge}, and PMF \cite{muegge2006pmf}.

\subsection{Similarity/Distance-based methods}

These methods assume that similar ligands bind to similar proteins, and vice versa and show similar properties and biological activities. \cite{johnson1990concepts}
In these methods, different similarity measures can be used, including, chemical/expression/ligand/annotaion/side effect-based for drugs \cite{perlman2011combining} and Smith-Waterman alignment algorithm for proteins \cite{pahikkala2015toward}.
One example of these approaches is QSAR uses machine learning methods to compare a candidate ligand with the set of all known ligands to infer its binding capability. \cite{butina2002predicting}
These methods have some shortcomings, for instance conformation complexity, or in ligand-based methods, not using protein information and just relying on the similarity between ligands, therefore limiting the search space to ligand space only.
Also, they need a large amount of ligands to form a good information space to use for the interaction prediction whereas there are not sufficient ligands for each particular protein.

\subsection{Structural methods}

Structure-based methods use the structural information of proteins or drugs, such as 3D coordinates or secondary structures or sequences, as input to represent protein and drug molecules. 

Some of these approaches use the 3D structure of proteins and ligands (which may not be available for all of them) and an interaction scoring function like molecular docking simulation to calculate the binding energy \cite{trott2010autodock, koes2013lessons}.
These 3D structures can be obtained from experimental methods like X-ray crystallography or NMR spectroscopy, or predicted using computational methods. \cite{maveyraud2020protein, howard1998protein}
Structural methods can be divided into three categories by the type of scoring function: the classic scoring function methods \cite{wang2018computational}, machine learning scoring function methods \cite{xie2018deep, wojcikowski2019building} and deep learning scoring function methods \cite{lee2019deepconv, tsubaki2019compound}.
The main part of these methods is to accurately model the three-dimensional structure of proteins and compounds.
The limitations of traditional structural approaches are the vastness of search space (therefore time consuming on large scale), unknown protein 3D structures, and unfeasiblity of applying methods like docking simulation on some cases. \cite{yildirim2007drug, opella2013structure}
Moreover, there are other non-machine learning methods available for computing affinity, such as FEP (Free-Energy Perturbation) \cite{jorgensen2008perspective} and MM/GBSA (or MM/PBSA) \cite{ccinarouglu2020comprehensive}.
These traditional methods are very slow and require significant amount of computation resources.
Drug compounds can directly dock to determine binding affinity when proteins have well-established structure information. 
However, numerous proteins lack such structural details, and even with extensive time dedicated to homology modeling, obtaining comprehensive structural information remains uncertain \cite{yadav2020homology}.

\subsection{Feature-based methods}

In these approaches, Feature vectors of drug-target pairs are obtained from their properties or by learning from raw data, and then fed into various classifiers or regressors \cite{sachdev2019comprehensive}.
Since both ligand-based and target-based aspects are considered in feature-based methods, they can be assigned to the so called "Chemo-genomic" approaches.

Chemo-genomics approaches use the information of proteins and drugs altogether and unify the chemical space of drugs (molecular weight, logP, hydrogen bond donors/acceptors, functional groups, etc) and genomic space of proteins (protein sequences, structures, functions, pathway information, etc) to make interaction predictions. 
chemo-genomics deals with the systematic analysis of chemical-biological interactions and can use extensive biological data that is readily available in public datasets.
These data often represent high-level biological relationships and associations as features. \cite{ezzat2019computational, yamanishi2008prediction}
The chemo-genomics methods can further be classified into feature-based and similarity-based methods \cite{mousavian2014drug}. 
The feature-based methods represent the drug-target pair with a vector of descriptors. 
The various properties of drugs, as well as the proteins, are encoded as corresponding features. 
The inputs of these methods are different feature vectors that may be produced by combining the properties of drug and targets and then they are fed into various machine learning models. 
On the other hand, a variety of similarity-based methods have also been developed to calculate the similarity between the drug compounds and target proteins. 
Various kernel functions have been defined that make use of similarity matrices which are computed using different techniques and measures.

\subsection{Machine learning and deep learning methods}

\begin{figure}
  \centering
  \includegraphics[width=\columnwidth]{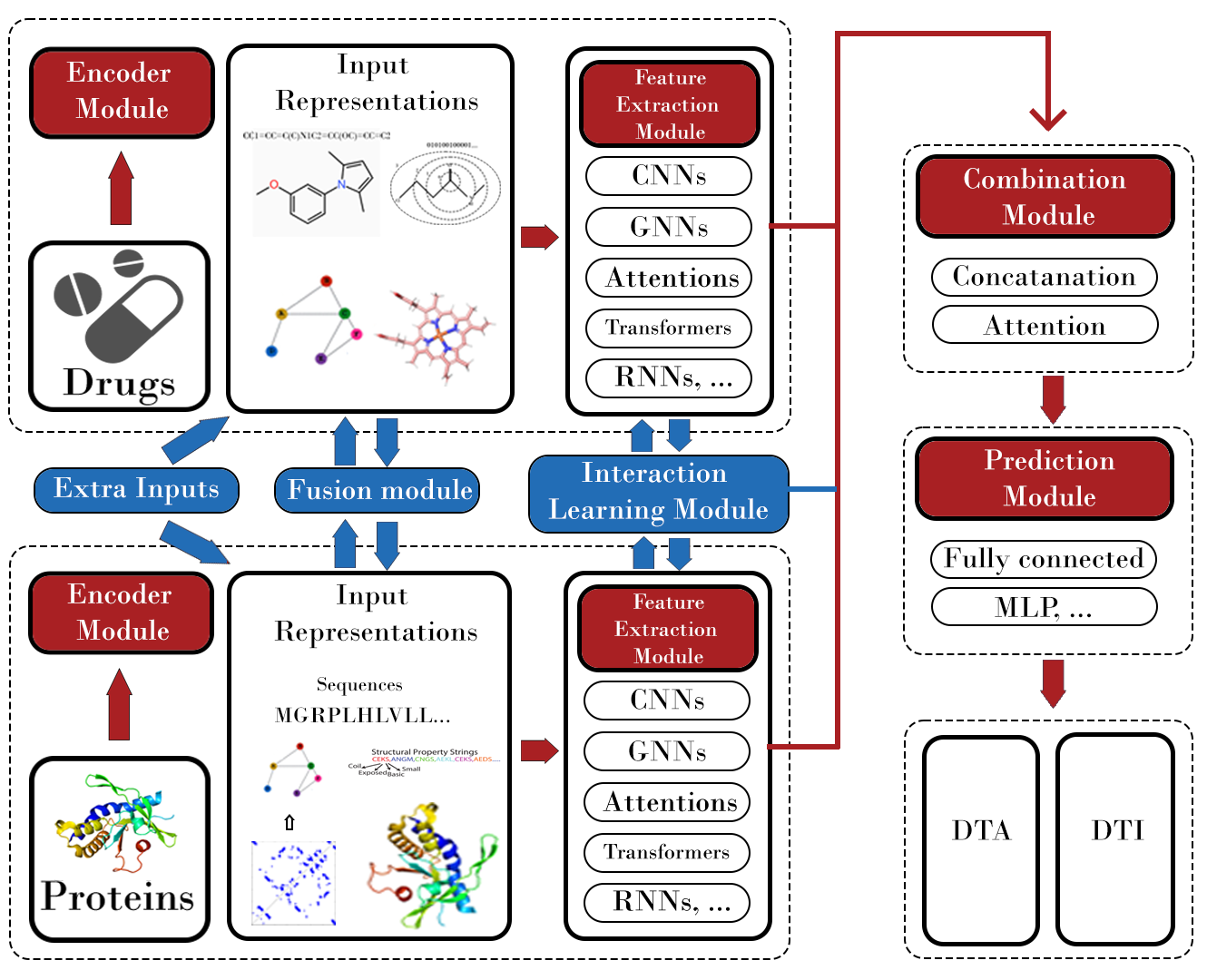}
  \caption{Overview of general frameworks used in deep learning methods. Red blocks are seen in most of the models. Blue blocks are modules that models have.}
  \label{fig:fig4}
\end{figure}

In most classic machine learning approaches, the hand-crafted features are used \cite{he2010predicting}. 
In recent years, however, it is shown that integrating feature learning capabilities into machine learning-based models improves prediction performance \cite{fu2016predicting, huang2020deeppurpose, yuan2016druge, lee2018identification}.
Some traditional machine learning method for drug-target interaction prediction are similarity-based (Bipartite local models (BLM) \cite{bleakley2009supervised}, K-nearest neighbor (KNN) \cite{shi2015predicting}, Kronecker Regularized Least Squares (KronRLS) \cite{pahikkala2015toward}, COSINE \cite{lim2016improved}, KRONRLS-MKL \cite{nascimento2016multiple}, BLM-NII \cite{mei2013drug}), feature-vector-based (Support Vector machine (SVM) \cite{faulon2008genome}, Adaptive Boosting Support Vector machine (Adaboost-SVM) \cite{wang2016virtual}), and semi-supervised (Restricted Boltzmann machines (RBMs) \cite{wang2013predicting}, Gaussian Interaction Profiles (GIP/RLS-Kron) \cite{van2011gaussian}, and Laplacian Regularized Least Squares (LapRLS) \cite{xia2010semi}).
Most of these methods cannot automatically extract high-level hidden features from drugs and proteins \cite{peng2015predicting}.

KronRLS \cite{pahikkala2015toward} uses kernels built from molecular descriptors of the drugs and targets within a regularized least squares regression (RLS) framework.
Following standards of kernel learning methods, they first defined an objective function where the kernel function indicated the similarity of two drug-target pairs in the Hilbert space. 
Then, they regarded the problem of learning a prediction function as finding a minimizer of the objective function with the aim of binding affinities prediction.

SimBoost \cite{he2017simboost} uses the affinity similarities among drugs and among targets to build new features.
After training gradient boosting machine based on final features of drug-target pairs, SimBoost can be applied to predict binding affinities of unknown drug-target pairs.

Predicting the potential space of binding sites and extracting the best features related to these sites is a key challenge in DTI/DTA.  
Many Deep learning-based methods tried to overcome the challenges with various models and mechanisms including The convolutional neural networks (CNNs) \cite{lu2023drug, wang2020ldcnn}, graph neural networks (GNNs) \cite{son2021development, wang2023agraphdta, szocinski2021awegnn, qi2024drug, wang2021drug, sun2024ingnn}, non-Euclidean-structured data in GNNs \cite{sun2020graph, wu2020comprehensive, yue2020graph}, equivariant graph neural networks (EGNNs) \cite{tu2024se, kumar2024caster}, feed forward neural networks (FNNs) \cite{rifaioglu2021mdeepred}, long short term memory networks (LSTMs), recurrent neural networks (RNNs) \cite{karimi2019deepaffinity}, attentions (multihead attentions, self attentions, cross attentions) \cite{huang2024predicting, deng2024multidta, koyama2020cross, li2023sagdti, cheng2024ert}, autoencoders \cite{wang2020novel, liu2021gadti, zhao2024drug, wang2024mdgae}, pre-training \cite{wu2023t}, transformers \cite{wang2020gnn, hu2023drugormerdti, monteiro2024tag, gao2024graphormerdti}, bidirectional encoder representations from transformer (BERT) \cite{chen2021predicting, wei2022mdl}, language models \cite{sharma2024bapulm, hua2025mmdg}, image inputs \cite{qian2022picture, qian2022cat}, Hierarchical learning and modeling \cite{zhang2024hieraffinity, gong2025multigrandti}, neural factorization machine (NFM) \cite{lei2022drug, ye2021unified}, constrastive learning \cite{hu2024heterogeneous}, and interactive learning \cite{zhu2024drug}. 
Some models try to be the fastest models to predict DTA \cite{liyaqat2023tem, boezer2023fastdti, veleiro2024gennius}, some try to overcome the limitations of models like CNNs which only works on a particular region of information, excluding comprehensive details, and some try to learn the best interaction by using interactive learning modules.

Except some novel studies \cite{gao2024qldti, yu2020fpsc}, as Figure \ref{fig:fig4} shows, deep-learning-based methods include data pre-processing and encoder, a drug feature extraction module, a protein feature extraction module, combination module, and a prediction module in general and unlike traditional methods, these methods can extract high-level features automatically \cite{iliadis2024comparison}.
Most methods in drug-target prediction which are called Y-shaped by this review \cite{shi2023review}, extract features from both drugs and proteins, then run deep learning model on them to get an embedding for each side, and concatenate them and apply head modules like a fully connected layer (FC) to get the final prediction output. 
Furthermore, deep learning methods in DTI or DTA sometimes have two different kinds of model as the type of the drug input or protein input can be different.
For example, GNN as a structural model can handle the feature extraction of molecular graphs of drugs while LSTM as a sequence model handles the feature extraction of string protein sequences.

Some studies \cite{cheng2022iifdti, yin2024fotf} tried to learn the interactive features of drugs and proteins alongside the two separate pipelines. 
Because most of the models are hybrids of other models as we are dealing with two spaces of problems (drug and protein space), therefore the categorization of the methods is better not to be based on model architecture type, instead it should be based on type of the inputs.
Therefore, approaches here are divided into three main categories: sequence-based methods, structure-based methods, and sequence-structure-based methods.
Sequence-based methods deal with drug and protein sequence inputs, structure-based methods get both drug and protein inputs as graphs, and sequence-structure-based methods are the models that get hybrid inputs of drug and protein graphs and sequences.
It is important to note that methods employing utility networks, while fitting into these categories based on their input types, represent a distinct and significant approaches within drug-protein relation prediction, therefore they are discussed in a separate section.

Some mentioned models are ligand-target or compound-target interaction prediction methods, and are used for predicting the interaction between a ligand and a target protein, 
Drug-target interactions are a specific type of protein-ligand interactions, where the ligand is a drug molecule.
Many of the fundamental principles and modeling techniques used in protein-ligand interaction studies are directly applicable to drug-target interactions.
Discussing protein-ligand interaction models can provide a broader context and deeper understanding of the underlying principles involved in drug-target interactions.
Most of the DTA models mentioned in the following sections can also be used for DTI task with applying a threshold on binding affinity datasets, changing the head of the framework and altering the loss of the model.

\begin{figure}
  \centering
  \includegraphics[width=\columnwidth]{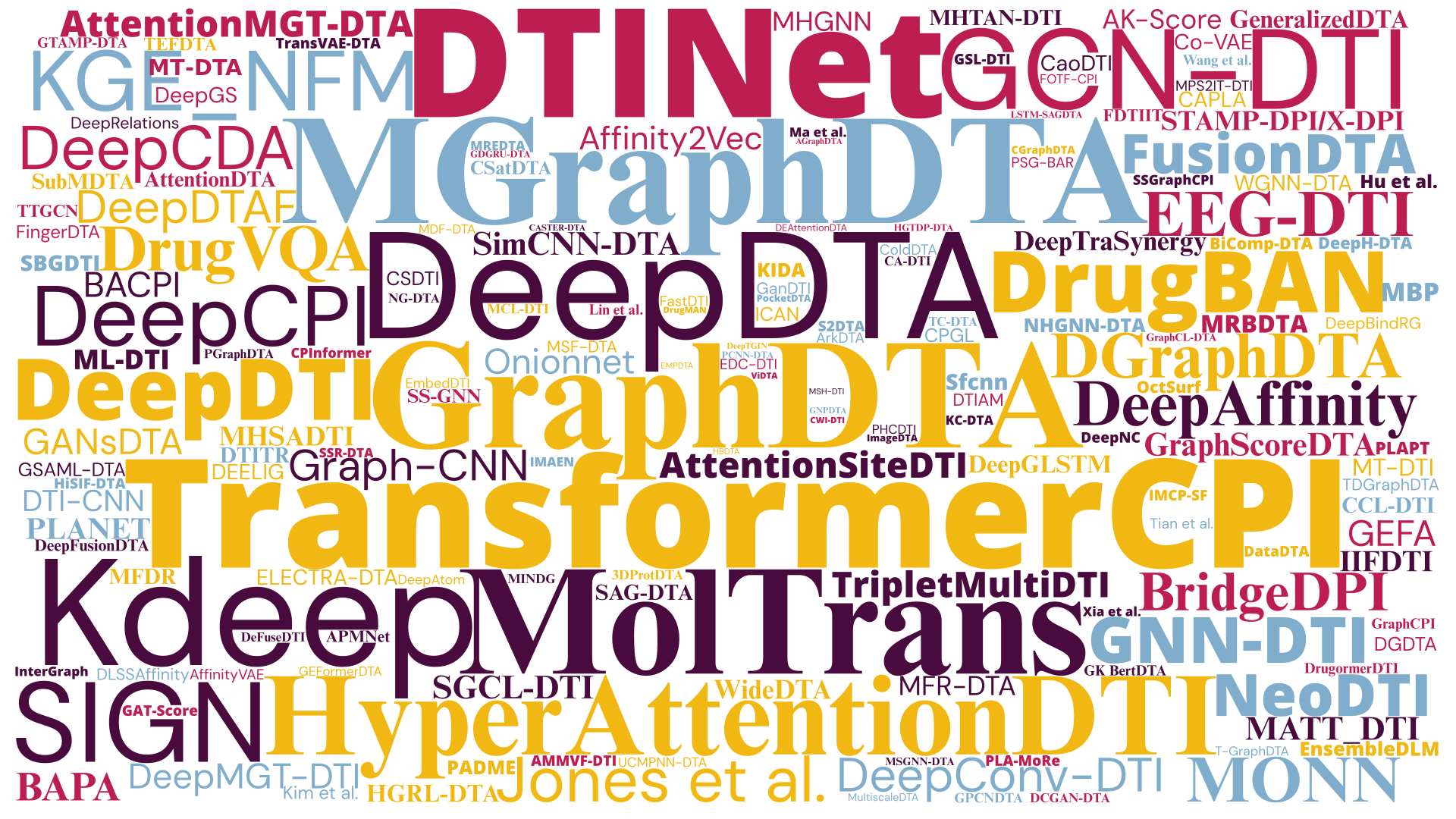}
  \caption{Importance of Models based on number of citations obtained from google scholar, recent years have stronger weights. Made by \url{https://www.wordclouds.com/}}
  \label{fig:fig5}
\end{figure}

Figure \ref{fig:fig5} shows the importance of Models based on number of citations obtained from google scholar.
% and Figure \ref{fig:fig6} demonstrate the timeline of the presented models from 2016 to late 2024

\subsubsection{Sequence-based methods}

Sequence-based deep learning methods mainly use the sequences of drug and proteins as inputs which can be SMILES and primary sequences \cite{yang2022modality}.
These methods mainly ignore the topologies of drugs and protein molecules, therefore, they cannot utilize the structure information of the molecules.
However, they are relatively less time-consuming and faster in feature engineering and model prediction modules than other methods because they deal only with some strings of sequences.
In the following paragraphs, an overview of some sequence-based methods is provided and tabulated in Table \ref{tab:sequence_based_models}.
Corresponding inputs for each model is shown in the table in Appendix C.

% \subsubsubsection{DTA methods}
\begin{itemize}

\item DeepDTA \cite{ozturk2018deepdta} employed two separate 1D-CNN modules to learn representations from the Protein sequences and SMILES strings (encoded as label encodings) and combine these two representations to feed into three fully connected layers block which was called DeepDTA.
They evaluate the effectiveness of CNN blocks by replacing one pipeline with similarity matrices (with S-W and PubchemSim algorithms) and testing different combinations of it.
DeepDTA compares its results to the KronRLS \cite{pahikkala2015toward} algorithm and  SimBoost \cite{he2017simboost} in which both used traditional machine learning algorithms and utilized 2D-representations of the compounds in order to obtain similarity information. 
In DeepDTA study, 1D representation of protein and drugs were preferred over 3D-structures of the binding complexes because there were limited to known protein ligand complex structures, with only 25 000 ligands reported in PDB in that time \cite{rose2016rcsb}.

\item WideDTA model \cite{ozturk2019widedta} is an extension of DeepDTA in which the sequences of the drugs and proteins are first summarized as higher-order features.
For example, the drugs are represented by the most common sub-structures (the Ligand Maximum Common Substructures (LMCS) \cite{wozniak2018linguistic}), while the proteins are represented by the most conserved sub-sequences (the Protein Domain profiles or Motifs (PDM) from PROSITE \cite{sigrist2010prosite}).
There are four inputs (protein sequence words, ligand sequence words, LMCS, PDM) to the 2 layered CNN models and output is followed by a concatanation and three FC layer.

\item Deepaffinity \cite{karimi2019deepaffinity} used a seq2seq auto-encoder model to learn protein and compound representation in an unsupervised manner. 
Next, the output is fed into an attention layer, and the output is given to a one-dimensional convolutional layer. 
The outputs of proteins convolutional layer and compounds convolutional layer are concatenated and fed into the fully connected layers.

\item MT-DTI \cite{shin2019self} propose a new MBERT drug molecule representation based on the self-attention mechanism and a MaxPool representation and CNNs module for proteins, followed by multi-layered interation dense layers to predict affinity scores.

\item Co-VAE (co-regulaized variational autoencoder) \cite{li2021co} used two variational autoencoders for generating drug SMILES and protein sequences, respectively, and a co-regularization part for generating the binding affinities.
This model tries to maximize the lower bound of the joint likelihood of drug, target and their affinity.

\item GANsDTA  \cite{zhao2020gansdta} comprises two GANs for feature extraction and a regression network for prediction.
Each GAN model gets drugs or proteins sequences and learn them with a discriminator module and a generator module which produces fake noisy data.

\item DeepCDA \cite{abbasi2020deepcda} uses two modules which are a combination of convolutional neural networks and long short-term memory networks, to analyze the structure of both the compound and the protein.
CNN captures local patterns while LSTM learns the global patterns and far dependent sections.
A key innovation is the use of a two-sided attention mechanism, which allows the model to be more interpretable and also focus on the most important parts of the compound and protein structures. 
Additionally, the model incorporates an adversarial discriminative domain adaptation method (ADDA) to improve its performance on new, unseen datasets and do the inference in cold setting.
The effectiveness of this new method is demonstrated through experiments on three widely used datasets: KIBA, Davis, and BindingDB.

\item DeepDTAF \cite{wang2021deepdtaf} integrates global and local features of proteins including protein sequence, protein-binding pockets, and secondary structural properties and SMILES of drugs. 
These features are then fed into embedding layers, dilated convolutions, and convolution layers for DTA prediction.
DeepDTAF utilizes the pockets obtained from the protein compound complex, which suggests that the pockets used in the DeepDTAF are the ones where the interaction occurs, therefore it cannot generalize to unseen drug-target complexes.

\item ML-DTI \cite{yang2021ml} formulated the DTA problem from a global perspective by adding mutual learning layers between the two pipelines. 
The mutual learning layers consist of multi-head attention and position-aware attention.

\item MATT-DTI \cite{zeng2021deep} uses a relative self attention for drug sequences and a Multi Head Attention to combine the results.
CNNs are used for FASTA sequences of proteins.

\item SimCNN-DTA \cite{shim2021prediction} calculates the Smith-Waterman and Tanimoto similarity of proteins and drugs, recpectively and uses 2D CNN for their outer product to predict DTA.

\item FingerDTA \cite{zhu2022fingerdta} utilizes CNNs to extract local patterns from drug and protein one-hot vectors and use generated fingerprints as a guide to prdict DTA.

\item MultiscaleDTA \cite{chen2022multiscaledta} uses multiscale CNNs and pooling with self attention for both drugs and protein sequences.
The extraction of different scale of features allow the model to learn the features in multiple levels.

\item CSatDTA \cite{ghimire2022csatdta} enhances the DTA prediction with CNNs by using a convolution model with self-attention mechanisms on molecular drug and target sequences.
Furthermore, a web server was used to implement the CSatDTA model. \footnote{\url{http://nsclbio.jbnu.ac.kr/tools/CSatDTA/}}

\item AttentionDTA \cite{zhao2022attentiondta} utilizes a label encoder, embedding layer and a 1D-CNN block to extract features from both drugs and proteins seperately.
A bilateral multi-head attention and a multi layer perceptron layer is used for the head of the framework to predict the affinity. 

\item MRBDTA \cite{zhang2022predicting} consists of three main parts including, embedding and positional encoding, Trans block, and the head.
This models utilizes transformers and skip connections for drug and protein sequences to predict DTA and offer some interpretablity as it uses attention weights. 

\item DTITR \cite{monteiro2022dtitr} proposes an end-to-end transformer-based architecture with cross attention transformer encoders for predicting DTA.

\item ELECTRA-DTA \cite{wang2022electra} incorporates an unsupervised learning mechanism to pre-train two ELECTRA-based contextual embedding models and uses it alongside a squeeze-and-excitation (SE) convolutional neural network block stacked over three fully connected layers.

\item FusionDTA \cite{yuan2022fusiondta} encodes drug molecules as SMILES strings, and proteins as word embeddings. 
The LSTM layers are designed to construct the basic blocks of the encoder layer. 
The intermediate carriers of drug molecules and proteins are imported into the fusion layer to obtain an output carrier representation of binding affinity.
Furthermore, FusionDTA leverages knowledge distillation to combine the predictions from multiple models.

\item AffinityVAE \cite{wang2023affinityvae} (Affinity variational autoencoder) uses interaction feature mapping and a variational autoencoder, which consists of A multi-objective model that combines the affinity prediction and ligand generation based on data augmentation to validate the prediction performance. 
The design of the adaptive autoencoder of target chemical properties and interaction feature maps allows them to generate new ligands similar to known ligands and add them to the original training set.
The protein sequences are first encoded on a pre-trained protein language model to fully capture their local and global structural information of the protein. 
For the ligand SMILES sequence, the first branch represents SMILES as an atom graph to extract the high-order structure information of the ligand, while the second branch feeds the ligand into a one-dimensional sequence feature extraction module in order to extract its atomic-level information. 
Furthermore, the extracted protein and ligand features are fed into the mutual attention module to capture the influence of different parts of the protein or ligand on the interaction. 
Based on the binding of proteins and ligands at the residue and atomic levels, innovative protein-ligand feature interaction module (ARCM) captures local features and constructs feature maps to identify the most important interaction features. 
A final prediction module employs the captured features to determine the affinity of the pairs of features.

\item ArkDTA \cite{gim2023arkdta} incorporates non-covalent interactions (NCIs) knowledge with multi-head attention modules to obtain better results.
This model use FASTA sequence of proteins with ESM and SMILES of drugs with Morgan fingerprint as inputs and two novel modules which are called Pooling by Multi-head Attention (PMA) and Multi-head Attention Block (MAB) as model to predict DTA.

\item CAPLA \cite{jin2023capla} employs the cross attention mechanism to capture the mutual effect of protein binding pockets and ligands.
Dilated convolution and 1D convolution is used for ligands/proteins and protein pockets, respectively.  

\item BiComp-DTA \cite{kalemati2023bicomp} argues that similarity-based information, such as Smith-Waterman scores have shown poor prediction performances, or depends on employing multiple information types.
Therefore, this model employs Lempel-Ziv-Markov chain algorithm (LZMA) alongside with Smith-Waterman to construct the unified measure for the DTA prediction task.
The novelty of this study is the BiComp encoding module which is the Hadamard product \cite{reams1999hadamard} of the normalized Smith-Waterman (SW) and normalized compression distance (NCD) \cite{cilibrasi2005clustering} similarity matrices.

\item MT-DTA \cite{zhu2023associative} utilizes mutual learning / Information interaction learning with self-attention between the drug and protein learning pipelines for DTA prediction.
A label encoding layer, an embedding layer, a CNN block and a Transformer block is used for both pipelines.

\item PCNN-DTA \cite{chen2023prediction} employs two feature pyramid network (FPN) with skip-connections and multi-head self-attentions (MHSA) as heads for drug and protein pipelines.
FPN fuses the features obtained in each layer of a multilayer convolution network to have more low-level feature information.

\item DGDTA \cite{zhai2023dgdta} uses a dynamic graph attention network for drugs and a bidirectional long short-term memory (Bi-LSTM) network with CNN for proteins to predict DTA.

\item FastDTI \cite{boezer2023fastdti} is a fast framework that uses pre-trained models like Grover \cite{rong2020self}, ProtBERT, and ChemBERTa to construct multimodal inputs and concatenates them via dense layers to predict DTA. 

\item TransVAE-DTA \cite{zhou2024transvae} consists of a VAE module, a transformer module, and an adaptive attention pooling (AAP) module.
AAP block for feature fusion significantly improves the prediction performance.

\item ImageDTA \cite{han2024imagedta} treats word vector-encoded SMILES sequences as images and applies multiple single-layer multi-scale 2D-CNNs horizontally.
CNN is used for protein sequences and BiLSTM is utilized as a final model to predict the task.

\item TEFDTA \cite{li2024tefdta} applies a transformer encoder on fingerprint representation (MACCS) of drugs and CNN on protein sequences (FASTA with label encoding) for bonded and non-bonded drug-target affinities.

\item DCGAN-DTA \cite{kalemati2024dcgan} inspires from generative adversarial networks (GAN) and applies a deep convolutional GAN (DCGAN) on drug and protein sequences.
This model uses CNN as discriminator and generator and tries to distinguish between real or fake (adding noise) SMILES and real or fake protein sequences.

\item PLAPT \cite{rose2024plapt} utilizes two pre-trained models (ProtBERT and ChemBERTa) and a branching neural network for DTA prediction.

\item DEAttentionDTA \cite{chen2024deattentiondta} is based on dynamic word embeddings and a self-attention mechanism.
SMILES, protein, and pocket sequences are fed into a dynamic word-embedding layer based on a 1D convolutional neural network.

\item MREDTA \cite{sun2024mredta} utilizes BERT and Transformer encoders in BERT-Trans block and Multi-Trans block to capture the relations between sequences in drugs and proteins seperately and concatanates them for DTA prediction.

\item MDF-DTA \cite{ranjan2024mdf} uses 1D, 2D, and 3D representations obtained from different pretrained models for both drugs (Mol2Vec, GIN, E3NN) and targets (ProVec, BERT, ESM-Fold).

\end{itemize}

% \subsubsubsection{DTI methods}
\begin{itemize}

\item MFDR \cite{hu2016large} used Auto-Encoders as building blocks of deep network for reconstructing drug and protein features to low-dimensional new representations, followed by a support vector machine module.

\item DeepDTI \cite{wen2017deep} models DTI using Deep Belief Network (DBN), which is a stack of Restricted Boltzmann Machines (RBMs) \cite{hinton2009deep, hinton2012practical}. 
It uses the concatenation of ECFP2, ECFP4, ECFP6 as the drug feature and uses protein surface classification for protein features.

\item DeepConv-DTI \cite{lee2019deepconv} performs convolution on various lengths of amino acids subsequences to capture local residue patterns of generalized protein classes.
For drugs it uses Morgan/Circular fingerprints and a fully connected layer to extract drug features.

\item Hu et al. \cite{hu2019predicting} used a 2D-CNN on drug-target augmented matrices of drug-target pairs to predict DTI.
Drugs were obtained from PaDEL descriptor program and targets were from different physicochemical properties of amino acids in Protein sequences.

\item DrugVQA \cite{zheng2020predicting} utilizes DynCNN and Sequential attention for 2D distance map of proteins and BiLSTM with Multi head attention for Smiles of drugs.
This study is the first to utilize protein 2D distance map for predicting drug-protein interaction.

\item MolTrans \cite{huang2021moltrans} utilized knowledge inspired sub-structural pattern mining algorithm (FCS) and interaction modeling module and an augmented transformer encoder to better extract and capture the semantic relations among sub-structures extracted from massive unlabeled biomedical data.

\item EnsembleDLM \cite{kao2021toward} utilizes different fingerprints and combination of 1Dconv and linears and integrate multiple models by taking the arithmetic mean on the predicted binding affinity scores of different models.

\item CA-DTI \cite{koyama2020cross} uses two Transformer encoders as feature extractors and the Cross Attention as a task executor, key regions of interest for novel drug candidates were found.

\item BridgeDPI \cite{wu2022bridgedpi} constructed a learnable drug-protein association network, which has drug nodes, protein nodes and new nodes which are called "bridge nodes".
A 3-layer graph-conv layer and 2 FFN layers are applied to predict DPI.

\item HyperAttentionDTI \cite{zhao2022hyperattentiondti} models complex non-covalent inter-molecular interactions among atoms and amino acids using an attention module and assign an attention vector to each atom or amino acid.

\item MPS2IT-DTI \cite{de2022novel} encodes protein and drug sequences to images by reshaping the normalized k-mer count vectors and pass them through 2D CNNs, separately.

\item AMMVF-DTI \cite{wang2023ammvf} employs two modules named ITM (interaction transformers) and NTN (neural tensor networks) for node level embeddings and graph level embeddings of drugs and proteins, respectively.
ITM consists of two parallel interacting transformer encoders. This module serves to extract node-level interaction features between drugs and targets.
NTN is employed to explore potential K major associations between drugs and targets as interactive eigenvector outputs.
To get the graph level embeddings of proteins and drugs, BERT and an attention layer is used for node level embedding of drugs which is acquired with word2vec model and GAT and an attention layer is utilized for node level embeddings of drugs.
Outputs of NTN and ITM is concatenated and fed into a MLP for the classification task.

\item MCL-DTI \cite{qian2023mcl} utilize both molecular image and chemical features of drugs. 
The image of the drug mainly has the structural information and spatial features of the drug, while the chemical information includes its functions and properties.
The framework consists of feature decoder and encoder block with Multi-head self attention (MSA) and Multi-head cross attention (MCA) modules.

\item FOTF-CPI \cite{yin2024fotf} uses encoders to extract features of fragmented proteins and drugs, learns each one with fully connected layers, and  also learns the interaction between these features with local affinity attention.
This study experiments with different methods such as character-based slicing (CS), BRICS \cite{degen2008art}, Byte Pair Encoding (BPE) \cite{sennrich2015neural}, FCS \cite{huang2021moltrans}, and FSOT (their method) and comes to a conclusion that FSOT is the better slicing strategy for fragmenting. 

\item DeFuseDTI \cite{feng2024defusedti} includes base encoder and detail encoder to extract locally aggregated features and detailed features of each target residue and drug atom. 
The detail encoder (utilizing Invertible Neural Networks for targets and graph transformers for drugs) can capture the features of each atom and residue. 
Protein output and output of GCN feeded with molcular drugs are fed into the Multi-view Fusion Attention learning module (MFA).

\item PHCDTI \cite{ye2024phcdti} handles combinatorial effects (i.e., high-order crossover between drugs and proteins) when multiple non-covalent interactions coexist by a multi-channel parallel high-order prediction model.
A tri-channel parallel model structure is used, allowing each channel to independently learn the interaction patterns (designed from low-order to high-order, incorporating linear, inner-product, and pair-interaction feature crossover methods). 
The stacking of residual connections enables the model to effectively model high-order crossover. 
Multi-head attention mechanisms and squeeze-excitation networks (SENET) is used for inner product and pair-interaction, recpectively.

\item CWI-DTI \cite{li2024accurate} employ an innovative auto-encoder framework which is called Stacked De-noising Sparse Auto-encoders (SDSAE) to fuse multiple drug/target topological similarity matrices.
This model compare its results in different datasets from Chinese and Western medicine.

\end{itemize}

\begin{table*}[htbp]
  \centering
  % \footnotesize
  \caption{Summary of sequence-based models}
  \tiny
  \label{tab:sequence_based_models}
  \begin{tabular*}{\textwidth}{@{\extracolsep{\fill}}lcccccr}
  \toprule
  Model & Task & Architecture & Drug/Ligand models & Protein/Target models & Extra models \\
  \midrule
  FastDTI & DTA & Comb-FC & Dense & Dense & Dense\\
  DeepDTA & DTA & Comb-FC & CNN & CNN & N/A\\
  WideDTA & DTA & Comb-FC & CNN & CNN & CNN\\
  MATT-DTI & DTA & MHA-FNN & SA, FNN and CNN & CNN & MHA\\
  TC-DTA & DTA & Comb-FC & CNN & Encoder blocks & N/A\\
  BiComp-DTA & DTA/DTI & Comb-FC & CNN & 3 FCs & N/A\\
  DeepDTAF & PLA & Comb-FC  & Dilated conv & Dilated conv & 3 Convs\\
  CAPLA & PLA & Comb-FC & 1D-Dilated Conv & 1D-Dilated Conv & CA\\
  DeepAffinity & CPA & Comb-FC & RNN-Attention-CNN & RNN-Attention-CNN & N/A\\
  ImageDTA & DTA & Comb-FC & Multiscale 2D-CNNs & 1D-CNN & BiLSTM, Skip concatanations\\
  FingerDTA & DTA & Comb-FC & Dense Convolutional Block CNN, FC & Dense Convolutional Block CNN, FC & N/A\\
  DCGAN-DTA & DTA & Comb-FC & DCGAN+CNN & DCGAN+CNN & BLOSUM encoding for protein\\
  SimCNN-DTA & DTA/DTI & 1pipe-FC & 2D-CNN & 2D-CNN & N/A\\
  MT-DTI & DTA/DTI & Comb-Multilayered FFN-FC & Transformer & CNN & Dense network\\
  ML-DTI & DTA & Comb-FC & 3Convs+maxpool & 3Convs+maxpool & Mutual learning blocks\\
  MT-DTA & DTA & Comb-FC & CNN + Transformer & CNN + Transformer & Mutual learning with SA\\
  MDF-DTA & DTA & Comb-FC & Mol2Vec, GIN, E2NN & ProVec, BERT, ESM-Fold & N/A\\
  MultiscaleDTA & DTA/DTI & Comb-FC & Multi-scale CNNs+SA & Multi-scale CNNs+SA & N/A\\
  AttentionDTA & DTA & Comb-BAH-FC & Encoder-CNN & Encoder-CNN & N/A\\
  ArkDTA & DTA & Comb-FC & MHA+Pooling, Pooling by MHA & MHA+Pooling, Pooling by MHA & N/A\\
  CSatDTA & DTA & Comb-FC & SA augmented Conv block & SA augmented Conv block & N/A\\
  MRBDTA & DTA & Comb-Attention-FC & Modified Transformer & Modified Transformer & N/A\\
  DTITR & DTA & Comb-FC & (CA) Transformer-encoder & (CA) Transformer-encoder & N/A\\
  DGDTA & DTA & Comb-FC & GATv2, GATv2+GCN & BiLSTM+CNN & Dense layers\\
  TransVAE-DTA & DTA & Comb-FC & Drug encoder & Target encoder & Adaptive attention pooling\\
  DEAttentionDTA & DTA & Comb-FC & Dynamic word embedding+MHA & Dynamic word embedding+MHA & SA\\
  GANsDTA & DTA/DTI & Comb-Convolutional regressor & GAN & GAN & convolutional regressor\\
  TEFDTA & DTA & Comb-FC & Transformer encoder  & CNN & N/A\\
  PLAPT & PLA & Comb-Linear, FC & ChemBERTa & ProtBert & N/A\\ 
  MREDTA & DTA & Comb-Linear & BERT, Transformer encoder & Transformer encoders & N/A\\
  DeepCDA & CPA/CPI & Comb-BAH & CNN-LSTM & CNN-LSTM & N/A\\
  FusionDTA & DTA & Comb-Attention-MLP & BiLSTM & BiLSTM & DNN\\
  Co-VAE & DTA & Comb-FC & VAE & VAE & Co-regularization part\\
  AffinityVAE & PLA & Mutual learning, Comb-Resnet & GRUMultiEncode & Pretrained Model, GRU & VAE\\
  ELECTRA-DTA & DTA & Comb-FC & Pretrained ELECTRA + SE & Pretrained ELECTRA + SE & N/A\\
  PCNN-DTA & DTA & Comb-FC & Feature pyramid network & Feature pyramid network & 2 MHA, RegressionNet\\
  \\
  BridgeDPI & DPI & Comb-FFN & Conv+POOL, FFN & Conv+POOL, FFN & 3Graph-Conv\\
  PHCDTI & DTI & Comb-FC & 1D-CNN & 1D-CNN & MHA, Pair-interaction layer\\
  DeepConv-DTI & DTI & Comb-FC & FC & CNN & N/A\\
  EnsembleDLM & DTI & Comb-FC & Linear and Conv1D & Linear and Conv1D & N/A\\
  Hu et al. & DTI & Representation Comb-1 pipe & 2D-CNN & 2D-CNN & N/A\\
  MCL-DTI & DTI & Comb-FC & CNN backbone+MSA, MSA+MCA & MSA+MCA & N/A\\
  CWI-DTI & DTI & Comb-FC & SDSAE & SDSAE & CNN\\
  HypAttentionDTI  & DPI & Comb-FC & CNN & CNN & Attention+Pooling\\
  MPS2IT-DTI & DTI/DTA & Comb-FC & Conv 2D layers & Conv 2D layers & Dense layers\\
  MFDR & DTI & 1pipe-FC & Stacked auto-encoder, SVM & Stacked auto-encoder, SVM & N/A\\
  DeepDTI & DTI & 1pipe-FC & Deep Belief Network & Deep Belief Network & N/A\\
  MolTrans & DTI & Interaction module-Decoder & Transformer & Transformer & FCS mining module\\
  AMMVF-DTI & DTI & Comb-MLP & BERT-Attention & GAT-Attention & NTN, ITM\\
  DeFuseDTI & DTI & Comb-MLP & GCN & CNN residual connection blocks & Multiview attention learning\\
  DrugVQA & DPI & Comb-FC & BiLSTM, MHA & DynCNN, Sequential attention & N/A\\
  FOTF-CPI & CPI & Comb-FC & Transformer encoder  & Transformer encoder  & Local affinity attention\\
  CA-DTI & DTI/DTA & Comb-FC & Transformer encoder & Transformer encoder & CA\\
  \bottomrule
  \end{tabular*}
\end{table*}

\begin{table*}[htbp]
  \centering
  \footnotesize
  \caption{Summary of structural-based models.}
  \tiny
  \label{tab:structure_based_models}
  \begin{tabular*}{\textwidth}{@{\extracolsep{\fill}}lcccccr}
  \toprule
  Model & Task & Architecture & Drug/Ligand models & Protein/Target models & Extra models \\
  \midrule
  DGraphDTA & DTA & Comb-FC & GNN(GCN,GAT) & GNN(GCN,GAT) & N/A\\
  WGNN-DTA & DTA/DTI & Comb-FC & GNN(GCN,GAT) & WGNN(GCN,GAT) & N/A\\
  HGRL-DTA & DTA & Comb-MLP readout-FC & GCN & GCN & GCN\\
  Ma et al. & DTA & Comb-FC & \makecell{GNNs (GCN, GAT, \\GraphConv, SAGE, SGC)} & \makecell{GNNs (GCN, GAT, \\GraphConv, SAGE, SGC)} & N/A\\
  GSAML-DTA & DTA & Comb-FC & GAT-GCN & GAT-GCN & N/A\\
  MSGNN-DTA & DTA & Comb-Attention-FC & GCN-GAT & GCN-GAT & GCN-GAT\\
  GEFA & DTA & EF-Comb-Linear & \makecell{GCN with residual blocks, \\Linear, Max pooling} & GCN with residual blocks, Linear & N/A\\de Fusion\\
  S2DTA & DTA & Comb-FC & GNN (GCN) & GNN (GCN) & 1D-CNN\\
  STAMP-DPI/X-DPI & DPA/DPI & Comb-FC & GCN, Mol2Vec & GCN, TAPE & Transformer decoder (BERT)\\
  HGTDP-DTA & DTA & Comb-FC & GCN & GCN & Transformer+Multi view fusion\\
  NG-DTA & DTA & Comb-FC & GNN & \makecell{GNN+Conformer\\+Attention weighted sum} & N/A\\
  GraphScoreDTA & PLA & Comb-Dense & Bi-transprort+GNN-GRU & Bi-transprort+GNN-GRU & \makecell{Distance dependant\\ GNN+SCs}\\
  CGraphDTA & DTA & Comb-FC & \makecell{NN, CNN with \\kernel X, Variant GNN} & \makecell{NN, CNN with \\kernel X, Variant GNN} & Mol2vec, PSSM, HMM\\
  GNPDTA & DTA & Comb-FC & Pre-trained GIN+2D-CNN & Pre-trained GIN+2D-CNN & N/A\\
  AttentionSiteDTI & DTI/DTA & Comb-MLP & Topology Adaptive Graph CNN & Topology Adaptive Graph CNN & SA\\
  AttentionMGT-DTA & DTA & Comb-attention-FC & Graph Transformer & Graph Transformer, cross CA & ESM-2\\
  GTAMP-DTA & DTA & Comb-attention-FC & Graph Transformer & Graph Transformer & ESM-2, Mol2Vec\\ 
  PLANET & PLA & Comb-FC & Graph encoder(JT-VAE) & EGCL & \makecell{CA, Conformation\\ and Interaction Modules}\\
  EMPDTA & DTA & Comb-DNN & RGCN, MolFormer & \makecell{Cloud Points Sampling\\+Quasi-Geodesic Convolution\\+MLP} & ESM-2, GearNet\\
  CASTER-DTA & DTA & Comb-FC & Geometric Vector Perceptron  & Geometric Vector Perceptron & CA\\
  \\
  Graph-CNN & DTI & Comb(Average)-FC & GCN & GCN & N/A\\
  \bottomrule
  \end{tabular*}
\end{table*}

\subsubsection{Structural-based methods}

Since the 1D sequence is not a natural representation for molecules, some important structural information of drugs could be lost, degrading model prediction performance.
Structural features can be acquired from 3D or 2D structures of drugs and proteins to predict the interaction between them.
Modeling the 3D structure of proteins is a challenging task, and the 3D structure of proteins is not always available.
For unknown or rare protein structures and functions, models prediction effectiveness may be limited.
Table \ref{tab:structure_based_models} shows the summary of structural-based models and corresponding inputs are shown in the table in Appendix C.

% \subsubsubsection{DTA methods}
\begin{enumerate}

\item DGraphDTA \cite{jiang2020drug} constructs target graphs (contact maps) from the corresponding protein sequences via the protein structure prediction methods and applies GNNs to mine structural information hidden in drug molecular graphs and target molecular graphs.

\item GEFA (Graph Early Fusion Affinity) \cite{nguyen2021gefa} transfers a drug molecule graph to a protein graph for learning via an attention mechanism which is called fusion.
This fusion enables the model to learn the interaction between drugs and proteins in a novel approach.
This model uses GCNs for molecular graph of drugs and contact map graphs of proteins.
After the fusion, GEFA concatenates the outputs and uses a FC layer for DTA prediction.

\item GSAML-DTA \cite{liao2022gsaml} used the molecular graph representation for drug Smiles and protein graphs created from contact maps to feed the separately into a GAT-GCN learning module.
The output features were concatenated and optimized via mutual information and fed into a FC head for affinity prediction.

\item HGRL-DTA \cite{chu2022hierarchical} utilized a hierarchical graph representation learning model for predicting DTA. 
Drug-target binding affinity data was represented as an affinity graph, with drugs and targets as vertices within the graph. 
Also, drugs and targets were represented as molecular graphs, respectively. 
GNN was used to learn global-level affinity relationship within the affinity graph and another GNN was also used to separately capture the local chemical structural features of drugs and proteins. 
Through a message propagation mechanism, the learned hierarchical graph information was integrated, and the structural features of drugs and targets were refined using GCN. 
Finally, features were combined and a FC network was used to predict DTA.

\item WGNN-DTA \cite{jiang2022sequence} constructs weighted protein and molecular graphs through sequence and SMILES that can effectively reflect their structures. 
To utilize the detail contact information of protein, graph neural networks (GAT,GCN) are used to extract features and predict the binding affinity/interaction.
Unlike DGraphDTA which required extensive database scanning, particularly during the sequence alignment step, WGNN-DTA uses Contact map of proteins which are carefully built by the ESM model. 

\item Ma et al. \cite{ma2022drug} used the graph structure of drugs and proteins (graph from contact maps acquired from Pconsc4 with PSSM features) and tested different graph models (S2GC, GCN, GAT, GraphConv, SAGE) for DTA prediction.

\item MSGNN-DTA \cite{wang2023msgnn} employed a multi-scale graph construction approach.
Two graphs (atomic level and motif level graph) were constructed for drugs and one graph (weighted graph from target sequence contact map using ESM-1b \cite{rives2021biological}) for proteins.
After learning each one with GNNs, attention was employed to fuse the multi-scale structural features and generate a join feature representation.
For the prediction of DTA, a FC head was used.

\item PLANET \cite{zhang2023planet} takes the graph represented 3D structure of the binding pockets on the target proteins as input of Equivariant Graph Convolutional Layer (EGCL) and the 2D chemical structure of the ligand molecules as input of a multi-head attention mechanism and updates the features using a protein ligand communication module (cross-attention).
This model is trained through a multi objective process with three tasks, including deriving the protein-ligand binding affinity, protein-ligand contact map, and intra-ligand distance matrix.

\item HiSIF-DTA \cite{bi2023hisif} is a hierarchical semantic information fusion framework for DTA prediction.
In this framework, a hierarchical protein graph is constructed that includes not only contact maps as low-order structural semantics but also protein-protein interaction (PPI) networks as high-order functional semantics.

\item CGraphDTA \cite{wang2023fusion} transforms the protein structure into graph with node features calculated by PSSM, HMM, and DSSP \cite{kabsch1983dictionary} and drug SMILES to molecular graphs with node features calculated by pre-training features from Mol2Vec.
Neural network layer (NNL), Convolutional neural network with kernel x (CNNX), and Variant GNN (VGNN) is used to predict DTA.

\item S2DTA \cite{zeng2023fusing} creates a heterogeneous graph from protein graph, pocket graph (enriched with 1D CNN semantic feature extraction from both Protein sequences and pocket sequences), and molecular graph of drugs which a GNN (GCN) is applied on it.
Final features are concatenated and a FC is used for DTA prediction.

\item NG-DTA \cite{tsui2023ng} takes molecular graphs of drugs and n-gram molecular sub-graphs of proteins as inputs which are then processed by graph neural networks and GNNs with a Conformer block, respectively.
A Conformer block consists of a feed-forward module, a multi-head self attention module, and a convolution module.

\item GraphscoreDTA \cite{wang2023graphscoredta} takes the combination of GNNs (GRU-GNN), bi-transport information mechanism, and physics-based distance vina terms into account.
This model uses 2D graphs of proteins (from contact maps), ligand-protein sub-graphs (from interaction contact maps), and graphs of drugs to predict the DTA task.

\item HGTDP-DTA \cite{xiao2024hgtdp} utilizes dynamic prompts (context-specific prompts for each drug-target pair) within a hybrid Graph-Transformer framework.

\item GTAMP-DTA \cite{tian2024gtamp} uses two separate graph transformer on molecular graph of drugs and pocket graph of proteins.
After using an attention layer, it concatenates features from pre-trained Mol2Vec and ESM-2 with its current output features and utilizes a MLP layer to predict the task.

\item AttentionMGT-DTA \cite{wu2024attentionmgt} is a method very similar to GTAMP-DTA which uses two attention modules alongside graph transformer module on pocket graph and molecular graph of drugs.

\item GNPDTA \cite{ye2024graph}  uses two pre-training models which are utilized to extract low-level features from drug atom graphs and target residue graphs. 
Then, two 2D convolutional neural networks are employed to combine the extracted drug atom features and target residue features into high-level representations of drugs and targets.

\item EMPDTA \cite{huang2024empdta} integrates protein pocket prediction and DTA prediction.
This framework consists of pocket online detection (POD), multi-modal representation learning for affinity prediction, and multi-task joint training.
EMPDTA uses MolFormer and RGCN for drug graphs, ESM-2b and GearNet for pocket graphs, and POD module (Cloud Points Sampling, Quasi-Geodesic Convolution, and MLP) for protein structures from Alpha-fold structures.

\item CASTER-DTA \cite{kumar2024caster} makes use of an SE(3)-equivariant graph neural network to learn more robust protein representations alongside a standard graph neural network.
This model uses Geometric Vector Perceptron (GVP) with GNN (Graph Isomorphism Network with Edge Enhancement) for drugs and GVP-GNN for proteins.
Finally, a cross attention module fuses two pipelines together for DTA prediction.

\end{enumerate}

% \subsubsubsection{DTI methods}
\begin{itemize}

\item Graph-CNN \cite{torng2019graph} uses protein pocket graphs and molecular graphs for drugs and two GCN layers and a FC as a head to predict DTI.

\item STAMP-DPI/X-DPI \cite{wang2021x, wang2022structure} constructed a structure-aware graph neural network method from the protein sequence by combining predicted contact maps and graph neural networks.
This method utilizes Mol2Vec for drugs to obtain drug drug associations, and a pre-trained BERT to learn embedded representations from a large set of unlabeled protein sequences provided by TAPE.
A transformer decoder head is applied on the output embeddings from drug and protein pipelines.

\item AttentionSiteDTI \cite{yazdani2022attentionsitedti} utilizes attention in molecular graphs of drugs and binding pocket graphs of proteins to find the important nodes and edges and predict DTI/DTA more accurately.

\end{itemize}

\subsubsection{Sequence-Structural-based methods}

These methods use the structural features and sequences of drugs and Proteins and try to capture the local and global patterns of the proteins and drugs.
The structural features allow the model to learn the relations in 2D/3D structure of the proteins and drugs, while the sequence features provide the model with the amino acid and atom level information.
Table \ref{tab:hybrid_based_models} shows the summary of hybrid-based models and corresponding inputs are shown in the table in Appendix C.

\begin{table*}[htbp]
  \centering
  \footnotesize
  \caption{Summary of structural-sequence-based models.}
  \tiny
  \label{tab:hybrid_based_models}
  \begin{tabular*}{\textwidth}{@{\extracolsep{\fill}}lcccccr}
  \toprule
  Model & Task & Architecture & Drug/Ligand models & Protein/Target models & Extra models \\
  \midrule
  ViDTA & DTA & Comb-FC & Graph transformer & 1D-CNN & Attention\\
  PADME & DTA & Comb-FC & GCN+MLP/MLP & MLP & N/A\\
  Wang et al. & DTA & Comb-FC & GNN(GCN,GAT,GIN,GAT-GCN) & CNN with TF-IDF & N/A\\
  GraphDTA & DTA & Comb-FC & GCN,GAT,GIN,GAT-GCN & CNN & N/A\\
  LSTM-SAGDTA & DTA & Comb-FC & GCN+GAT+SAG-pool & LSTM+1D-CNN & N/A\\
  T-GraphDTA & DTA & Comb-FC & GAT-GCN & Pretraining module & N/A\\
  SAG-DTA & DTA/DTI & Comb-two FC & \makecell{GCN+SAGpool\\(hiearchichal and global)} & 1D-CNN & N/A\\
  DeepGLSTM & DTA & Comb-FC & multiblock GCN & BiLSTM & GCN, Global pooling\\
  DeepNC & DTA & Comb-FC & GEN, HGC-GCN & 1D Convs & N/A\\
  Xia et al. & DTA & Comb-FC & GCN & 2D-CNN with word2Vec & N/A\\
  GPCNDTA & DTA & Comb-FC & Residual CensNET & Residual EW-GCN & SA, CA\\
  MGraphDTA & DTA/DTI & Comb-MLP & Deep multi-scale GNN & Multi-scale CNN & N/A\\
  GanDTI & DTA/DTI & Comb-MLP & Residual GNN & Attention on other pipelines & N/A\\
  Tian et al. & DTA & Comb-FC & GNN & LSTM & BiGRU\\
  SSR-DTA & DTA & Comb-FC & MLP, Multilayer GNN with GAT & Multilayer GNN with BIGNN & SSA-Fusion block\\
  PocketDTA & DTA & Comb-FC & FCN, GraphMVP-Decoder & FCN, GVP-GNN layers & MHBAN\\
  ColdDTA & DTA & Comb-Attention head-FC & GNN & CNN & N/A\\
  KC-DTA & DTA & Comb-FC & GNN & 2D-CNN, 3D-CNN & FC\\
  AGraphDTA & DTA & Comb-FC & GNN & GNN, CNN & N/A\\
  PGraphDTA & DTA & Comb-FC & GAT & \makecell{PLMs (DistilProtBert,\\ ProtBert, Seq2Vec)} & N/A\\
  EmbedDTI & DTA/DTI & Comb-FC & \makecell{Graph attention layer\\ and GCN+Pooling} & 1D-TextCNN & GloVe for proteins\\
  DeepRelations & CPA/CPI & \makecell{Inter-attention\\ in Rel-CPI-FC} & GCN, Attention & RNN, Attention & N/A\\
  PSG-BAR & PLA & Comb-Linear & RGATs+Pool & RGATs+Attention weighted pooling & Interaction attention module\\
  GEFormerDTA & DTA & Comb-FC & Attention, Maxpool & GCN, res blocks, graph pooling & Linear\\
  GraphCL-DTA & DTA & Comb-FC & GCN & 1D-CNN & Graph constrastive learning\\
  DeepFusionDTA & DTA & Comb-FC-LightGBMs & \makecell{StruM(Dilated-CNN),\\ SeqM(Fusion of\\ Dilated-CNN and BiLSTM)} & N/A\\
  BACPI & CPI/CPA & Comb-BiAttention-FC & GAT & CNN & Bidirectional Attention\\
  SSGraphCPI & CPA & Comb-FC & \makecell{RNN encoding,\\ Attention 1D-CNN} & \makecell{RNN encoding,\\ Attention 1D-CNN} & GCNN+Attention\\
  DEELIG & PLA & Comb-FC/1pipe-FC & FC & 3D-CNN & N/A\\
  DLSSAffinity & DTA & Comb-FC & CNNs & CNNs & FCNNs\\
  GDGRU-DTA & DTA & Comb-FC & Gated graph+TransformerConv & GRU/BiGRU & N/A\\
  DeepGS & DTA & Comb-FC & GAT, BiGRU with Smi2Vec & CNN with Prot2Vec & N/A\\
  DeepH-DTA & DTA/DTI & Comb-FC & Bi-Directional ConvLSTM & Dense+SE block & HeteroGAT\\
  PLA-MoRe & PLA & Comb-FC & GIN & Transformer encoder  & Auto encoder\\
  DeepTGIN & PLA & Comb-FC & GIN & Transformer & Transformer\\
  NHGNN-DTA & DTA & Comb-FC & BiLSTM, Multilayer GIN & BiLSTM, Multilayer GIN & MHSA\\
  HBDTA & DTA & Comb-FC & \makecell{MHGAT+GCNs/\\MHGAT+GenConv} & multi-layer BiLSTM & Attention\\
  GK BertDTA & DTA & Comb-MLP & GIN, KB-BERT & DenseSENet & yes\\
  MFR-DTA & DTA/DTI & Comb-Attention-FC & BioMLP Block & BioCNN Block & Elem feature fusion block\\
  3DProtDTA & DTA & Comb-FC & GNNs+Dense network & GNNs+Dense network & Dense layers\\
  SubMDTA & DTA & Comb-FC & pretraining GIN encoder & BiLSTMs & N/A\\
  DataDTA & DTA & Comb-Attention-MLP & \makecell{Linear, Residual\\ dilated gated CNN} & Linear, Residual dilated gated CNN & Dual-interaction module\\
  UCMPNN-DTA & DTA & Comb-FC & Undirected-CMPNN & MLM and CPCprot & Attention\\
  IMAEN & DTA/DTI & Comb-FC &  MSAM & ISCEM & N/A\\
  TDGraphDTA & DTA & Comb-FC & \makecell{Diffusion-based graph\\ optimization+graph conv} & Convs+skip maxpools & Multi scale interaction block\\

  \\
  DrugBAN & DTI & Comb-FC & GCN & CNN & \makecell{Bilinear interaction attention\\ module+pooling, CDAN}\\
  Lin et al. & DTI & EarlyComb-FC & GCN & GCN & N/A\\
  GraphCPI & CPI & Comb-FC & GNN (GCN/GAT/GIN) & Prot2Vec + CNN & N/A\\
  TransformerCPI & CPI & \makecell{Comb-Transformer\\ Decoder-FC} & GCN & (Conv1D) with gated linear units & N/A\\
  FDTIIT & DTI & Comb-FC & GCN & multilayer CNN & \makecell{Mutual attention fusion network,\\ Interaction information tradeoff}\\
  Kim et al. & DPI & Comb-FC & GraphNet & Transformer+CNN & N/A\\
  DeepMGT-DTI & DTI & Comb-FC & MCGCN+Transformer & CNN+FC & N/A\\
  IIFDTI & DTI & Comb-FC & GAT & CNN & Bidirectional encoder decoder\\
  MINDG & DTI & Comb-FC & MPNN+CNN & MPNN+CNN & HOAGCN\\
  DTIAM & DTI/DTA/MoA & Comb-FC & \makecell{Transformer,\\ MaskedLM+DescriptorPred+\\FunctionalPred} & \makecell{Transformer,\\ MaskedLM+ContactPred} & N/A\\
  DeepCPI, GNN-CPI & CPI & Comb-Attention-Softmax & GNN & CNN & Attention\\
  CPGL & CPI & \makecell{Vertical \\Comb-FC-Attention} & GAT & BiLSTM & Two-sided attention\\
  MHSADTI  & DTI & Comb-FC & GAT & GAT & MHSA\\
  CPInformer & CPI & Comb-Conv-FC & SA & Prob SA & N/A\\
  CaoDTI & DTI & Comb-FC & GraphSAGE & Transformer & Co-attention module\\
  DrugormerDTI & DTI & Comb-FC & Drug encoder & Protein encoder & Transformer decoder\\
  CSDTI & DTI & Comb-FC & Drug molecule aggregator & Encoder (Conv1D) & CA\\

  \bottomrule
  \end{tabular*}
\end{table*}

% \subsubsubsection{DTA methods}
\begin{itemize}

\item Interpretable drug-target Prediction Using Deep Neural Representation \cite{gao2018interpretable} uses LSTMs for protein sequences, and modified CNNs for Drug 2D molecular graph, and a FC for head.

\item PADME \cite{feng2018padme} introduces two model architectures named PADME-ECFP and PADME-GraphConv which are trained with each drug-target pairs.
PADME-ECFP uses ECFP and PSC with MLP and PADME-GraphConv uses embeddings from Molecular Graph Convolution (MGC) and PSC with MLP to predict DTA.

\item Wang et al. \cite{wang2020dipeptide} propose a DTA model based on dipeptide frequency of word frequency encoding and a hybrid graph convolutional network. 
Word frequency characteristics of natural language are used to improve the frequency characteristics of peptides to express target proteins. 
The obtained protein features and drug graph structure are used as the input of convolution neural network and the input of graph convolution neural network, respectively.

\item DeepGS \cite{lin2020deepgs} utilizes Smi2vec with BiGRU, and molecular graphs with GATfor drugs, and Prot2Vec with CNN for proteins.
All of these outputs are concatenated together and were fed into a FC layer for prediction.
The contextual information contained in protein and drug sequences are considered using Prot2Vec and Smi2Vec.

\item DeepRelations / DeepAffinity+ \cite{karimi2020explainable} use protein sequences and compound graphs in DeepRelations model with three novel module called Rel-CPI to integrate different relations in drug-protein interaction prediction learning.
These relations, including predicted solvent exposure, intermediate predicted k-mer binding residues, intermediate predicted binding residues, and predicted protein contact map, were used as inputs to different Rel-CPI modules.

\item DeepH-DTA \cite{abdel2020deeph} propose using a heterogeneous graph attention (HGAT) for topological information of compound molecules, bidirectional ConvLSTM layers for modeling spatio-sequential information in SMILES, and a squeezed-excited dense convolutional network (SE) for protein sequences.

\item GanDTI \cite{wang2021gandti} employs a residual graph neural network for the compound molecular fingerprint data and forms a vector that could project product-based attention on the protein sequence to determine the binding importance on the sequence. 
Then the two parts of the data are concatenated and a MLP is used for prediction.
The results are reported for both DTI and DTA tasks.

\item DeepFusionDTA \cite{pu2021deepfusiondta} utilizes sequence (SMILES and Protein sequences) and structure information (protein secondary structure, molecular fingerprint) of protein and drugs to generate fusion feature map of candidate protein and drug pair through StruM (Dilated-CNN) and SeqM (fusion of Dilated-CNN and BiLSTM) modules. 
Then it applies bagging-based ensemble learning strategy for regression prediction with Light Gradient-Boosting Machines. 
Similar framework to DeepFusionDTA is DFDTA-MultiAtt \cite{jha2024original} which uses a dilated-CNN block and Bi-LSTM with a multi-attention module.

\item GraphDTA \cite{nguyen2021graphdta} leveraged the structural features of drugs meaning molecular graphs and one-hot encoding of protein sequences.
A 3-layer GNN (GCN, GAT, GIN, GAT-GCN) for the extraction of structural features for drugs, a 3-layer 1D-CNN for Protein sequences and FC as head was used to predict DTA.

\item SAG-DTA \cite{zhang2021sag} introduces global or hierarchical pooling with GCN to aggregate node representations weightedly for molecular drugs. 
1D-CNNs are used for proteins and two FC layers for the final output.

\item EmbedDTI \cite{jin2021embeddti} leverages GloVe for pre-training amino acid feature embeddings, which are then fed into a 1D-TextCNN. 
Both an atom graph and a substructure graph is constructed to capture structural information at different levels for drugs, used by GCN.
Finally, the outputs are concatenated and FC layers are applied to predict DTA and labels.

\item DEELIG \cite{ahmed2021deelig} used different features for pocket protein structure and ligands.
This model employed atomic (one pipeline) and composite (two pipelines) models with protein-ligand and ligand features.

\item SSGraphCPI \cite{wang2022ssgraphcpi} is composed of RNNs with an attentional mechanism and GCNNs on SPS, SMILES and molecular graph data.

\item PLA-MoRe \cite{li2022pla} applies an autoencoder model on bio-active property of drugs, GIN on molecular graph of drugs and transformer encoder on protein sequences.

\item MGraphDTA \cite{yang2022mgraphdta} also, used the graph of drugs and target sequences in form of an embedded vector. 
However, it enhanced the global structural features extraction by employing a deeper multi-scale GNN (MGNN) (GCN with 27 layers), inspired by dense connections. 
Simultaneously, multi-scale CNN (MCNN) was applied to extract multi-scale features from target sequences. 
The resulting fused features were concatenated to form a combined representation of drug-target pair. 
Finally, the combined representation was fed into MLP to predict DTA/DTI.
Drug-Online is an online platform for drug-target interaction that uses MGraphDTA and other models. \cite{zeng2024drug}

\item Tian et al. \cite{tian2022predicting} improved GraphDTA model and changed the model's channel to a triple-channel.
They interpreted the protein sequences as time series and used LSTM, used GNNs for drug molecular graphs and BiGRU for local chemical background of drugs.

\item DeepNC (Deep Neural Computation) \cite{tran2022deepnc} utilized Generalized Aggregation Networks (GENConv), Graph Convolutional Networks (GCNConv), and Hypergraph Convolution-Hypergraph Attention (HypergraphConv). 
This framework learns the features of drugs and targets by the layers of GNN and 1D convolution network, respectively. 
Then, representations of the drugs and targets are fed into fully-connected layers to predict the binding affinity values.
This study also proposes Allergy dataset to test their model further.

\item DeepGLSTM \cite{mukherjee2022deepglstm} uses a parallel GCN module composed of three GCNs with different layers for molecular graphs. Alongside the drug module, it adopts a bi-LSTM for protein sequences. 

\item GDGRU-DTA \cite{zhijian2022gdgru} utilizes GRU/BiGRU for proteins and GNNs for compounds.
The GRU model processes the same task more quickly than LSTM because it has a different and more simple internal structure.

\item BACPI \cite{li2022bacpi} uses a bidirectional attention module on the outputs of GAT for molecular graphs for drugs and CNN for amino acid sequences of proteins in 3 character words.

\item Xia et al. \cite{xia2022drug} proposes a DTA model based on graph neural networks and word2vec. 
In this model, the word embedding method is used to convert proteins sequence into sentences containing words to capture the local chemical information. 
SMILES is used to convert drug molecules into graphs. After feature fusion, DTA is predicted by graph convolutional networks.

\item PSG-BAR (Protein Structure Graph-Binding Affinity Regression) \cite{pandey2022ligand} utilizes 3D structural information of proteins, ESM of Protein sequences, and 2D graph of ligands.
Framework consists of residual graph attention layers (RGAT), pooling for drugs, attention weighted pooling for graphs of protein contact maps, and an interaction attention module learning the interaction between two pipelines.
Output of RGAT for drugs, RGAT for protein graphs, and ESM for protein sequences are concatenated and are used for linear layers to predict PLA.

\item DLSSAffinity \cite{wang2022dlssaffinity} uses the pocket-ligand structural pairs as the local information to predict short-range direct interactions and uses the full-length protein sequence and ligand SMILES as the global information to predict long-range indirect interactions.

GeneralizedDTA \cite{lin2022generalizeddta} introduces pre-training tasks for drugs and proteins and a multi-task learning model with dual adaptation mechanism (MAML-based updating strategy \cite{finn2017model} ) for addressing the forgetting problem of pre-training parameters.

\item T-GraphDTA \cite{wu2023t} developed a novel protein pre-training method (PTR) for protein representation learning, then proposed a DTA prediction framework based on PTR and hybrid graph neural network (GAT-GCN).

\item AGraphDTA \cite{wang2023agraphdta} employs GNNs to extract graph features of drugs and proteins. 
Then it employs CNN to extract amino acid sequence features. 
Also, a fusing block is employed to generate the fusion feature that represent the complex information of proteins.

\item ColdDTA \cite{fang2023colddta} enhanced the model's generalization performance by data augmentation and attention-based feature fusion techniques.
It used structural knowledge of drugs and target sequence information.
Firstly, a new drug-target pair was generated by removing a sub-graph from the original graph of drug. 
Next, the structural features of drug and the sequence features of target were extracted using GNN and CNN, respectively.
These extracted features were then fused via an attention-based fusion block and the fused features were fed into MLP to predict DTA. 

\item DataDTA \cite{zhu2023datadta} uses four different inputs, including Protein sequences and drug SMILES strings, as well as binding pocket descriptors (acquired from a pre-trained scorer) and AG-FPs. 
Then, it is concatenated after an embedding module with CNN sub-modules and the fusion strategy with a highway block and a multihead attention block (dual-interaction module) was designed to integrate and capture multi-scale interaction information.
Finally, a MLP is used for the DTA prediction.

\item MFR-DTA \cite{hua2023mfr} uses amino acid embedding and word embedding for protein feature representation, and uses functional-connectivity fingerprints (FCFPs) and graph neural network (GNN) features for drugs. 
The proposed BioMLP module assists the model in extracting individual features of sequence elements and an Elem feature fusion block to refine the extracted features. 
A Mix-Decoder block was designed to extract drug-target interaction information and predicts their binding regions.

\item UCMPNN-DTA \cite{xia2023drug} uses an undirected cross graph message passing neural network (CMPNN) for molecular embedding and combines CPCProt and MLM models for protein embedding.
This model also uses an attention to find the important parts of proteins.
MLP head was used to predict the final result and framework is evaluated with MSE and CI.

\item KC-DTA \cite{jiang2023deep} utilizes sequence information learning and converts protein sequences to 2D and 3D matrices using k-mer analysis and Cartesian product calculation.
This model applies GNN for drug molecular graphs and 2D and 3D CNNs for the calculated matrices.

\item SubMDTA \cite{pan2023submdta} pre-trains a GIN encoder using constrastive learning method for drug feature extraction.
This model also uses 3 Bidirectional LSTMs for 2,3, and 4 gram protein sequences.

\item PGraphDTA \cite{bal2023pgraphdta} utilizes GAT and CNN for molecular graphs of drugs and proteins, respectively. 
It also uses several protein language models for protein sequences and adds contact maps as an extra input to improve its results.

\item TDGraphDTA \cite{zhu2023drug} employs diffusion-based optimization for drug molecular graphs and uses Convs for protein sequences.
This model utilizes a multi head (two) cross-attention module which uses a drug query matrix, a protein key matrix and a protein value matrix for one cross attention and the other cross attention uses the rest of the matrices. 

\item GPCNDTA \cite{zhang2023gpcndta} utilizes GNN module, the linear projection and self attention layer to extract features from graphs of drugs (from SMILES) and proteins (from protein data bank and Alpha-fold structure predictions) and sequences of drugs (pharmacophore, SMILES) and proteins (FASTA).
This model used residual CensNet and the residual EW-GCN to correspondingly extract features of drug and protein graphs and devised intramolecular and intermolecular cross-attention to respectively fuse and interact features of drugs and proteins.

\item 3DProtDTA \cite{voitsitskyi20233dprotdta} uses structures obtained from Alpha-fold model for proteins, sequences (Morgan fingerprints) and structures of drugs with GNNs.

\item NHGNN-DTA \cite{he2023nhgnn} feed the text embeddings into an adaptive feature generator including a tokenizer, BiLSTM and LayerNorm layers and the output is fed into the multi-head self-attention layer.
This model combines the graphs of proteins and drugs with a central node and integrates adaptive node features, which are then fed into a multilayer GIN.
The sequence-based and graph-based prediction results are integrated to obtain the final DTA prediction.

\item GraphCL-DTA \cite{yang2024graphcl} employs a graph constrastive learning framework for drugs which allows the model to preserve the semantic information of drugs without data augmentation methods i.e. dropout-based augmentation strategies.
GCN and 1DCNN is used for drug graphs and protein sequences, respectively.

\item IMAEN \cite{zhang2024imaen} employs a molecular augmentation mechanism in its Molecular Structure Enhancement Module (MSAM) to enhance drug molecular structures by fully aggregating molecular node neighborhood information. 
It then uses multi-scale GCN and CNN in its interpretable stack convolutional encoding module (ISCEM) for drug and protein processing, respectively. 

\item ViDTA \cite{li2024vidta} adds virtual nodes to drug molecular graphs and applies a graph-transformer-based encoder.
This model uses 1D-CNNs for protein sequences and fuses the features using an attention-based linear feature fusion block.

\item DeepTGIN \cite{wang2024deeptgin} uses graph isomorphism network (GIN) encoder and transformer encoder on molecular graphs and protein/pocket residue sequences, respectively.

\item PocketDTA \cite{zhao2024pocketdta} enhances the generalization performance by using pre-trained models like ESM-2 and GraphMVP.
This framework utilizes both sequences and structural information alongside pocket information for precise DTA prediction.
GVP-GNN layers are used for pocket inputs and a multi-head bilinear attention network is used for generating and learning the interaction matrix.

\item GEFormerDTA \cite{liu2024geformerdta} proposes a framework that considers bond encoding, degree centrality encoding, spatial encoding of drug molecule graphs, and the structural information of proteins such as secondary structure and accessible surface area.
This method uses a graph early fusion module that fuses drugs and protein graphs and applies GCN to obtain graph-based representation of proteins.
Information in this framework is passed between protein and drug pipelines to learn the interaction better.

\item LSTM-SAGDTA \cite{qiu2024lstm} utilizes LSTM and CNN for protein sequences from seqvec (ELMo) model and two encoders consisting GCN, GAT, and SAG-Pool in different order for molecular graphs of drugs.

\item HBDTA \cite{wu2024hbdta} employs multi-head graph attention networks (Multi-head GAT), generalized aggregation networks (GENConv), and graph convolutional networks (GCNConv) for drugs.
This model utilizes a multi-layer bi-directional long short-term memory (MBLSTM) with residual blocks to extract protein features.

\item GK BertDTA \cite{qiu2024gk} utilized GIN for molecular graphs of drugs, DenseSENet for proteins, and knowledge-based BERT semantic model on SMILES of drugs to obtain rich pre-trained semantic embeddings.

\item SSR-DTA \cite{liu2024ssr} uses BIGNN and a multi-layer graph network capable of adapting to diverse structural sizes, which enables the capture of molecular motifs across different scales, ranging from atomic to macro-cyclic motifs.
\end{itemize}

% \subsubsubsection{DTI Methods}
\begin{itemize}

\item DeepCPI/GNN-CPI \cite{tsubaki2019compound} used a GCN to learn compound features and used CNN to learn protein sequence features. 
They used an attention mechanism to compute the attention coefficients, which considers the affinity, and the protein's feature vector is obtained using the weighted sum of protein sub-sequence's features along with attention coefficients.

\item GraphCPI \cite{quan2019graphcpi} employed 3-gram encoding with pretrained Word2Vec (Prot3Vec) to process protein sequences, followed by a CNN to handle the protein embeddings.
Drug molecular graphs are fed into a GCN to extract their features, and the outputs are combined and a FC layer is used. 
This work claims that they were first to combine the local chemical context and topological structure to learn the interaction between compound protein pairs.

\item TransformerCPI \cite{chen2020transformercpi} used word2vec for protein sequences and transformer encoder (1Dconv) and GCN for molecular representation of drugs.
This model aims to learn the interaction with transformer decoder module. 

\item Kim et al. \cite{kim2021bayesian} used transfer learning in encoding protein sequences with a pre-trained model, which trains general sequence representations in an unsupervised manner. 
GraphNet model is used for molecular graphs of drugs and 1D CNNs is utilized for proteins.
They used a Bayesian neural network to make a robust model by estimating the data uncertainty.

\item MHSADTI \cite{cheng2021drug} employs GAT model for drugs and proteins with multi-head self-attention mechanism.

\item CaoDTI \cite{huang2022coadti} incorporates the Co-attention mechanism to learn the interaction between drugs and targets.
This method uses graphSAGE for R-radius sub-graphs from SMILES of drugs and a transformer for protein sequences.

\item IIFDTI \cite{cheng2022iifdti} considers interactive features of local substructures of drug-target pairs with bidirectional encoder decoder and independent features for both targets and drugs with CNN and GAT, respectively.

\item Lin et al. \cite{lin2022detecting} constructed a drug-target interaction graph based on fusing drug and target similarity graphs using similarity network fusion (SNF) method. 
Graph isomorphic network (GIN) and TextCNN were used for drugs and proteins, respectively and GCN is used for the constructed DTI graph.

\item CPGL \cite{zhao2022cpgl} uses GAT and Bidirectional LSTM for drug substructures and 3-grams of Protein sequences, respectively.
CPGL also utilizes a two-sided attention mechanism to capture the interaction between drugs and proteins and utilizes a FC to predict DTI based on concatenated features of proteins and compounds.
This model experiments on label reversal datasets to ensure its robustness and generalization.

\item CPInformer \cite{hua2022cpinformer} extracts heterogeneous compound features, including structural graph features and functional class fingerprints, to reduce prediction errors caused by similar structural compounds. 
Then, they combine local and global features using dense connections and apply ProbSparse self-attention to protein features, under the guidance of compound features, to eliminate information redundancy.

\item DeepMGT-DTI \cite{zhang2022deepmgt} utilizes a transformer network incorporating multilayer graph information for drug graphs and a CNN for protein embeddings.

\item CSDTI \cite{pan2023csdti} utilizes a molecular aggregator to learn the structural representation of drug graphs and uses a protein encoder for protein representations acquired from Prot2Vec model.
This model fuses these representations using a cross attention block, receiving protein features as key and values and drug features as queries.

\item DrugormerDTI \cite{hu2023drugormerdti} uses Graph Transformer on the input molecule graph and protein embeddings from Resudual2vec.

\item DTIAM \cite{lu2023dtiam} proposes a unified framework that uses transformer encoder and pre-training for drug molecular substructures and protein sequences and predicts DTI, DTA, and mechanism of action/activation/inhibition (MoA).

\item DrugBAN \cite{bai2023interpretable} employs GCN and CNN blocks to encode molecular graph and protein sequences, respectively. 
Then they use a bilinear attention network module to learn local interactions between the representations of drugs and proteins and generate an interaction map.
This model also tries to generalize to out of domain datasets and integrates a Conditional Adversarial Domain Adaptation (CDAN) module to its framework to better predict DTA. 

\item FDTIIT \cite{he2024flexible} uses flexible mutual attention to extract interaction information about drugs and targets, and then limit the dependence between them with an interactive information trade-off module to avoid redundant information.
This model uses GCN and multiplayer CNN for drugs and proteins, respectively.

\item MINDG \cite{yang2024mindg} utilizes a mixed deep network (MPNN and CNN) to extract sequence features of drugs and proteins, a higher-order graph attention convolutional network (HOAGCN) is proposed to better extract and capture structural features, and a multi-view adaptive integrated decision module (concatenation).

\end{itemize}

\subsubsection{Complex-based models}

Deep learning methods have been introduced to predict DTIs by direct use of 3D protein-compound complexes \cite{wee2021forman, cang2018representability, ballester2010machine}.
Studies like PLIG \cite{moesser2022protein}, PaxNet \cite{zhang2022efficient}, GLI \cite{zhang2022predicting_b}, GIGN \cite{yang2023geometric}, IGN \cite{jiang2021interactiongraphnet}, PotentialNet \cite{feinberg2018potentialnet}, SchNet\cite{schutt2017schnet}, FGNN \cite{dong2023ligand} are examples of these methods.
In some studies, input features were based on 3D matrix defined around pocket-ligand complexes instead of the whole complex.
These methods generated a large number of input variables, and had the problem of limited number of training set.
To overcome this problem, some studies used voxelization, which limits a regular cube range to reduce the number of input variables.
Most of the complex-based models rely heavily on the time-consuming docking process, but they provide more interpretability than complex-free models.
The protein structure prediction models like AlphaFold \cite{tunyasuvunakool2021highly, abramson2024accurate} and RoseTTA \cite{baek2021accurate} can provide better and more accurate templates for target proteins, resulting in better DTI/DTA prediction.
Table \ref{tab:complex_based_models} summarizes the complex-based models.

\begin{itemize}

\item AtomNet is a deep CNN for modeling bioactivity and chemical interactions \cite{wallach2015atomnet}. 

\item Atomic convolutional neural network (ACNN) \cite{gomes2017atomic} was developed for binding affinity by generating new pooling and convolutional layers specific to atoms. 

\item 3D-CNN \cite{jimenez2018k} used 3D CNN with molecular representation of 3D voxels assigned to various physicochemical property channels.

\item Cang and Wei \cite{cang2017topologynet} represented 3D structures in novel 1D topology invariants in multiple channels for CNN.

\item TopologyNet \cite{cang2017topologynet} uses protein ligand complex 3D structures and the element-specific persistent homology (ESPH) method and CNNs to generate representations of these structures for prediction.

\item Pafnucy \cite{stepniewska2018development} utilizes a 3D convolution neural network to produce a feature map of protein ligand 3D structures, followed by dense layers for predicting affinity values.
Also, a voxel representation of protein and ligand complexes is used, which allows the visualization of protein and ligand structures and interactions.
Using a regularization technique, their designed network focused on capturing the general properties of interactions between proteins and ligands.

\item DeepAtom \cite{li2019deepatom} designed a light-weight model based on 3D-CNN which includes Atom information integration block, 3D shuffle groups (3D PW/DwConvs), 3D pointwise Conv, and FCs to minimize two losses for PLA prediction.

\item GNN-DTI \cite{lim2019predicting} employed a GNN which directly incorporates the 3D structure of a protein-ligand complex.
They also applied a distance-aware graph attention algorithm with gate augmentation for DTI prediction.

\item OnionNet-1 \cite{zheng2019onionnet, wang2021onionnet} used features of all inter molecular contacts in circles which are called onions to predict DTA usign a CNN network.
Also, in OnionNet-2, the protein-ligand interactions are characterized by the number of contacts between protein residues and ligand atoms in multiple distance shells.

\item DeepBindRG \cite{zhang2019deepbindrg} applied a 2D CNN and ResNet on protein-ligand complexes in the 2D format of interface spatial information.

\item AK-score \cite{kwon2020ak} utilizes ensemble of several residual networks with multiple channels of 3D convolutional neural network layers to improve the DTA prediction tasks (scoring, ranking, and docking).

\item Jones et al. \cite{jones2021improved} fused spatial graph convolutional neural network (SG-CNN) for distance of pairs and 2D molecular graphs of drug-targets and 3D-CNN for complexes in two levels (mid and late).

\item SIGN \cite{li2021structure} attempts to model the 3D structural complex and protein-ligand spatial interactions with representing complexes as a complex interaction graph and feeding it into polar-inspired graph attention layer (PGAL) and pairwise interactive pooling layer (PiPool).
PGAL can propagate the node's and edge's embeddings alternately with learning the spatial distance and angle information.
PiPool performs on the edges' representations to obtain the atomic type-based interaction matrix of the complexes which tries to approximate the overall interactions.

\item GAT-Score \cite{yuan2021protein} designed a dynamic feature mechanism to enable the model to deal with bond features for graph of drug-target complexes.
Also, a virtual super node is introduced to aggregate node-level features into graph-level features, so that the model can be used in the graph-level regression problems.

\item OctSurf \cite{liu2021octsurf} uses a novel space partitioning structure which has a flexible search hierarchy \cite{riegler2017octnet}.
This paper shows the higher performance of this new representation against dense voxel representation in DTA prediction via 3D-CNN, ResNet, and VGG models.

\item BAPA \cite{seo2021binding} employs 1D conv layers and attention layer for 1D vector of intermolecular descriptors of protein-ligand complexes.
Descriptors were generated focused on contacted protein and ligand atom pairs using nine heavy atoms commonly observed in protein-ligand complexes,.

\item APMNet \cite{shen2021cascade} integrate the ARMA graph convolution (based on an auto-regressive moving average filter) \cite{bianchi2021graph} method and the graph convolution layer in the MPNN module \cite{gilmer2017neural}.  

\item Sfcnn \cite{wang2022sfcnn} employed 3D-CNN to generate a score function for DTA prediction. 
Drug-target complexes were transformed into a 3D grid representation. 
This grid served as input to 3D-CNN, which learned high-level structural features. 
Finally, multiple density layers were applied to the extracted features for DTA prediction.

\item IMCP-SF \cite{wang2022protein} uses profiles of intermolecular contacts (IMCP, IMCPiDB, IMC, IMCiDB, ECIF, APIF, SPLIF, PLEC FP) of proteins and ligands with random forest to score the interaction.

\item KIDA \cite{lu2023improving} builds a complex-free model inspired by knowledge distilation framework and implements a method for modeling protein atomic features by fusing atomic chemical and spatial position features.
This model needs ball query graph and pocket 2D graph of proteins, drug 2D graphs and bipartite graph of complexes to use EGNN, MPNN, student block and teacher block to predict the drug-target affinity.

\item SS-GNN \cite{zhang2023ss} constructs a single undirected graph based on a distance threshold for protein-ligand interactions and applies GNNs and MLP layers.
The novelty of this framework causes the scale of the graph data to be greatly reduced.

\item InterGraph \cite{mekni2023encoding} modeled protein-ligand complexes as 3D topological multi-graphs (generated interaction spheres around each ligand atom, specifically to enclose proximity-based contacts) and fed them to a four-layer graph convolutional neuronal network.

\item MBP \cite{yan2024multi} proposes a pre-training framework for structure-based PLA prediction which constructs a pre-training dataset called ChEMBL-Dock with more than 300k experimentally measured affinity labels and about 2.8M docked 3D structures. 
By introducing multi-task pre-training to treat the prediction of different affinity labels as different tasks and classifying relative rankings between samples from the same bio-assay.

\end{itemize}

\begin{table*}[htbp]
  \centering
  \footnotesize
  \caption{Summary of complex-based models.}
  \tiny
  \label{tab:complex_based_models}
  \begin{tabular*}{\textwidth}{@{\extracolsep{\fill}}lcccccr}
  \toprule
  Model & Task & Architecture & Representation & Model \\
  \midrule
  Kdeep & PLA & 1pipe-FC & Complex & 3D-CNN+Subsampling\\
  DeepBindRG & PLA & 1pipe & Complex (interface spatial info in 2D format) & 2D-CNN model and ResNet\\
  GNN-DTI & DTI & 1pipe-FC & Complex & Embedding and distance-aware graph attention mechanism\\
  Onionnet & PLA & 1pipe-FC & 3D-Complex & CNN\\
  DeepAtom & PLA & 1pipe & Voxelized complex & 3D PWConvs and 3D DWConvs\\
  AK-Score & PLA & 1pipe-Dense & Complex & Ensemble-based residual network 3D convolutional\\
  OctSurf & PLA & 1pipe-FC & Protein pocket ligand complex & 3D-CNN/ResNet/VGG\\
  Jones et al. & PLA & Comb-FC & Complex & 3D-CNN, SG-CNN\\
  APMNet & PLA & 1pipe-Linear & Complex & Linear, ARMA graph conv, MPNN\\
  SIGN & PLA & 1pipe-MLP & Complex interaction graph & PGAL, PiPool\\
  GAT-Score & PLA & 1 pipe & Graph of complex & GAT\\
  BAPA & PLA & 1pipe-FC & Complex using intermolecular descriptors & 1D Conv layer, attention layer\\
  IMCP-SF & PLA & Comb-FC & Profiles of intermolecular contacts & RF\\
  Sfcnn & PLA & 1pipe & 3D & 3DCNN\\
  KIDA & DTA & Comb-FC & Drug 2D graph, protein pocket as 2D chemical graphs, ball query graph & MPNN, EGNN+student and teacher blocks\\
  SS-GNN & DTA & 1Pipe-MLP & Complex & MLP, GNN(GIN)+global graph add pool\\
  MBP & PLA & 1pipe & Graph & encoder+interacting module+readout\\
  InterGraph & PLA & 1pipe & Graph using pairwise atomic distances in complexes & GCN\\
  \bottomrule
  \end{tabular*}
\end{table*}

\begin{table*}[htbp]
  \centering
  \footnotesize
  \caption{Summary of utility-based models.}
  \tiny
  \label{tab:utility_based_models}
  \begin{tabular*}{\textwidth}{@{\extracolsep{\fill}}lcccccr}
  \toprule
  Model & Task & Architecture & Representation & Model \\
  \midrule
  DTINet & DTI/DRep & 1pipe & Heterogenous graphs & RWR+Diffusion component analysis (DCA)\\
  NeoDTI & DTI & 1pipe & \makecell{DSS, DDI, PSS,\\ PPI, DDA, DiDA,\\ side effect, DPI, PDiA} & neighborhood information aggregation\\
  GCN-DTI & DTI & Comb-DNN & DPP graph & GCN+DNN\\
  SBGDTI & DTI & 1pipe-MLP & BiPartite Graph+G17 & Weisfeiler-Lehmanalgorithm, MLP\\
  DTI-CNN & DTI & Comb-Autoencoder-CNN-FC & \makecell{DSide, DDI, DDi\\ and similarities, PPI,\\ PDi and similarities} & \makecell{Jaccard similarity,\\ RWR, deniosing autoencoder,\\ CNN}\\
  EEG-DTI & DTI & Comb-FC & Heterogenous graph & GCN\\
  KGE-NFM & DTI & Comb-MLP & DDi, DDI, PPI, PDi, DPI, other relations & DistMult, NFM, Bi-Interaction layer\\
  Affinity2Vec & DTA/DTI & Comb-FC & \makecell{SMILES, Protein sequences,\\ DTBA Matrix and weighted graph} & ML regressor+ Min-Max normalization\\
  SGCL-DTI & DTI & Comb-FC & Metapaths in heterographs & GCN\\
  MHGNN & DTI & Comb & Metapath-based graphs & GAT, GCN\\
  MSF-DTA & DTA/DTI & Comb-FC & Molecular graphs, PPI, PSeqS & GCNs+Maxpool+FC, VGAE+FC\\
  MHTAN-DTI & DTI & Comb-FC & Heterogenous graphs & Transformer, Attention\\
  EDC-DTI & DTI & Comb-FC & \makecell{Tanimoto coefficient, \\DDI, Smith-Waterman score,\\ PPI, similarity of disease DAGs, DsideA,\\ DDiA, DPA, PDiA} & GAT with memory units+MLP+2DConv\\
  TripletMultiDTI & DTI & Comb-Triplet loss-FC & SMILES, Protein sequences, DDI and PPI & CNN, Node2Vec\\
  DeepTraSynergy & \makecell{DTI,\\ toxic effect,\\ and drug synergy} & Comb-MLP-3 output heads & SMILES, PPI, DTI & transformer, Node2Vec, Synergy module, MHSA\\
  TTGCN & DPI & Comb-FC & \makecell{Neighbour graph based on \\metapaths from hetero networks} & \makecell{ResGCN, GAT+Attention,\\ inductive matrix completion}\\
  CCL-DTI & DTI & Comb-FC & SMILES, protein sequences, PPI, DDI & CNN, Node2Vec\\
  GSL-DTI & DTI & Comb-MLP & MetaPath subgraph & Attention, filter gate, GCN\\
  MSH-DTI & DTI & Comb-FC & \makecell{Molecular graph from SMILES, Target sequences,\\ PPI, DDI, DDS, PDI} & InfoGraph/CPCProt+GCN+Attention\\
  DrugMAN & DTI & Comb-FC & \makecell{DDi, DSide, Transcriptome-based drug similarity,\\ Structure-based drug similarity, Gene-disease,\\ Gene-pathway, Gene-chromosomal location,\\ Transcriptome-based gene sim, PPI,\\ Gene-coexpress, Protein sequence sim} & GAT, MHSA+FC+skips\\
  \bottomrule
  \end{tabular*}
\end{table*}

\subsubsection{Utility networks and heterogeneous graph-based models}

Network-based methods make their prediction based on various biological networks such as drug-drug interaction graph (DDI), protein-protein interaction graph (PPI), or drug-disease graphs (DDi) \cite{olayan2018ddr, hao2017predicting}.
Most of traditional approaches cannot learn the complex relations between ligands and proteins as they are learning just from utility networks in a non-detailed way, however, different methods do not have this limitation, recently.
These methods may use or combine the utility networks \cite{zong2017deep, yan2019prediction, yue2021dti, xuan2020prediction, ning2025dmhgnn, chen2024drug} or make a new heterogeneous graph from ligands and proteins \cite{yildirim2007drug, yu2022hgdti, peng2024dti, zhang2023mhtan, liu2024ssldti}.
Methods can be similarity-based (DTINet \cite{luo2017network}, deepDTNET \cite{zeng2020target}, MEDTI \cite{shang2021prediction}, NEDTP \cite{an2021heterogeneous}, MultiDTI \cite{zhou2021multidti}, HAS-DTI \cite{wang2022sparse}), knowledge graph-based (DistMult \cite{jiang2021heterogeneous}, KGE-NFM \cite{ye2021unified}), GNN-based (NeoDTI \cite{wan2019neodti}, IMCHGAN \cite{li2021imchgan}, DTI-MGNN \cite{li2022drug}, SGCL-DTI \cite{li2022supervised}, MAGNN \cite{fu2020magnn}, MHGNN \cite{li2023metapath}, MLGANN \cite{lu2024multi}, MVGCN \cite{fu2022mvgcn}).
Some methods apply hierarchical learning (MHADTI \cite{tian2022mhadti}) or use metapaths (AMGDTI \cite{su2024amgdti}, DTI-HAN \cite{yu2024drug}) or hybrids of them (MHTAN-DTI \cite{zhang2023mhtan}).
Table \ref{tab:utility_based_models} summarizes the utility based models.

\begin{itemize}

\item DTINet \cite{luo2017network} learns low-dimensional representations of drugs and targets by Random Walk with Restart (RWR) and diffusion component analysis (DCA). 
Then, DTINet finds the optimal projection from the drug space to the target space and predicts new drug-target interactions based on the geometric proximity of the mapping vector in the unified space.

\item NeoDTI \cite{wan2019neodti} in close relation of DTINet \cite{luo2017network} integrates various information from different heterogeneous networks including drug structure similarities, drug side effect associations, drug-protein interactions, drug drug interactions, drug disease associations, protein sequence similarities, protein drug interactions, protein disease associations, protein protein interactions, disease protein associations, disease drug associations, side effect drug associations.
This model learns topology-preserving representations of drugs and targets to predict DTIs by adopting a neighborhood information aggregation operation. 

\item DTI-CNN \cite{peng2020learning} utilizes seven graphs including, drug side-effect graph, drug-drug interaction graph, drug-disease association graph and drug-disease similarity graph, protein-protein graph, protein-disease association graph, and protein-disease similarity graph.
This model applies two jacquard similarity, two Random walk with restart (RWR), and a de-niosing auto-encoder for features extraction and a CNN module for drug-target interaction prediction.

\item SBGDTI \cite{eslami2020drug} constructed a semi-bipartite graph by known DTIs and drug-drug and protein-protein similarity.
They used Weisfeiler-Lehman algorithm in graph labeling process and employed deep neural network to learn the complex pattern of interacting pairs from embedded graphs.  

\item GCN-DTI \cite{zhao2021identifying} constructs DPP networks and assigns equal weights to directly and indirectly related DPP pairs.
GCN is used to learn the topological structure and obtain features and DNN is utilized for prediction.

\item EEG-DTI \cite{peng2021end} constructs a heterogeneous network with multiple entities (i.e. drug, protein, disease, side-effect) and multiple types of edges.
They propose a HGCN-based method to learn the drug and target feature representation based on a heterogeneous network.
Finally, they concatenate the two outputs and predict the DTI task.

\item Affinity2Vec \cite{thafar2022affinity2vec} constructed a weighted heterogeneous graph that integrates data from several sources including similarities and predict binding affinity using s several computational techniques from feature representation learning, graph mining, and machine learning.

\item SGCL-DTI \cite{li2022supervised} learns representations of drugs and proteins in heterogeneous networks through meta-paths using attention. 
The topology and semantic structure of the DPP network are then constructed, and a novel contrastive optimization module is used to generate a collaborative contrastive loss of the two views, and guide the model optimization in a supervised manner.

\item MHGNN \cite{li2023metapath} simulates complex biological entity inter-relationships and uses them for DTI prediction by way of capturing the contextual relationships of the meta-paths. 
The meta-paths of the MHGNN are heterogeneous graphs, and thus can learn complex biological relationships well.
GAT is used for graphs and GCN in utilized for the DTP correlation graph at the end.

\item KG-DTI \cite{wang2022kg} propose a novel knowledge graph-based deep learning method for DTIs predictions. 
Specifically, a knowledge graph of 29,607 positive drug-target pairs is constructed by DistMult embedding method. 
A Conv-Conv module is proposed to extract features of drug-target pairs, which is followed by a FC layer for DTIs prediction.

\item TripletMultiDTI \cite{dehghan2023tripletmultidti} uses DDI and PPI networks and feed them into node2vec modules, then combines it with the output of CNN blocks for drug and proteins with a triplet loss function.

\item DeepTraSynergy \cite{rafiei2023deeptrasynergy} is a model for drug combinations which proposes a new architecture that effectively combines drug-target interaction, PPI, and cell target interaction to incorporate drug synergy prediction using transformers.
Their approach is a multitask approach that predicts three outputs including the drug-target interaction, its toxic effect, and drug combination synergy.
Drug combination synergy is the main task and the two other ones are the auxiliary tasks that help the approach to learn a better model.

\item MSF-DTA \cite{ma2023predicting} gathers additional information for the proteins from its biologically related neighboring proteins in protein-protein interaction and sequence similarity networks to get prior knowledge. 
The representation was learned using a graph pre-training framework, VGAE, which could not only gather node features but also learn topological connections, therefore contributing to a richer protein representation. 

\item EDC-DTI \cite{yuan2023edc} employs ontology-based feature construction (Tanimoto coefficient, Smith-Waterman score, similarity of disease DAGs) and graph topology-based feature construction (DDI, PPI, drug-side-effect, drug-disease, drug-protein and protein-disease associations) and learns them via graph attention network with memory units, MLP and 2DConv.

\item DrugMAN \cite{zhang2024drug} used utility networks such as Drug-Disease, Drug side effect, transcriptome-based drug similarity, structure-based drug similarity, gene-disease, gene-pathway, gene-chromosomal location, transcriptome-based gene similarity, protein-protein, gene-coexpress, protein sequence similarity networks.
This model applies GAT on all of them and concatenates the drug output with protein output and feed it to 5 blocks of multi-head self attention and FC.

\item GSL-DTI \cite{zixuan2024gsl} constructs a heterogeneous network by integrating data from various sources, employs a GCN to learn representations of proteins and drugs by utilizing meta-paths that encode relationships between various types of nodes, constructs the network of Drug-Protein Pairs (DPPs) using a filter gate on the affinity scores of DPPs, and learns low-dimensional representations for each node in this network using another GCN.

\item MSH-DTI \cite{zhang2024msh} uses self-supervised learning methods (CPCProt and InfoGraph) and heterogeneous aggregation from DDI, PPi, and DDS networks with GCN and attention layers.

\item TTGCN \cite{wei2024predicting} utilizes ResGCN and GAT and attention on drug and target neighbor graphs obtained from meta-paths from diverse heterogeneous networks.
Novelty of this framework is that it uses an inductive matrix completion technique to forecast DTIs while preserving the network's node connectivity and topological structure.

\item KGE-NFM \cite{ye2021unified} learns a low-dimensional representation for various entities in the KG, and then integrates the multi-modal information via neural factorization machine (NFM).

\item MHTAN-DTI \cite{zhang2023mhtan} is a meta-path-based hierarchical transformer and attention network which applies meta-path instance-level transformer, single-semantic attention and multi-semantic attention to generate low-dimensional vector representations of drugs and proteins.
Meta-path instance-level transformer employs internal aggregation on the meta-path instances, and models global context information to capture long-range dependencies. 
Single-semantic attention learns the semantics of a certain meta-path type, introduces the central node weight and assigns different weights to different meta-path instances to obtain the semantic specific node embedding. 
Multi-semantic attention captures the importance of different meta-path types and performs weighted fusion to obtain the final node embedding.

\item CCL-DTI \cite{dehghan2024ccl} used multi-modal knowledge (DDI, PPI, sequences) as input and proposed an attention-based fusion technique.
They also investigated how utilizing constrastive loss functions (max-margin contrastive loss function, triplet loss function, Multi-class N-pair Loss Objective, and NT-Xent loss function) could help the prediction. 

\end{itemize}

\subsubsection{Matrix factorization-based models}

The approach for these models often involves decomposing the drug-target interaction matrix into two lower dimensional matrices using matrix factorization method.
NRLMF \cite{liu2016neighborhood} focuses on modeling the drug-target interaction probability by logistic matrix factorization, where the properties of drugs and targets are represented by drug-specific and target-specific latent vectors, respectively.
This model proposes regularization to incorporate neighborhood information for the local structure of the drug-target interaction data.
NRLMF$\beta$ \cite{ban2019nrlmfbeta} is NRLMF with beta distribution rescoring which addresses the issue of limited sample information leading to inaccurate predictions.
KBMF2K \cite{gonen2012predicting} employs Bayesian matrix factorization on kernel matrices obtained from drug and target information.
SPLCMF \cite{xia2019improved} uses collaborative matrix factorization techniques and utilizes a self-paced learning strategy based on weighted low-rank approximation to progressively select training samples and control the feature learning process.

\subsubsection{Interpretability}

While research into DTI and DTA prediction is fairly mature, generalizability and interpretability are not always addressed in the existing works in this field.
By leveraging explainable AI techniques, researchers can develop more accurate, reliable, and interpretable AI models for DTA and DTI prediction.
Also, it is necessary in some cases for researchers in drug discovery to understand the reasoning of the models and important parts of proteins and drugs which are responsible for the prediction results.
It is difficult to understand how deep learning black box models reach their prediction results and the contribution of each input feature to the results due to the large number of hidden layers and parameters. 
Currently, the main methods of interpretation are general methods for neural networks that cannot be easily adapted to biology. 

In order to ensure the model could focus on the important parts of the proteins, the attention mechanism is usually used, some intuitive explanations can be provided by attention mechanisms.
Some methods that use attention modules can check the weights and find the binding cites that are predicted through their model.
The attention mechanism is not always able to provide a clear explanation of the model's decision-making process.

Gradient-weighted Affinity/Class Activation Mapping (Grad-AAM/Grad-CAM) is utilized in some research like MGraphDTA \cite{yang2022mgraphdta} to identify hot-spots for binding affinity prediction task.
This technique detects the sum of the magnitude of gradients that activated the nodes of the last layer and highlights the important parts of a molecule that contribute to the predicted drug-target affinity.
The model's gradients are calculated with respect to the input molecule's features, the gradients are used to weight the activation maps of the model's layers, and the weighted activation maps are visualized as heat-maps.
However, the binding of proteins and ligands is based on the potential contact of amino acids and atoms. 
Therefore, this mechanism for providing interpretability through attention alone has a low accuracy.

MONN (Li et al., 2020) is a multi-objective model for the prediction of non-covalent interactions and binding affinity between proteins and ligands and provides some interpretability for the final affinity prediction results. 
AttentionSiteDTI \cite{yazdani2022attentionsitedti} which utilizes a graph-based attention mechanism to identify the protein binding sites that contribute most to the drug-target interaction.
ICAN (Interpretable Cross-Attention Network) \cite{kurata2022ican} which employs a cross-attention mechanism to capture the complex interactions between drug molecules and target proteins and analyzes the interpretability via attentions.
BindingSiteAugmentedDTA \cite{Yousefi2022.08.30.505897} reduces the search space of potential binding sites of the protein and can provide a deeper understanding of its underlying prediction mechanism by mapping attention weights back to protein binding sites.
The limitations of affinity prediction in terms of interpretability are tackled in AffinityVAE \cite{wang2023affinityvae} by proposing the concept of a protein-ligand interaction feature map.

SHAP (SHapley Additive exPlanations) \cite{zhang2024msh} and LIME (Local Interpretable Model-Agnostic Explanations) can also be used to explain and optimize the existing methods.
Ru et al. \cite{ru2023optimization} utilized SHAP values and the incremental feature selection approach to obtain the high-quality features.

\section{Datasets and Evaluation Metrics}

\subsection{Datasets and databases}

\begin{figure}[ht]
  \centering
  \includegraphics[width=\columnwidth]{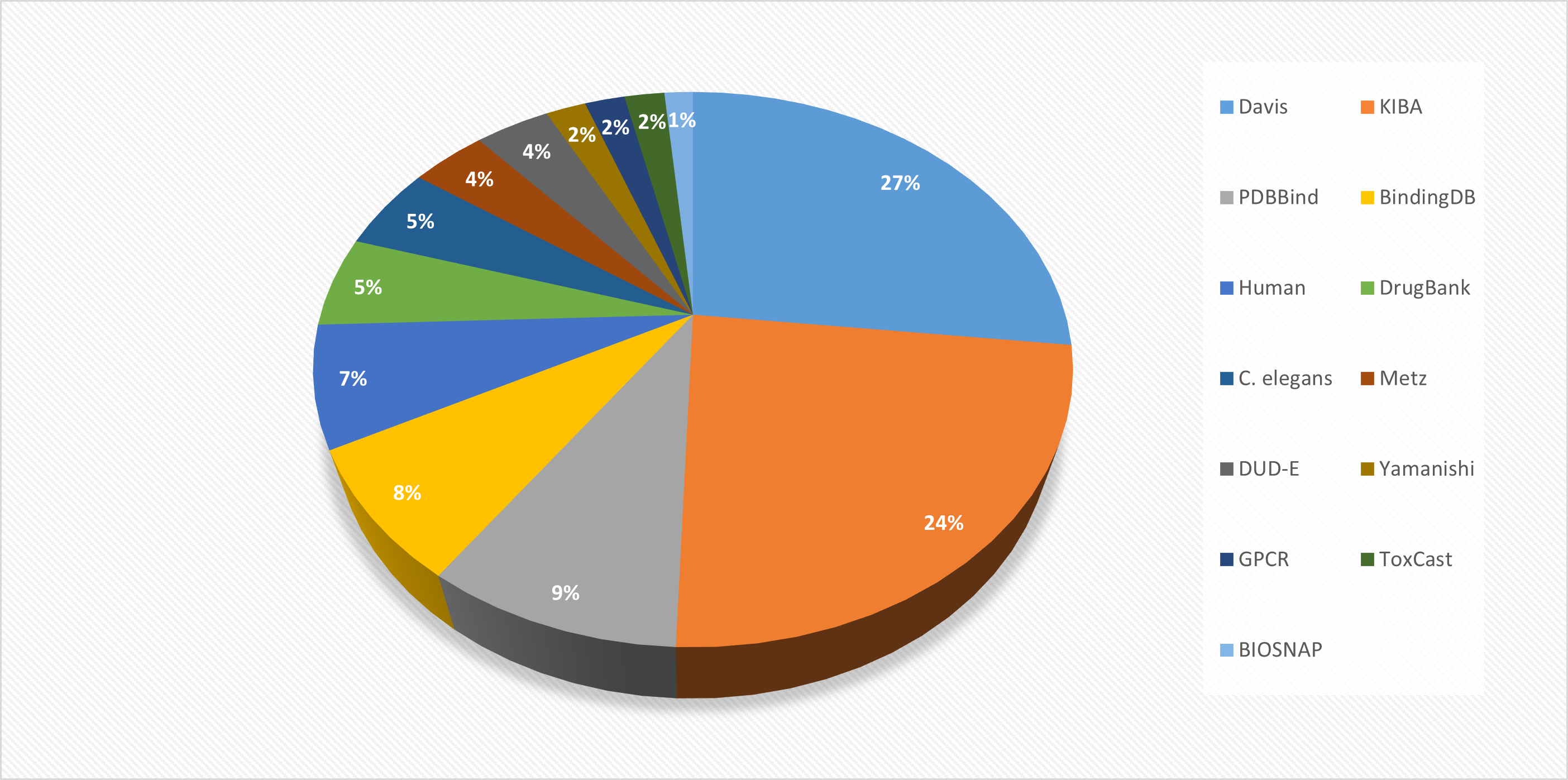}
  \caption{Portion of dataset occurrence in examined models.}
  \label{fig:fig7}
\end{figure}

There are currently lots of various datasets available for the tasks of DTI or DTA.
The most used recources are Davis \cite{davis2011comprehensive}, KIBA \cite{tang2014making}, PDBbind \cite{wang2005pdbbind}, BindingDB \cite{liu2007bindingdb}, Metz \cite{metz2011navigating}, Human \cite{liu2015improving}, and C.elegans \cite{tsubaki2019compound}.
Other used datasets are ToxCast \cite{judson2012us}, DUD-E \cite{mysinger2012directory}, DTC  (drug-target Commons) \cite{tang2018drug}, Allergy \cite{tran2022deepnc}, and GPCR \cite{chen2020transformercpi}.
In learnings with untility networks HPRD \cite{keshava2009human}, SIDER \cite{kuhn2016sider}, and CTD \cite{davis2019comparative} are used.
Human and C. elegans datasets are specifically used in DTI task while other datasets are usually used in DTA Tasks.
Datasets used for DTA tasks are all convertable to datasets for DTI task by selecting an appropriate threshold on affinity score.
Usually, In two-pipline methods datasets are combined and gathered from multiple recources as there are more than one area to consider (proteins and drugs). 
Figure \ref{fig:fig7} shows the portion of dataset occurrence in examined models.

Some datasets can have missing information or there may be a need for more information on drugs or proteins, therefore supplementary databasess are commonly used in the literature.
Some of these databases are UniProt \cite{uniprot2017uniprot}, UniRef \cite{suzek2015uniref}, PDB \cite{berman2000protein}, STITCH \cite{kuhn2007stitch}, and ZINC \cite{zinc_database}. 

\begin{table}[ht]
  \centering
  \footnotesize
  \tiny
  \caption{Statistics of some Drug-Target affinity/Interaction Datasets/databases. 
  Values can vary as they get updated overtime.}
  \label{tab:drug_target_datasets}
  \begin{tabular}{lccc}
  \toprule
  Dataset Name & Drugs/Components & Proteins/Targets & Total Samples/interactions \\
  \midrule
  Davis \cite{davis2011comprehensive} & 68 & 442 & 30,056 \\
  Filtered Davis \cite{rifaioglu2021mdeepred} & 68 & 379 & 9,125 \\
  KIBA \cite{tang2014making} & 52,498 & 467 & 246,088 \\
  Filtered KIBA \cite{he2017simboost} & 2111 & 229 & 118,254 \\
  Metz \cite{metz2011navigating} & 1,423 & 170 & 35,259 \\
  ToxCast \cite{judson2012us} & 9,076 & 1,192 & 530,000 \\
  Human \cite{liu2015improving} & 2,726 & 2001 & 6728 \\
  C. elegans \cite{tsubaki2019compound} & 1767 & 1,876 & 7786 \\
  GPCR \cite{chen2020transformercpi} & 5,359 & 356 &  15,343 \\  
  \href{http://www.pdbbind.org.cn/}{PDBind2020} & 41,847 & 1,469 & 194,443 \\
  \href{https://bindingdb.org/}{BindingDB} & 1,160,301 & 9,021 & 2,701,247 \\
  \href{https://bindingdb.org/}{BindingDB (IC50\_labeled)} & 265,627 & 2,793 & 376,751 \\
  \href{https://bindingdb.org/}{BindingDB (Kd\_labeled)} & 5,895 & 812 & 12,589 \\
  \href{https://bindingdb.org/}{BindingDB (Ki\_labeled)} & 93,437 & 1,619 & 144,525 \\
  \href{https://bindingdb.org/}{BindingDB (EC50\_labeled)} & 31,970 & 513 & 37,896 \\
  \bottomrule
  \end{tabular}
\end{table}

\begin{itemize}

\item The Davis dataset contains selectivity assays of the kinase protein family and the relevant inhibitors with their respective dissociation constant ($K_{d}$) values. 
It comprises all interactions of 442 proteins and 68 ligands, which is 30056 interactions.
All drug-target pairs that cannot be experimentally measured for bio-activity are assigned a $pK_{d}$ of 5 (bio-activity value of 10 $\mu$M).
Filtered Davis dataset is generated by omitting pairs with $pK_{d}$ of 5, as there are lots of these interactions which their $pK_{d}$ is 5.

\item The KIBA database originated from an approach called KIBA, in which kinase inhibitor bio-activities from different sources such as $k_{i}$, $k_{d}$, and $IC50$ were combined (KIBA score ranges from 0.0 to 17.2). 
KIBA scores were constructed to optimize the consistency between $k_{i}$, $k_{d}$, and $IC50$ by utilizing the statistical information they contained. 
The KIBA dataset originally comprised 467 targets and 52498 drugs.
Filtered KIBA contains only drugs and targets with at least 10 interactions yielding a total of 229 unique proteins and 2111 unique drugs. 

\item PDBbind involves a collection of experimentally measured binding affinity data ($k_{i}$, $k_{d}$, and $IC50$) for biomolecules complexed from the Protein Data Bank (PDB), which also have multiple versions (PDBbind2008-2021) with default version and refined version.
This dataset consist of drug SMILES, target sequences, 3D structure of targets, target pocket information.
Refined versions exclude several cases including, ligands, ternary complexes, steric hindrance complexes with resolution higher than 2.5 \AA (Angstrom), instances of covalent bonding, and complexes missing $k_{i}$ or $k_{d}$, or $k_{d}$ lower than 1 pM (picomolar).
PDBbind version 2020 provides binding affinity data for a total of 23,496 biomolecular complexes in PDB, including protein-ligand (19,443), protein-protein (2,852), protein-nucleic acid (1,052), and nucleic acid-ligand complexes (149).
PDBbind2020 general set consist of 14127 samples and 5316 samples in its refined set.
Moreover,some CASF \cite{li2014comparative, su2018comparative} core datasets are derived from PDBbind refined sets. 

\item The BindingDB is a public and web-accessible dataset with drug-target pairs for $k_{i}$, $k_{d}$, $IC50$, and $EC50$ labels which can provide drug SMILES and target sequences.

\item The Metz dataset provides the $pK_{i}$ as a measure of binding affinity. 
It consist of 1,423 drugs and 170 targets with 35,259 pairs.

\item ToxCast contains toxicology data obtained from in vitro high-throughput screening of drugs (i.e., chemicals). 
Several companies have done ToxCast curation with 61 different measurements of binding affinity scores.

\item C. elegans, a simple nematode worm, has emerged as a powerful model organism for drug discovery and repurposing research. 
Its genetic simplicity, rapid life cycle, and conserved biological pathways with humans make it an ideal system for studying the effects of drugs on various biological processes. 
C. elegans drug datasets, which often include information on drug concentration, exposure time, and phenotypic outcomes, provide valuable insights into drug efficacy, toxicity, and potential mechanisms of action. 
C. elegans can be acquired from ChEMBL and DrugBank and is provided with highly confident negative samples. 

\item Human drug dataset provide comprehensive information on the efficacy, safety, and pharmacokinetics of various drugs, derived from clinical trials, observational studies, and electronic health records. 
By analyzing this dataset, researchers can identify potential drug-targets, optimize drug dosing, and predict adverse drug reactions.
Human dataset, like C. elegans dataset, is also provided with highly confident negative samples. 

\item UniProt is a comprehensive database of over 220 million protein sequences and functional information.
It has the ability to add new protein entries and update publicly available annotation information. 

\item The ZINC database is a free massive collection of commercially available compounds, often used in drug discovery and virtual screening.
It provides both 2D and 3D representations of molecules, making it a valuable resource for researchers working with molecular data.
It contains over 230 million purchasable compounds in ready-to-dock, 3D formats.

\item UniRef is a database of protein sequences that are clustered based on sequence similarity. 
It provides three levels of clustering: UniRef100, UniRef90, and UniRef50. 
These clusters help to reduce redundancy in protein sequence data and improve search efficiency.
For example, UniRef50 is built by clustering UniRef90 seed sequences that have at least 50 percent sequence identity to.

\item The Protein Data Bank (PDB) is a global repository of experimentally determined 3D structures of biological macromolecules, such as proteins and nucleic acids. 
PDB entries contain detailed information about the molecule's structure, including atomic coordinates, bond lengths, and angles. 
This data allows researchers to visualize and analyze the molecule's shape, interactions, and potential binding sites.

\item \href{http://stitch.embl.de/}{STITCH} (Search Tool for Interactions of Chemicals) includes information on interactions between 43,000 compounds and 9,643,763 proteins from 2,031 species.
Each interaction in STITCH database is assigned a score value.
STITCH also provides information on compounds that are similar to the drug of target, along with their similarity scores.

\item The \href{https://www.ebi.ac.uk/chembl/}{ChEMBL} database contains compound bio-activity data against drug-targets. 
Bio-activity is reported in $k_{i}$, $k_{d}$, $IC50$, and $EC50$.
ChEMBL version 2 (ChEMBL02) included 2.4 million bio-assay measurements covering 622,824 compounds, including 24,000 natural products.
ChEMBL08 was launched with over 2.97 million bio-assay measurements covering 636,269 compounds.
ChEMBL10 saw the addition of the PubChem confirmatory assays, in order to integrate data that is comparable to the type and class of data contained within ChEMBL.

\item PubChem is a public chemical information database maintained by the National Center for Biotechnology Information (NCBI). 
It provides a vast repository of chemical substances, including small molecules, proteins, and bio-assays. 
Researchers can use PubChem to search for compounds by name, structure, or properties. 
It also offers information on the biological activities of compounds, including their interactions with proteins and their effects on cells. 

\item \href{https://go.drugbank.com/}{DrugBank} is a comprehensive, freely accessible online database which combines detailed drug (chemical, pharmacological, and pharmaceutical) data with comprehensive drug-target (sequence, structure, and pathway) information.
It provides information on Drugs, targets, drug interactions and drug side effects.

\item The Therapeutic Target Database (\href{https://db.idrblab.net/ttd/}{TTD}) provides information on known and explored therapeutic protein and nucleic acid targets, the targeted diseases, pathway information, and the corresponding drugs directed at each of these targets.

\item DUD-E benchmark consists of 102 targets, a set of active compounds known to bind these targets, and a lot of decoys for each active one. 
The DUD-E is a huge dataset consisting of over one million compounds and every target has a different number of active compounds to bind it.

\item BioSNAP \cite{huang2021moltrans, zitnik2018biosnap} is a collection of diverse biomedical networks, including protein-protein interaction networks, single-cell similarity networks, and drug-drug interaction networks. 
The BioSNAP dataset is created from the DrugBank database by \cite{huang2021moltrans} and \cite{zitnik2018biosnap}, consisting of 4,510 drugs and 2,181 proteins.
It is a balanced dataset with validated positive interactions and an equal number of negative samples randomly obtained from unseen pairs.
The MINER DTI dataset from the BIOSNAP collection \cite{zitnik2018biosnap} includes 13,741 DTI pairs. 

\item DTC/TDC serves as a valuable repository for researchers investigating DTIs. 
This platform provides a curated collection of standardized datasets encompassing various DTI-related aspects, such as binding affinity, IC50 values, and Ki constants.

\item GPCR dataset is constructed from GLASS database \cite{chan2015glass}.
GPCR dataset contains 15343 interactions between 5359 drugs and 356 targets and has a 6.0 threshold for negative and positive sample separations.

\end{itemize}

\subsubsection{Binding affinity measures}

Binding affinity indicates the strength of the interaction between drug-target pair.
It is usually expressed in measures such as dissociation constant ($K_{d}$), inhibition constant ($K_{i}$) or the half maximal inhibitory concentration ($IC_{50}$) \cite{cer2009ic}.
Since all of these three measures are concentrations, the lower the measure number, the higher the affinity is.
($K_{d}$) and ($K_{i}$) values are usually represented in terms of ($pK_{d}$) or ($pK_{i}$), the negative logarithm of the dissociation or inhibition constants.
For example Equation. \ref{eq:0} shows the negative logarithm of the dissociation constant.

\begin{equation} \label{eq:0}
pk_{d} = -log_{10}\left ( \frac{k_{d}}{1e9} \right )
\end{equation}

\begin{itemize}
\item Dissociation Constant ($K_{d}$) represents the concentration of drug at which half of the target sites are occupied.
Lower Kd values indicate higher affinity, meaning the drug binds more tightly to the target.
Dissociation Constant can be acquired via various experimental techniques, including Radio-ligand Binding Assays, Surface Plasmon Resonance (SPR), Fluorescence Spectroscopy, and Isothermal Titration Calorimetry (ITC).

\item Inhibition constant ($K_{i}$) \cite{lehninger2005lehninger} represents the concentration of inhibitor required to reduce the enzyme activity by half when the substrate concentration is constant.
Inhibition constant is essentially an equilibrium dissociation constant for the inhibitor-target complex.
Inhibition constant can be determined by Dixon Plot, Lineweaver-Burk Plot or Direct Binding Assays like ITC and SPR.

\item Half maximal inhibitory concentration ($IC_{50}$) is a measure of a substance's potency in inhibiting a specific biological function. 
It represents the concentration of a drug or compound required to inhibit a specific biological process or component by half.
IC50 can be influenced by factors like the experimental conditions, the specific assay used, and the mechanism of action of the compound.
IC50 can be acquired from different assays, including Enzyme Inhibition, Cell-Based, and Biochemical assays.

\item Half maximal effective concentration $EC50$ and Half Maximal Active Concentration $AC50$ represent the concentration of a drug that induces a 50 percent response compared to the maximum possible effect.
A lower EC50 value indicates a more potent drug, as it requires a lower concentration to achieve the same effect.

\end{itemize}

\subsection{Common evaluation metrics}

One major challenge in drug-target interaction research is the absence of a unified evaluation framework and benchmarks. 
This makes it difficult to compare and assess the performance of different models and methods, therefore, in this paper, models are not compared with each other with numbers.
However, Information regarding the datasets, baselines, and evaluation metrics used in the models are provided in the supplementary data.
In this section, several common evaluation metrics for both regression and classification tasks are mentioned.

\subsubsection{Drug-target affinity prediction evaluation metrics}

Common evaluation metrics for drug-target regression task are Mean Absolute Error, Mean squared error, Root mean squared error, R-squared, Pearson correlation coefficient, Spearman correlation coefficient, and Confidence interval count.

\begin{itemize}

\item Mean Absolute Error (MAE) Equation. \ref{eq:1} indicates the size of actual prediction error.
It measures the absolute mean of differences between predicted and real values. 
$n$ is the number of samples, $y_{i}$ is the vector of actual value, $\hat{y}_{i}$ is the prediction vector.

\begin{equation} \label{eq:1}
MAE = \frac{1}{n}\sum_{i=1}^{n}\left|\hat{y}_{i} - y_{i} \right|
\end{equation}

\item MAPE (Mean Absolute Percentage Error) is Equation. \ref{eq:2} and sMAPE (Symmetric Mean Absolute Percentage Error) is Equation. \ref{eq:3}.
These formulas are designed for situations where both $\hat{y}_{i}$ and $y_{i}$ are non-negative.
sMAPE is generally preferred over MAPE because it avoids issues when $y_{i}$ is close to or equal to zero.

\begin{equation} \label{eq:2}
MAPE = \frac{100}{n} \sum_{i=1}^{n} \left| \frac{y_{i} - \hat{y}_{i}}{y_{i}} \right|
\end{equation}

\begin{equation} \label{eq:3}
sMAPE = \frac{100}{n} \sum_{i=1}^{n} \frac{2 \left| y_{i} - \hat{y}_{i} \right|}{\left| y_{i} \right| + \left| \hat{y}_{i} \right|}
\end{equation}

\item Mean squared error (MSE) Equation. \ref{eq:4} is defined as the average squared difference between the predicted and actual binding affinity scores. 

\begin{equation} \label{eq:4}
MSE = \frac{1}{n}\sum_{i=1}^{n}\left ( \hat{y}_{i} - y_{i} \right ) ^{2}
\end{equation}

\item Root mean squared error (RMSE) Equation. \ref{eq:5} is defined as the root of MSE.
In this equation and euqtions before, the smaller the value is, the better the prediction because the final score shows the error rate of the model. 

\begin{equation} \label{eq:5}
RMSE = \sqrt{\frac{1}{n}\sum_{i=1}^{n}\left ( \hat{y}_{i} - y_{i} \right ) ^{2}}
\end{equation}

\item Standard deviation (SD) of the regression Equation \ref{eq:6} can also be used. 
Terms a and b are the slope of interception of the linear regression line of predicted and measured y datapoints.

\begin{equation} \label{eq:6}
SD = \sqrt{\frac{1}{n-1}\sum_{i=1}^{n}\left (  (a*\hat{y}_{i} + b)- y_{i} \right ) ^{2}} 
\end{equation}

\item R-squared ($R^{2}$) Eq. \ref{eq:7} / Squared correlation coefficient (Regression towards the Mean Index) ($r_{m}^{2}$ index) \cite{roy2013some} Equation. \ref{eq:9} are measures used to show how well the prediction value fits the actual value (the degree of match).
The more closer the score is to 1, the better the model is.

$\bar{y}$ is the average value of all actual values $y_{i}$.
$r_{2}$ is the squares of the correlation coefficients with intercept and $r_{0}^{2}$ is the squares of the correlation coefficients without intercept.
\begin{equation} \label{eq:7}
R^{2} = 1 - \frac{\sum_{i=1}^{n}\left (  y_{i}-\hat{y}_{i} \right ) ^{2}}{\sum_{i=1}^{n}\left (  y_{i}-\bar{y} \right ) ^{2}}
\end{equation}

\begin{equation} \label{eq:8}
  R = \frac{\left (  \hat{y}_{i}-\overline{\hat{y}_{i}} \right )(y_{i}-\overline{y_{i}})}{SD_{\overline{\hat{y}_{i}}}.SD_{\overline{y_{i}}}}
\end{equation}

\begin{equation} \label{eq:9}
r_{m}^{2} = r^{2} \times \left ( 1 - \sqrt{r_{2} - r_{0}^{2}} \right )
\end{equation}

\item Pearson correlation coefficient ($PCC$) Equation. \ref{eq:10} is used to measure the mutual relationship (linear correlation) between two variables X and Y.
PCC score ranges from -1 to +1, showing the strength of correlation.
$cov(X, Y)$ represents the covariance of two variables, $\sigma_{X}$ is the standard deviation of X, and $\sigma_{Y}$ is the standard deviation of Y.
If the score is positive, the variables are positively correlated, if the score is negative, the variables are negativelt correlated, and if the score is zero, the variables are not correlated.

\begin{equation} \label{eq:10}
\rho_{XY} = \frac{cov(X, Y)}{\sigma_{X}\sigma_{Y}}
\end{equation}

\item Spearman correlation coefficient ($SCC$) Equation. \ref{eq:11} is another correlation metric which is a nonparametric measure of the dependence of two variables.
$d_{i}$ is the difference between prediction value and actual value of each group.
Like pearson corelation the closer score is to -1 or +1, the stronger the correlation is.

\begin{equation} \label{eq:11}
\rho = 1 - \frac{6\sum d_{i}^{2}}{n(n^{2})}
\end{equation}

\item Confidence interval count ($CI$) Equation. \ref{eq:12} evaluates the ranking performance of the model in discrimination.
$CI$ ranges from 0 to 1, and the closer the score to 1, the better the model fit is.
Where $b_{i}$ is the prediction value for affinity $\sigma_{i}$, $b_{j}$ is the prediction value for affinity $\sigma_{j}$, $Z$ is a normalization constant. 
For function $\varphi(x)$, it is 1 if the value of x is greater than 0, 0.5 if the value of x is equal to 0, and 0 if the value of x is less than 0.

\begin{equation} \label{eq:12}
CI = \frac{1}{Z} \sum_{\sigma_{i}>\sigma_{j}} \varphi(b_{i} - b_{j})
\end{equation}

\end{itemize}

\subsubsection{drug-target interaction prediction evaluation metrics}

Common evaluation metrics for drug-target classification task are accuracy, precision, recall, specificity, F1-score, and area under the precision recall curve.

\begin{itemize}

\item Accuracy \ref{eq:13} is the proportion of correct predictions.

\begin{equation} \label{eq:13}
Accuracy = (TP + TN) / (TP + TN + FP + FN)
\end{equation}

\item Precision \ref{eq:14} is the proportion of positive predictions that are actually positive.

\begin{equation} \label{eq:14}
Precision = TP / (TP + FP)
\end{equation}

\item Recall (Sensitivity) \ref{eq:15} is the proportion of actual positive cases that are correctly identified.

\begin{equation} \label{eq:15}
Recall = TP / (TP + FN)
\end{equation}

\item Specificity \ref{eq:16} is the proportion of actual negative cases that are correctly identified.

\begin{equation} \label{eq:16}
Specificity = TN / (TN + FP)
\end{equation}

\item F1-Score \ref{eq:17} is the harmonic mean of precision and recall.

\begin{equation} \label{eq:17}
F1-Score = 2 \times (Precision \times Recall) / (Precision + Recall)
\end{equation}

\item matthews correlation coefcient (MCC) Equation \ref{eq:18}

\begin{equation} \label{eq:18}
    MCC = (TP * TN) - (FP * FN) / \sqrt{(TP + FP)(TP + FN)(TN + FP)(TN + FN)}
\end{equation}

\item Area Under the Precision-Recall Curve (AUPR/AUCPR/PRC) is a measure of the trade-off between precision and recall.
It is The area under the curve plotted with precision on the y-axis and recall on the x-axis.

\item Area Under the Receiver Operating Characteristic Curve (AUROC) is the area under the plot of the true positive rate (sensitivity) against the false positive rate (1-specificity) for various threshold values.

\end{itemize}

The AUPR assesses a binary model by averaging the precision across all recall values. 
Some datasets in DTI are imbalanced and have skewed distribution. 
Therefore, AUPR should be chosen as the right evaluation metric because the precision-recall curve is more appropriate than the receiver operating characteristic (ROC) curve for imbalanced data. 
AUPR is suitable for tasks in which there is a significant skew in the class distribution.

\section{Conclusion and Future work}

\subsection{Conclusion}

Researchers are rapidly developing and proposing new powerful frameworks for predicting the interaction between proteins and drugs.
This paper aimed to provide a comprehensive review of the methods for drug-target interaction prediction to help the researchers become familiar with the current state-of-the-art methods and to provide a reference for future research.
Moreover, the paper aimed to provide a comprehensive overview of the different common representations for proteins and drugs, the different settings, the datasets used in the literature, and the evaluation metrics used to assess the performance of the models.
In this paper, more than 180 models in three main categories (sequence-based, structural-based, and sequence-structural-based methods) for DTI and DTA prediction tasks were reviewed.
Matrix factorization methods, 3D complex-based methods, and interpretability considerations of models were also discussed.
We hope that by providing this comprehensive review, we paved the way for future research in this field.

\subsection{Future work and challenges}

After being studied after so many years, the field of drug-target interaction prediction still has many challenges and opportunities for future research.
A major challenge in DTI prediction lies in accurately capturing the conformational flexibility of both drugs and proteins. Proteins are dynamic entities that undergo constant conformational changes, and drugs can also exhibit multiple conformations.
Current methods often rely on rigid or simplified representations of these molecules, limiting their ability to accurately predict binding affinities and understand the nuances of the interaction. 
Future work should focus on developing more sophisticated computational methods that can effectively sample the conformational space of both drug and protein molecules, incorporating techniques like enhanced sampling methods and machine learning approaches. 
Additionally, advancements in experimental techniques that can provide more detailed information about the dynamic behavior of protein-ligand complexes will be crucial for refining and validating these models.

New interaction and binding information can be extracted from the protein-ligand complex structures.
Pocket information is mostly ignored in the current DTA/DTI prediction framework because binding pocket prediction is regarded as another prediction task than drug-target prediction.
This task can be integrated into the DTI/DTA prediction framework to improve the performance of the models.
Also, finding the best combination of informative features for the models is still a challenge.
One major challenge in drug-target interaction research is learning the mutual information between proteins and drugs.
Another challenge is how to find and highlight the important atoms of the drugs and residues of the proteins.

The Alpha-fold 3 \cite{abramson2024accurate} model, with its remarkable accuracy in predicting protein structures within drug-target complexes, has significantly impacted the field of DTI prediction. 
This will lead to improvements in DTI prediction models, as they can now leverage highly accurate protein structures as input, leading to more reliable predictions of binding affinities. 
By integrating Alpha-fold 3 predictions with existing DTI prediction models, researchers can develop more accurate and reliable models for drug discovery and re-purposing.

Evaluations for models in DTI/DTA predictions should be standardized to allow for more accurate comparisons between different models.
Many models in the literature show high performance but perform poorly in real datasets, indicating high bias towards a specific dataset.
Performance should be evaluated on data that significantly differs from the training data in terms of drug and target properties and should be in cold setting.
Also, models should be assessed on their ability to predict interactions for newly discovered drugs.
Lastly, interpretability of the models should be examined as they can ensure that predicted interactions are not only accurate but also biologically plausible and potentially useful for drug discovery.

\bibliographystyle{unsrtnat}
\bibliography{DTI_survey_arxiv}

%---------------------------- Appendix -------------------------------
\appendix
\section{Abbreviations}

\textbf{CNN}: Convolutional Neural Network\\
\textbf{Adaboost}: Adaptive Boost\\
\textbf{ADDA}: Adversarial discriminative domain adaptation\\
\textbf{AG-FP}: algebraic graph-based ﬁngerprints\\
\textbf{AGL-Score }: Algebraic Graph Learning Score\\
\textbf{AI}: Artificial Intelligence\\
\textbf{Att}: Attention\\
\textbf{BERT}: Bidirectional Encoder Representations from Transformer\\
\textbf{Bi}: Bidirectional\\
\textbf{BiGRU}: Bidriectional Gated recurrent unit\\
\textbf{BiLSTM}: Bi directional Long short term memory network\\
\textbf{BLA}: Biologics License Application\\
\textbf{BLM}: Bipartite local model\\
\textbf{BPE}: Byte Pair Encoding\\
\textbf{CA}: Cross Attention\\
\textbf{CNN}: Convolutional Neural Network\\
\textbf{Comb}: Combinator\\
\textbf{Conv}: Convolution\\
\textbf{CPA}: component-protein affinity prediction\\
\textbf{CPI}: component-protein interaction prediction\\
\textbf{CS}: Character-based slicing\\
\textbf{CTD}: Composition, Transition and Distribution\\
\textbf{DBN}: Deep Belief Network\\
\textbf{Dda}: Drug-drug assossiation\\
\textbf{DDI}: Drug-drug interaction\\
\textbf{DDi}: Drug-disease\\
\textbf{DNA}: Deoxyribonucleic Acid\\
\textbf{DPA}: Drug-protein affinity prediction\\
\textbf{DPI}: Drug-protein interaction prediction\\
\textbf{DPP}: Drug-protein pair\\
\textbf{Dside}: Drug side effect\\
\textbf{DTA}: Drug-target affinity prediction\\
\textbf{DTI}: Drug-target interaction prediction\\
\textbf{E}: Enzyme\\
\textbf{ECFP}: Extended-Connectivity Fingerprint\\
\textbf{EF}: Early fusion\\
\textbf{EGCL}: Equivariant Graph Convolutional Layer\\
\textbf{EGNN}: Equivariant graph neural network\\
\textbf{ErG}: ElectroRetinoGraphy\\
\textbf{ESM}: Evolutionary Scale Modeling\\
\textbf{FC}: Fully Connected\\
\textbf{FCFP}: Functional-Class Fingerprint\\
\textbf{FDA}: Food and Drug Administration\\
\textbf{FEP}: Free-Energy Perturbation\\
\textbf{FNN}: Feed forward neural network\\
\textbf{FP}: Fingerprint\\
\textbf{GAN}: Generative Adversarial Network\\
\textbf{GAT}: Graph Attention Network\\
\textbf{GCN}: Graph Convolutional Network\\
\textbf{GIN}: Graph Isomorphism Network\\
\textbf{GIP}: Gaussian Interaction Profile\\
\textbf{GNN}: graph neural network\\
\textbf{GPCR}: G-protein-coupled receptor\\
\textbf{Grad-AAM/Grad-CAM}: Gradient-weighted Afﬁnity/Class Activation Mapping\\
\textbf{GRU}: Gated recurrent unit\\
\textbf{HMM}: Hidden Markov Matrix\\
\textbf{IC}: Ion-Channel\\
\textbf{ID}: Identification Number\\
\textbf{IND}: Investigational New Drug Application\\
\textbf{KNN}: K-nearest neighbour\\
\textbf{KronRLS}: Kronecker Regularized Least Square\\
\textbf{LapRLS}: Laplacian Regularized Least Square\\
\textbf{LM}: Language model\\
\textbf{LMCS}: Ligand Maximum Common Substructure\\
\textbf{LSTM}: Long short term memory network\\
\textbf{LZMA}: Lempel Ziv Markov chain algorithm\\
\textbf{MAB}: Multihead Attention Block\\
\textbf{MACCS }: MELCOR Accident Consequence Code System\\
\textbf{MCA}: Multihead cross attention\\
\textbf{MFA}: Multiview Fusion Attention\\
\textbf{MGC}: Molecular Graph Convolution\\
\textbf{MHSA/MSA}: Multihead Self Attention Block\\
\textbf{MLP }: Multi Layer Perceptron\\
\textbf{MM/GBSA}: Molecular Mechanics/Generalized Born Surface Area\\
\textbf{MM/PBSA}: Molecular Mechanics Poisson-Boltzmann Surface Area\\
\textbf{MPNN}: Message Passing Neural Network\\
\textbf{MSA}: Multiple sequence alignment\\
\textbf{N/A}: Not Applicable\\
\textbf{NDA}: New Drug Application\\
\textbf{NFM}: Neural factorization machine\\
\textbf{NLP}: Natural Language Processing\\
\textbf{NMR}: Nuclear Magnetic Resonance\\
\textbf{NNL}: Neural network layer\\
\textbf{NR}: Nuclear receptor\\
\textbf{NTN}: Neural Tensor Network\\
\textbf{PDi}: Preotein-disease\\
\textbf{PDM}: Protein Domain proﬁles or Motif\\
\textbf{PLA}: Protein-ligand affinity prediction\\
\textbf{PLI}: Protein-ligand interaction prediction\\
\textbf{PPa}: Protein-protein assossiation\\
\textbf{PPI}: Protein-protein interaction\\
\textbf{PSC}: Protein Sequence Composition\\
\textbf{Pside}: Protein side effect\\
\textbf{PSSM}: Position-specific score matrix\\
\textbf{QSAR}: Quantitative StructureActivity Relationship\\
\textbf{RBM}: Restricted Boltzmann machine\\
\textbf{ResGCN}: Residual GCN\\
\textbf{RGCN}: Relational GCN\\
\textbf{RNN}: recurrent neural network\\
\textbf{SA}: Self Attention\\
\textbf{SC}: Skip Connection\\
\textbf{SDG}: structure diagram generation\\
\textbf{SE}: Squeeze-and-excitation\\
\textbf{SELFIES}: SELF-referencIng Embedded String\\
\textbf{SHAP}: SHapley Additive exPlanation\\
\textbf{SMARTS}: SMILES Arbitrary Target Speciﬁcation\\
\textbf{SMILES}: Simpliﬁed Molecular Input Line Entry System\\
\textbf{SPS}: Structural Property String\\
\textbf{SSE}: Secondary Structure Element\\
\textbf{SVM}: Support Vector machine\\
\textbf{S-W}: Smith-Waterman\\
\textbf{TAPE}: Tasks Assessing Protein Embedding\\
\textbf{TSR}: Triangular Spatial Relationship\\

\newpage

\section{Available official codes of models}

\begin{table}[htbp]
  \centering
  \footnotesize
  \label{tab:model_codes}
  \caption{Available official codes of Models. Models which do not have official codes are not mentioned here. In order to access the codes click on the corresponding models.}
  \begin{tabular}{|c | c | c | }
  \toprule
  & Official code links &  \\
  \midrule
  \href{https://github.com/vtarasv/3d-prot-dta.git}{3DProtDTA} & \href{https://github.com/peizhenbai/DrugBAN}{DrugBAN} & \href{https://github.com/wowowoj/MCL-DTI.git}{MCL-DTI}\\
  \href{https://github.com/MahaThafar/Affinity2Vec}{Affinity2Vec} & \href{https://github.com/lipi12q/DrugMAN}{DrugMAN} & \href{https://github.com/Amitranjan71/MDF-DTA.git}{MDF-DTA}\\
  \href{https://github.com/frankchenqu/AMMVF}{AMMVF-DTI} & \href{https://github.com/joannacatj/drugormerDTI}{DrugormerDTI} & \href{https://github.com/JU-HuaY/MFR}{MFR-DTA}\\
  \href{https://github.com/dmis-lab/ArkDTA}{ArkDTA} & \href{https://github.com/CSUBioGroup/DTIAM}{DTIAM} & \href{https://github.com/guaguabujianle/MGraphDTA}{MGraphDTA}\\
  \href{https://github.com/JK-Liu7/AttentionMGT-DTA}{AttentionMGT-DTA} & \href{https://github.com/MedicineBiology-AI/DTI-CNN.git}{DTI-CNN} & \href{https://github.com/Zora-LM/MHGNN-DTI}{MHGNN}\\
  \href{https://github.com/yazdanimehdi/AttentionSiteDTI}{AttentionSiteDTI} & \href{https://github.com/luoyunan/DTINet}{DTINet} & \href{https://github.com/ranzhran/MHTAN-DTI}{MHTAN-DTI}\\
  \href{https://github.com/CSUBioGroup/BACPI}{BACPI} & \href{https://github.com/larngroup/DTITR}{DTITR} & \href{https://github.com/jnuaipr/MINDG}{MINDG}\\
  \href{https://github.com/Blue1993/BAPA}{BAPA} & \href{https://github.com/DTI-dream/EDC-DTI}{EDC-DTI} & \href{https://github.com/guaguabujianle/ML-DTI.git}{ML-DTI}\\
  \href{https://github.com/mahmood83/BiComp-DTA}{BiComp-DTA} & \href{https://github.com/MedicineBiology-AI/EEG-DTI}{EEG-DTI} & \href{https://github.com/kexinhuang12345/moltrans}{MolTrans}\\
  \href{https://github.com/SenseTime-Knowledge-Mining/BridgeDPI}{BridgeDPI} & \href{https://github.com/IILab-Resource/ELECTRA-DTA}{ELECTRA-DTA} & \href{https://github.com/lishuya17/MONN}{MONN}\\
  \href{https://github.com/Layne-Huang/CoaDTI}{CaoDTI} & \href{https://github.com/Aurora-yuan/EmbedDTI}{EmbedDTI} & \href{https://github.com/LiZhang30/MRBDTA}{MRBDTA}\\
  \href{https://github.com/lennylv/CAPLA}{CAPLA} & \href{https://github.com/BioCenter-SHU/EMPDTA}{EMPDTA} & \href{https://github.com/songxuanmo/MSGNN-DTA}{MSGNN-DTA}\\
  \href{https://github.com/dehghan1401/CCL-DTI}{CCL-DTI} & \href{http://lanproxy.biodwhu.cn:9099/mszjaas/FingerDTA.git}{FingerDTA} & \href{https://github.com/Beiyi0719/MSH-DTI}{MSH-DTI}\\
  \href{https://github.com/CSUBioGroup/CGraphDTA}{CGraphDTA} & \href{https://github.com/NTU-MedAI/FOTF-CPI}{FOTF-CPI} & \href{https://github.com/Lamouryz/Code/tree/main/MT-DTA}{MT-DTA}\\
  \href{https://github.com/doragu/coldDTA}{ColdDTA} & \href{https://github.com/yuanweining/FusionDTA}{FusionDTA} & \href{https://github.com/FangpingWan/NeoDTI}{NeoDTI}\\
  \href{https://github.com/RobinDoyle/CPGL}{CPGL} & \href{https://github.com/shuyu-wang/GanDTI}{GanDTI} & \href{https://github.com/hehh77/NHGNN-DTA}{NHGNN-DTA}\\
  \href{https://github.com/WANG-BIN-LAB/CWIDTI}{CWI-DTI} & \href{https://github.com/zty2009/GCN-DNN}{GCN-DTI} & \href{https://github.uconn.edu/mldrugdiscovery/OctSurf}{OctSurf}\\
  \href{https://github.com/YanZhu06/DataDTA}{DataDTA} & \href{https://github.com/ngminhtri0394/GEFA}{GEFA} & \href{http://github.com/zhenglz/onionnet}{Onionnet}\\
  \href{https://github.com/mojtabaze7/DCGAN-DTA}{DCGAN-DTA} & \href{https://github.com/CellNest/GEFormerDTA}{GEFormerDTA} & \href{https://github.com/simonfqy/PADME}{PADME}\\
  \href{https://github.com/whata}{DEAttentionDTA} & \href{https://github.com/Frank-39/GeneralizeDTA}{GeneralizedDTA} & \href{https://github.com/yijia-xiao/pgraphdta}{PGraphDTA}\\
  \href{https://github.com/asadahmedtech/DEELIG}{DEELIG} & \href{https://github.com/AmbitYuki/G-K-BertDTA}{GK-BertDTA} & \href{https://github.com/YuQingYe1/PHCDTI}{PHCDTI}\\
  \href{https://github.com/Shen-Lab/DeepAffinity}{DeepAffinity} & \href{https://github.com/yeqing0713/GNPDTA}{GNPDTA} & \href{https://github.com/QingyuLiaib/PLA-MoRe}{PLA-MoRe}\\
  \href{https://github.com/haiping1010/DeepBindRG}{DeepBindRG} & \href{https://github.com/LiZhang30/GPCNDTA}{GPCNDTA} & \href{https://github.com/ComputArtCMCG/PLANET}{PLANET}\\
  \href{https://github.com/LBBSoft/DeepCDA}{DeepCDA} & \href{https://github.com/thinng/GraphDTA}{GraphDTA} & \href{https://github.com/trrt-good/WELP-PLAPT/tree/main}{PLAPT}\\
  \href{https://github.com/GIST-CSBL/DeepConv-DTI}{ DeepConv-DTI} & \href{https://github.com/CSUBioGroup/GraphscoreDTA}{GraphScoreDTA} & \href{https://github.com/zhaolongNCU/PocketDTA}{PocketDTA}\\
  \href{https://github.com/masashitsubaki}{DeepCPI/GNN-CPI} & \href{https://github.com/catly/GSL-DTI}{GSL-DTI} & \href{https://github.com/diamondspark/PSG-BAR}{PSG-BAR}\\
  \href{https://github.com/hkmztrk/DeepDTA/}{DeepDTA} & \href{https://github.com/Zhaoyang-Chu/HGRL-DTA}{HGRL-DTA} & \href{https://github.com/dldxzx/S2DTA}{S2DTA}\\
  \href{http://github.com/KailiWang1/DeepDTAF}{DeepDTAF} & \href{https://github.com/zhaoqichang/HpyerAttentionDTI}{HyperAttentionDTI} & \href{https://github.com/bioinfocqupt/Sfcnn}{Sfcnn}\\
  \href{https://github.com/Bjoux2/DeepDTIs}{DeepDTI} & \href{https://github.com/kuratahiroyuki/ICAN}{ICAN} & \href{https://github.com/catly/SGCL-DTI}{SGCL-DTI}\\
  \href{https://github.com/guofei-tju/DeepFusionDTA}{DeepFusionDTA} & \href{https://github.com/czjczj/IIFDTI}{IIFDTI} & \href{https://github.com/PaddlePaddle/PaddleHelix/tree/dev/apps/drug_target_interaction/sign}{SIGN}\\
  \href{https://github.com/MLlab4CS/DeepGLSTM.git}{DeepGLSTM} & \href{https://github.com/zhangjing-dmu/IMAEN}{IMAEN} & \href{https://github.com/xianyuco/SS-GNN}{SS-GNN}\\
  \href{https://github.com/jacklin18/DeepGS}{DeepGS} & \href{https://github.com/neuhanli/ImageDTA}{ImageDTA} & \href{https://github.com/Xinyan-Xia/SSR-DTA}{SSR-DTA}\\
  \href{https://github.com/Hawash-AI/deepH-DTA}{DeepH-DTA} & \href{https://github.com/debbydanwang/BAP}{IMCP-SF} & \href{https://github.com/1q84er/SubMDTA}{SubMDTA}\\
  \href{https://github.com/zhangpl109/DeepMGT-DTI}{DeepMGT-DTI} & \href{https://github.com/NedraMekni/InterGraph}{InterGraph} & \href{https://github.com/Lamouryz/TDGraph}{TDGraphDTA}\\
  \href{https://github.com/thntran/DeepNC}{DeepNC} & \href{https://github.com/llnl/fast}{Jones et al.} & \href{https://github.com/lizongquan01/TEFDTA}{TEFDTA}\\
  \href{https://github.com/zhc-moushang/DeepTGIN}{DeepTGIN} & \href{https://github.com/syc2017/KCDTA}{KC-DTA} & \href{https://github.com/lifanchen-simm/transformerCPI}{TransformerCPI}\\
  \href{https://github.com/fatemeh-rafiei/DeepTraSynergy}{DeepTraSynergy} & \href{https://zenodo.org/record/5500305}{KGE-NFM} & \href{https://github.com/dehghan1401/TripletMultiDTI}{TripletMultiDTI}\\
  \href{https://github.com/fbbgood/DefuseDTI.git}{DeFuseDTI} & \href{https://github.com/RuiqiangLu/KIDA}{KIDA} & \href{https://github.com/XiaLeiming/UCMCDTA}{UCMPNN-DTA}\\
  \href{https://github.com/luojunwei/DGDTA}{DGDTA} & \href{https://github.com/QHwan/PretrainDPI}{Kim et al.} & \href{https://github.com/595693085/WGNN-DTA}{WGNN-DTA}\\
  \href{https://github.com/595693085/DGraphDTA}{DGraphDTA} & \href{https://github.com/blah957/blah957-lstm-sagDTA}{LSTM-SAGDTA} &  \\
  \href{https://github.com/HWangLab-github/DLSSAffinity}{DLSSAffinity} & \href{https://github.com/jiaxianyan/MBP}{MBP} &  \\
  
  \bottomrule
  \end{tabular}
\end{table}

\section{Other tables and figures}

\begin{table*}[htbp]
  \centering
  \footnotesize
  \caption{Summary of seuqence-based models input representations}
  \tiny
  \label{tab:sequence_based_representations}
  \begin{tabular*}{\textwidth}{@{\extracolsep{\fill}}lcccccr}
  \toprule
  Model & Drug/Ligand representation & Protein/Target representation & Extra representation/input \\
  \midrule
  DeepAffinity & SMILES & Structural property strings & N/A\\
  DeepCDA & SMILES & Protein sequences & N/A\\
  BridgeDPI & SMILES, Morgran FP, Physiochemical features & Protein sequences, onehot, 1/2/3-mer features & Bridge graph\\
  DeepDTA & SMILES & Protein sequences & N/A\\
  WideDTA & SMILES (8 words) & Protein sequences (3 words) & LMCS, Protein Domains and Motifs\\
  MATT-DTI & SMILES & FASTA & N/A\\
  ML-DTI & SMILES & Protein sequences & N/A\\
  Co-VAE & SMILES & Protein sequences & N/A\\
  FusionDTA & SMILES & Protein sequences & N/A\\
  ELECTRA-DTA & SMILES & FASTA & N/A\\
  AttentionDTA & SMILES & Protein sequences & N/A\\
  DTITR & SMILES & Protein sequences, FCS+BPE & N/A\\
  CSatDTA & SMILES & Protein sequences & N/A\\
  MRBDTA & SMILES & FASTA & N/A\\
  FingerDTA & SMILES onehot & AminoAcid onehot & Drug and target fingerprints (ECFP4)\\
  PCNN-DTA & SMILES & Protein sequences & N/A\\
  MT-DTA & SMILES & Protein sequences & N/A\\
  ArkDTA & SMILES and Morgan FP & FASTA with ESM & N/A\\
  DGDTA & SMILES & Protein sequences & N/A\\
  TEFDTA & SMILES to MACCS FP & FASTA & N/A\\
  TransVAE-DTA & SMILES & Protein sequences & N/A\\
  DCGAN-DTA & SMILES & Protein sequences & N/A\\
  DEAttentionDTA & SMILES & Protein Sequences & Pocket sequences\\
  ImageDTA & SMILES & Protein Sequences & N/A\\
  TC-DTA & SMILES & Protein Sequences & N/A\\
  MultiscaleDTA & SMILES & Amino Acid strings & N/A\\
  GANsDTA & SMILES & Protein Sequences & N/A\\
  SimCNN-DTA & Tanimoto drug-drug similarity & Smith Waterman similarity & N/A\\
  BiComp-DTA & SMILES & \makecell{Protein sequences with LZMA,\\ Smith-Waterman and BiCOMP encoding} & N/A\\
  MFDR & 881 dim fingerprints & Protein Sequences & N/A\\
  DeepDTI & Extended-Connectivity Fingerprints (ECFP) & Protein sequence composition descriptors (PSC) & N/A\\
  DeepConv-DTI & Morgan/Circular drug fingerprints & Protein Sequences & N/A\\
  MolTrans & SMILES & Protein Sequences & N/A\\
  EnsembleDLM & Daylight, ErG, Morgan FP, SMILES 1-char embeddings & AAC, character-based embedding of FASTA & N/A\\
  MT-DTI & SMILES & FASTA & N/A\\
  DeepDTAF & SMILES & Protein Sequences & \makecell{SSE, physicochemical chars,\\ sequence+structural property}\\
  CAPLA & SMILES & \makecell{AA sequences, SSEs, physicochemical\\ of residues of protein and pockets} & pocket input\\
  AffinityVAE & SMILES & Protein Sequences & N/A\\
  PLAPT & SMILES & Amino Acid strings & N/A\\
  Hu et al. & PaDEL-Descriptor & Physicochemical properties of Amino Acids & data augmentation\\
  DrugVQA & SMILES & 2D pairwise distance map & N/A\\
  CA-DTI & Compound feature from molecular graph  & Protein Sequences & N/A\\
  HyperAttentionDTI & SMILES & Amino Acid sequences & N/A\\
  MPS2IT-DTI & SMILES & Protein Sequences & N/A\\
  FastDTI & Chemical properties, Grover, ChemBERTa & Protein properties, ProtBert & Yes\\
  MCL-DTI & Image and the chemical features text & FASTA k-gram & Yes\\
  AMMVF-DTI & SMILES & Protein Sequences & N/A\\
  FOTF-CPI & Fragmented SMILES & Fragmented Protein Sequences & N/A\\
  MDF-DTA & SMILES & Protein Sequences & N/A\\
  DeFuseDTI & Molecular graph from SMILES & Protein Sequences & N/A\\
  PHCDTI SMILES & FASTA & N/A\\
  MREDTA & SMILES & Protein Sequences & N/A\\
  CWI-DTI & SMILES and FPs & protein sequences & Similarity of drug and target structures\\
  
  \bottomrule
  \end{tabular*}
\end{table*}

\begin{table*}[htbp]
  \centering
  \footnotesize
  \caption{Summary of structural-based models input representations}
  \tiny
  \label{tab:structure_based_representations}
  \begin{tabular*}{\textwidth}{@{\extracolsep{\fill}}lcccccr}
  \toprule
  Model & Drug/Ligand representation & Protein/Target representation & Extra representation/input \\
  \midrule
  Graph-CNN & Molecular graph from SMILES & Protein pocket graph & N/A\\
  STAMP-DPI/X-DPI & Molecular graph+Mol2vec embedding & Contact map graph+TAPE embedding & N/A\\
  AttentionSiteDTI & Molecular graph from SMILES & Binding pocket graphs & N/A\\
  DGraphDTA & Molecular graph from SMILES & Protein Graph from contact maps & N/A\\
  GEFA & Molecular graph from SMILES & TAPE embeddings and Protein Graph from contact maps & N/A\\
  WGNN-DTA & Molecular graph from SMILES & Weighted Protein Graph from contact maps & N/A\\
  HGRL-DTA & Molecular graph from SMILES & Graph from contact maps & Affinity graph\\
  Ma et al. & Molecular graph from SMILES & Graph of contact maps (Pconsc4) and PSSM features & N/A\\
  GSAML-DTA & Molecular graph from SMILES & Graph from contact maps & N/A\\
  MSGNN-DTA & Drug Atom-Level Graph from SMILES & Weighted Protein Graph from contact maps & Drug Motif-Level Graph\\
  GraphScoreDTA & Molecular graphs & Graph from Contact maps & subgraphs from contact maps, Vina terms\\
  PLANET & Graph in 2D space & Protein pocket graph & N/A\\
  CGraphDTA & Molecular graph & PSSM and HMM graphs & Secondary structure features using DSSP\\
  AttentionMGT-DTA & Molecular graphs & Protein pocket graphs & N/A\\
  GTAMP-DTA & Molecular graphs & Protein pocket graph & N/A\\
  S2DTA & Graphs & Protein graphs (PSSM and HMM) & Pocket graphs\\
  NG-DTA & Molecular graphs from SMILES & N-gram graphs from protein sequences & N/A\\
  EMPDTA & Molecular graphs & Protein structures from ALPHAFOLD & Pocket graphs\\
  HGTDP-DTA & Molecular graph from SMILES & Contact map graphs & Affinity graph\\
  GNPDTA & Graphs & Graphs & N/A\\
  CASTER-DTA & Molecular SMILES & Protein graphs from PDBank or AlphaFOLD2 & N/A\\
  \bottomrule
  \end{tabular*}
\end{table*}

\begin{table*}[htbp]
  \centering
  \footnotesize
  \caption{Summary of structural-sequence-based models input representations}
  \tiny
  \label{tab:hybrid_based_representations}
  \begin{tabular*}{\textwidth}{@{\extracolsep{\fill}}lcccccr}
  \toprule
  Model & Drug/Ligand representation & Protein/Target representation & Extra representation/input \\
  \midrule
  DTIAM & Substructures of Molecular graphs & Protein sequences & N/A\\
  DeepCPI & R-radius subgraphs & Protein sequences & N/A\\
  GraphCPI & Molecular graph from SMILES & 3-gram encoding with pretrained Word2Vec & N/A\\
  TransformerCPI & Molecular graph from SMILES & Amino Acid sequence & N/A\\
  Kim et al. & molecular graphs & Protein sequences & N/A\\
  DeepMGT-DTI & Molecular graphs & Sequence embeddings & N/A\\
  CPInformer & Molecular graphs, FCFPs from SMILES & Protein sequences & yes\\
  CaoDTI & R-radius subgraphs from SMILES & Protein sequence & N/A\\
  CPGL & R-radius subgraphs & 3-gram from Protein sequences & N/A\\
  DrugormerDTI & Molecular graphs & Residual2vec embeddings & N/A\\
  CSDTI & Molecular graphs from SMILES & Protein sequences & N/A\\
  MINDG & SMILES, & Protein sequences & Binding affinity graph\\
  FDTIIT & Molecular graphs from SMILES & FASTA & N/A\\
  PADME & SMILES, ECFP/graphs & Protein sequences (PSC) & N/A\\
  Wang et al. & Molecular graphs  & Protein sequences & N/A\\
  DeepGS & SMILES (Smi2Vec), Molecular graph from SMILES & Amino Acid sequences (Prot2vec) & yes\\
  DeepH-DTA & SMILES & Protein sequences & Molecular graphs\\
  GraphDTA & Molecular graph from SMILES & Protein sequences-Onehot encoding & N/A\\
  DeepRelations & Molecular graphs & Protein sequences & \makecell{Predicted solvant exposure,\\ intermediate predicted \\k-mer binding residues,\\ intermediate predicted\\ binding residues,\\ predicted protein\\ contact map}\\
  GanDTI & Molecular graph from SMILES & Protein sequences (3 words) & N/A\\
  DEELIG & \makecell{PaDEL-Descriptor, FPs,\\ QikProp to get ADMET} & \makecell{Atomic level, Amino acid\\ level protein-pocket 3D} & N/A\\
  DeepFusionDTA & SMILES & Protein sequences & \makecell{Protein secondary structures,\\ Molecular fingerprints}\\
  SAG-DTA & Molecular graph from SMILES & Amino Acid sequences & N/A\\
  EmbedDTI & Atom graph and substructure graph & Protein sequences & Substructure network for drugs\\
  MGraphDTA & Molecular graph from SMILES & Protein sequences-integer encoding & N/A\\
  BACPI & Molecular graph from SMILES, ECFPs & Protein sequences (3 word) & N/A\\
  Tian et al. & Molecular structures & Protein sequences & Chemical background sequences\\
  SSGraphCPI & SMILES, Molecular graph from SMILES & Structure property sequences & yes\\
  DLSSAffinity & SMILES & Complete Protein sequences & Pocket-ligand structural pairs\\
  DeepGLSTM & Molecular graphs  & Amino Acid sequences & Molecular graphs paths\\
  GDGRU-DTA & Molecular graph from SMILES & Protein sequences & N/A\\
  DeepNC & Molecular graph from SMILES & Protein sequences-one hot & N/A\\
  Lin et al. & constructed DDI graphs & similarity graph with TEXTCNN & N/A\\
  IIFDTI & Molecular graph from SMILES & Protein sequences & Smi2Vec, Prot2Vec\\
  PSG-BAR &  Attributed graphs & Attributed graphs from contact maps & ESM with Protein sequences\\
  Xia et al. & Molecular graphs  & Amino Acid sequence & N/A\\
  PLA-MoRe & Molecular graph & Protein sequences & Bioactive proprety of drugs\\
  MFR-DTA & FCFPs and GNN features & \makecell{Amino Acid embedding (AAE)\\ and word embedding (WE)} & yes\\
  DrugBAN & Molecular graph from SMILES & Protein sequences & N/A\\
  3DProtDTA & Graph, Morgan FP & Graphs from AlphaFold & N/A\\
  NHGNN-DTA & SMILES, Molecular graphs & Protein sequences, Pconsc4 graphs & yes\\
  ColdDTA &  Graph  & Protein sequences & N/A\\
  SubMDTA & Molecular graph from SMILES & Protein sequences (2,3,4 grams) & N/A\\
  DataDTA & AG-FP(3D structures), SMILES & Pockets, Protein sequences & yes\\
  UCMPNN-DTA & Molecular graph from SMILES & Protein sequences & N/A\\
  GPCNDTA & Molecular graph from SMILES & Protein graphs from distances & Pharmacophore, SMILES, protein FASTA\\
  PGraphDTA & Molecular graph from SMILES & Protein sequences & Contactmaps with Pconsc4 and DiffDock\\
  IMAEN & Molecular graph from SMILES & Protein sequences & N/A\\
  TDGraphDTA & Molecular graph from SMILES & Protein sequences & N/A\\
  KC-DTA & Molecular graph from SMILES & 2D and 3D Protein sequences & yes\\
  GraphCL-DTA & Molecular graph & Protein sequences & N/A\\
  AGraphDTA & Graph & Graph, sequences & yes\\
  HBDTA & molecular graph from SMILES & Protein sequences & N/A\\
  LSTM-SAGDTA & Molecular graph from SMILES & SeqVec model (ELMo) & N/A\\
  T-GraphDTA & Molecular graph from SMILES & Protein sequences & N/A\\
  GEFormerDTA & ESC from SMILES, graph & Graph from distance map & N/A\\
  GK BertDTA & Molecular structures using SMILES & Amino Acid sequence & SMILES\\
  SSR-DTA & Molecular fingerprints and molecular maps, ECFP & 3D-structural information of proteins & yes\\
  PocketDTA & Molecular Morgan fingerprint, 2D molecular graphs from SMILES & Protein sequences with ESM-2, pocket graphs & yes\\
  ViDTA & Molecular graph from SMILES with virtual node & Protein sequences & N/A\\
  DeepTGIN & molecular graph & Protein residue sequences & pocket residue sequences\\
  \bottomrule
  \end{tabular*}
\end{table*}

\begin{figure}[htbp]
  \centering
  \includegraphics[width=\textwidth]{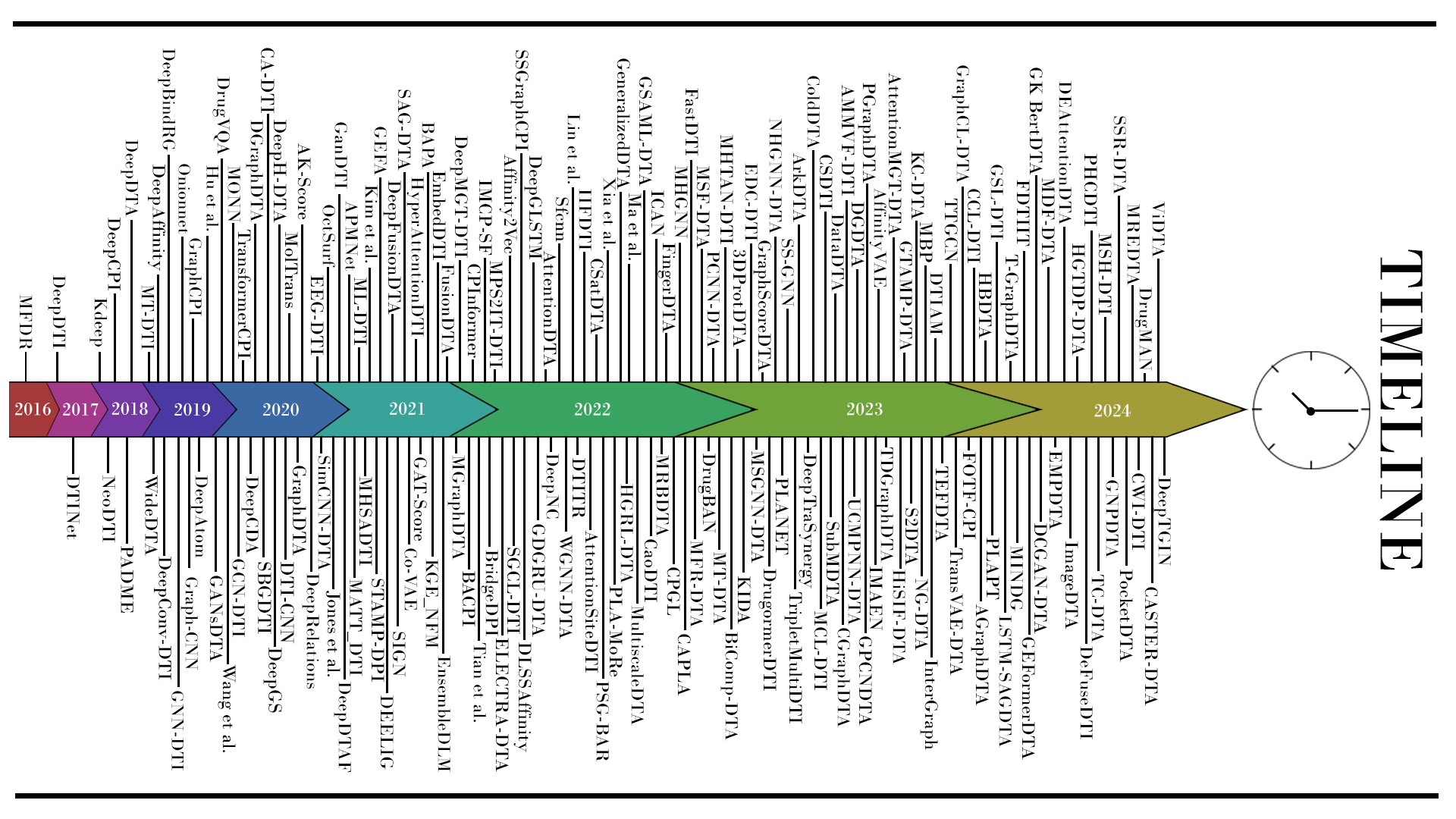}
  \caption{Published date of models timeline}
  \label{fig:fig6}
\end{figure}

\end{document}